\documentclass[aip,reprint,amsmath,amssymb]{revtex4-1} 
\usepackage{graphicx}
\usepackage{dcolumn}
\usepackage{bm}
\usepackage{amsmath}
\usepackage{amssymb}
\usepackage{txfonts}
\usepackage{url}
\usepackage{braket}
\usepackage{multirow}
\usepackage{here}
\usepackage[usenames,dvipsnames]{color}
\definecolor{Gray}{gray}{0.0}
\definecolor{lightGray}{gray}{0.35}
\newcommand{\Pf}[1]{ {\text{Pf}\left( #1 \right)} }
\newcommand{\tvb}{\textsc{TurboRVB}} 
\newcommand{\tvbg}{\textsc{Turbo-Genius}} 
\newcommand{\bvec}[1]{\ensuremath{\mathbf{#1}}}
\newcommand{\bsym}[1]{\ensuremath{\boldsymbol{#1}}}

\makeatletter
\def\Hline{%
\noalign{\ifnum0=`}\fi\hrule \@height 1pt \futurelet
\reserved@a\@xhline}
\makeatother

\begin{document}
\title{\tvb: a many-body toolkit for {\it ab initio}  
electronic simulations by quantum Monte Carlo}
%
%
\author{Kousuke Nakano}
\email{kousuke\_1123@icloud.com}
\affiliation{International School for Advanced Studies (SISSA), Via Bonomea 265, 34136, Trieste, Italy}
\affiliation{Japan Advanced Institute of Science and Technology (JAIST), Asahidai 1-1, Nomi, Ishikawa 923-1292, Japan}
\author{Claudio Attaccalite}
\affiliation{Aix-Marseille Universit\'e, CNRS, CINaM UMR 7325, Campus de Luminy, 13288, Marseille, France}
\author{Matteo Barborini}
\affiliation{CNR-NANO, Via Campi 213/a, 41125 Modena, Italy}
\author{Luca Capriotti}
\affiliation{New York University, Tandon School of Engineering, 6 MetroTech Center, Brooklyn, NY 11201, United States of America}
\affiliation{Department of Mathematics, University College London, Gower Street, London WC1E 6BT, United Kingdom}
\author{Michele Casula}
\email{michele.casula@upmc.fr}
\affiliation{Institut de Min{\'e}ralogie, de Physique des Mat{\'e}riaux et de Cosmochimie (IMPMC), Sorbonne Universit{\'e}, CNRS UMR 7590, IRD UMR 206, MNHN, 4 Place Jussieu, 75252 Paris, France}
\author{Emanuele Coccia}
\affiliation{Department of Chemical and Pharmaceutical Sciences, University of Trieste, Via L. Giorgieri 1, 34127, Trieste, Italy}
\author{Mario Dagrada}
\affiliation{Forescout Technologies, John F. Kennedylaan 2, 5612AB, Eindhoven, The Netherlands}
\author{Claudio Genovese}
\affiliation{International School for Advanced Studies (SISSA), Via Bonomea 265, 34136, Trieste, Italy}

\author{Ye Luo}
\affiliation{Computational Science Division, Argonne National Laboratory, 9700 S. Cass Avenue, Lemont, IL 60439, United States of America}
\affiliation{Argonne Leadership Computing Facility, Argonne National Laboratory, 9700 S. Cass Avenue, Lemont, IL 60439, United States of America}
\author{Guglielmo Mazzola}
\affiliation{IBM Research Zurich, S\"aumerstrasse 4, 8803 R\"uschlikon, Switzerland}%
\author{Andrea Zen}
\affiliation{Department of Earth Sciences, University College London, Gower Street, London WC1E 6BT, United Kingdom}
\affiliation{Thomas Young Centre and London Centre for Nanotechnology, 17-19 Gordon Street, London WC1H 0AH, United Kingdom}
\author{Sandro Sorella}
\email{sorella@sissa.it}
\affiliation{International School for Advanced Studies (SISSA), Via Bonomea 265, 34136, Trieste, Italy}


\date{\today}
\begin{abstract}
\tvb\, is a computational package for  {\it ab initio} Quantum Monte
Carlo (QMC) simulations of both molecular  and bulk electronic
systems. The code implements two types of well established QMC
algorithms: Variational Monte Carlo (VMC),  
and Diffusion Monte Carlo 
in its robust and efficient lattice regularized variant.
A key feature of the code is the possibility of using 
strongly
correlated many-body wave functions, capable of describing several materials  with  very high  
accuracy,  even when standard  
mean-field approaches ({\it e.g.}, density functional theory (DFT)) fail. 
The electronic wave function (WF) 
is obtained by applying a Jastrow factor, which takes into account dynamical correlations,
to the most general mean-field ground state, written either as an antisymmetrized geminal power with spin-singlet pairing, or as a Pfaffian, including both singlet and triplet correlations.
This WF can be viewed as an efficient implementation of the so-called resonating valence bond (RVB) {\it Ansatz}, first proposed by Pauling and Anderson in quantum chemistry [L. Pauling, {\it The Nature of the Chemical Bond} (Cornell University Press, 1960)] and condensed matter physics [P.W. Anderson, Mat. Res. Bull 8, 153 (1973)], respectively.
The 
RVB ansatz 
implemented in \tvb\, 
has a large variational freedom, including the Jastrow
correlated Slater determinant as its simplest, but nontrivial case.
Moreover, it has the remarkable advantage of remaining with  an affordable  
computational cost, proportional 
to the one spent for the evaluation of 
a  single Slater determinant. Therefore, its application to large systems is computationally feasible.
The WF is expanded in a localized basis set. Several basis set
functions are implemented, such as Gaussian, Slater, and mixed types,
with no restriction on the choice of their contraction. The code
implements the adjoint algorithmic differentiation 
that enables a very  efficient evaluation of energy derivatives,
comprising the ionic forces.
Thus, one can perform structural optimizations
and molecular dynamics in the canonical NVT ensemble at the VMC
level. For the electronic part, a full WF optimization (Jastrow and
antisymmetric parts together) is made possible thanks to  
state-of-the-art 
stochastic algorithms
for energy minimization.
In the optimization procedure, the first guess can be obtained at the mean-field level by a built-in DFT driver.  
The code has been efficiently 
parallelized by  using a hybrid MPI-OpenMP protocol, that is also an  ideal environment  for exploiting the computational power 
of modern Graphics Processing Unit (GPU) accelerator. 
\end{abstract}
\maketitle

%
%


%
%


%
%

\tableofcontents

%
%

\section{Introduction}

The solution of the many-body Schr\"{o}dinger equation, which describes the interaction between electrons and ions at the quantum mechanical level,
represents  a fundamental challenge in computational chemistry, condensed matter, and materials science.
Since about a century, there has been a relentless theoretical and computational effort to find an accurate solution to this problem,
which also features many recent interdisciplinary applications in
machine learning~{\cite{2017RAM}} and materials informatics~{\cite{2005RAJ}}.
While computationally, the scaling of the problem is exponential with the number of electrons,  several numerical approximate methods have been put forward in the last decades.
Among them, the Density Functional Theory (DFT) method,
proposed by Kohn and Sham~{\cite{1965KOH}} in 1965, is one of the most successful approaches.
In this framework, the original interacting 3$N$ many-body problem ($N$ being the number of electrons in a system) is mapped to a non-interacting electron system, defined by an effective mean-field potential, to be determined self-consistently~{\cite{2004MAR}}. 
While DFT is an exact theory in principle, 
the exact form for 
 the exchange-correlation functional, which is an essential part of the DFT mean-field potential, remains unknown.
  Unfortunately, the progress in generating increasingly successful approximations of this functional is rather slow~{\cite{2017MED}}, partly because there is no established strategy 
for systematic improvement{\cite{2001PER}}, while maintaining an efficient scaling with the system size.
The commonly adopted  approximations for the exchange-correlation DFT 
functional 
have well-known limitations
especially in describing weak dispersive interactions, strongly correlated materials, and extreme environments ({\it e.g.}, high pressure){\cite{2012BUR,2012COH}}.

\vspace{1mm}
Alternative strategies are represented by the so-called WF-based approaches popular in
quantum chemistry applications{~\cite{1982SZA}}.
In these methods, electronic correlations are captured either
 variationally or perturbatively by post-Hartree-Fock theories, such as 
the M{\o}ller-Plesset perturbation theory (MP)~{\cite{1934MOL}}, configuration interaction (CI) and Full-CI (FCI)~{\cite{1984KNO}}, multi-configurational self-consistent field (MCSCF)~{\cite{1980ROO}}, Coupled-Cluster theory (CC)~{\cite{1966CIZ}}, to name a few.
Among them, 
coupled-cluster with single, double, and perturbative triple excitations, or CCSD(T),
is considered to be the \emph{gold standard} in quantum chemistry as it typically provides 
results in good agreement with experiments, and a reasonable balance between accuracy and computational affordability (despite its cost grows as the seventh power of the number of electrons).
%
While quantum chemistry methods have the advantage of treating  electronic exchange and correlation effects in a systematically improvable fashion, they are also much more computationally demanding compared to DFT,
and their applicability to large or periodic systems is often computationally prohibitive.

\vspace{1mm}
Another way to tackle the problem of the electron  correlation and 
the huge  dimension of a many-body WF,  exponentially  large in the number of electrons,  is by means of stochastic approaches, in  this context  referenced with  
the widely used expression of 
quantum Monte Carlo (QMC) methods\cite{2001FOU,2017BEC}.
Since the  invention of Markov Chain Monte Carlo (MCMC) in the 1940s,\cite{1949ULA}  {\it stochastic} approaches to numerical algorithms had a pervasive influence in a wide range of fields, from physics engineering to finance.
In this respect, the QMC framework is qualitatively different from the {\it deterministic} one mentioned above,
and it represents, therefore, an original and alternative approach for the solution of the Schr\"{o}dinger equation, overcoming some 
of the drawbacks of DFT and the deterministic WF-based approaches of quantum chemistry.
In particular, QMC does not rely on uncontrolled approximations; 
its accuracy can be systematically improved, the scaling with system size is good, and it is straightforwardly applicable to both isolated and periodic systems.
While the scaling with the system size is comparable with standard DFT methods, the prefactor is typically much larger.
However, methods based on stochastic sampling are well suited for 
massively parallel computing architectures, as the algorithms can sustain an almost ideal scaling with the number of cores.
As a result, the feasibility and popularity of QMC are expected to increase with the foreseeable substantial improvements in high-performance computing (HPC) facilities.


%
\vspace{1mm}
\tvb\ includes two of the most popular ground-state QMC algorithms: variational Monte Carlo (VMC) and diffusion Monte Carlo (DMC){~\cite{2001FOU}}.
In VMC, a systematically improvable approximation for the ground state is obtained  by direct minimization of the energy, evaluated by a  parametrized many-body WF.
In an explicitly correlated WF,  the electron coordinates are not separable, and the expectation value of the Hamiltonian  has to be calculated with Monte Carlo integration, hence the name.
%
In realistic calculations, 
a faithful  and nevertheless compact parametrization for the trial state is essential for a successful energy optimization.
The first property allows an unbiased treatment of the electronic correlations across different regimes,
 such as bond dissociation and electronic phase transitions.
The second is needed for a stable optimization of the variational parameters, which may  become  a  too difficult task when  the chosen  parametrization is redundant or unnecessarily detailed.
 Therefore, a central effort, in the \tvb\ project, has been devoted to the 
 development of efficient and systematic 
 parametrizations of correlated WFs.

\vspace{1mm}
As we will see in the following, an accurate trial WF is also fundamental in DMC,
which is the second main algorithm present in \tvb.
DMC is an imaginary-time projection technique{~\cite{2017BEC}}, that, when  combined with the so-called \emph{fixed-node} (FN) approximation, represents a powerful route to electronic ground-state calculations.
In this approximation, the nodal-surface, which is  kept fixed during the projection, can be determined by an accurate variational optimization.


\vspace{1mm}
So far, several groups have implemented the above algorithms and established excellent QMC codes 
such as QMCPACK{~{\cite{2018KIM}}}, CASINO{~{\cite{2009NEE}}}, QWALK{~{\cite{2009WAG}}}, CHAMP{~{\cite{2019UMR}}}, and HANDE-QMC{\cite{2019SPE}}.
We also remark that
other QMC algorithms have also 
been developed
recently, such as the auxiliary field quantum Monte Carlo (AF-QMC){~{\cite{1997ZHA, 2018MOT}}}, 
full-configuration interaction quantum Monte Carlo (FCI-QMC){~\cite{2009BOO,2013BOO}}, coupled cluster Monte Carlo (CCMC){~\cite{2010THO, 2016SPE}}, density matrix quantum Monte Carlo (DMQMC)~{\cite{2014BLU, 2015MAL}}, model space quantum Monte Carlo (MSQMC){~{\cite{2013TEN, 2015OHT, 2017TEN}}}, clock quantum Monte Carlo (CQMC){\cite{2015MCC}}, or driven-dissipative quantum Monte Carlo (DDQMC){~\cite{2018NAG}}. 

\vspace{1mm}
A typical workflow for a QMC calculation performed with \tvb\ is shown in Fig.~\ref{fig:workflow}.
It shall be noticed that the choice of the most suited ansatz for the WF, as well as the optimization of its parameters, is a prerequisite of both VMC and DMC evaluations. Ionic forces, if evaluated, can be used to perform QMC-based structural relaxations, Langevin molecular dynamics (MD) or path integral molecular dynamics (PIMD). Moreover, several post-processing tools are available to analyze the outcomes. 
The \tvb\ code   
is peculiar because it relies on the assumption that a complex electronic problem, 
though,  until now, cannot be solved exactly with a black-box  computer software, 
can be described very accurately  by approriate variational 
wavefunctions, guided  by  robust physical and chemical requirements and human ingenuity.  
In particular  \tvb\ is different from other codes  
 in the following features:
$\rm(\hspace{.18em}i\hspace{.18em})$ 
It employs the Resonating Valence Bond (RVB) WF,
first proposed by Pauling and Anderson in 
quantum chemistry~{\cite{1960PAU}
and condensed matter physics~{\cite{1973AND}}, 
respectively.
It includes 
static and dynamical correlation effects 
beyond the commonly used Slater determinant, 
while keeping
the computational cost 
at the single-determinant level,
thanks to its efficient
implementation.
$\rm(\hspace{.08em}ii\hspace{.08em})$ 
The code implements 
a VMC algorithm based on 
localized orbitals ({\it e.g.}, Gaussians) and 
state-of-the-art optimization methods,
such as 
the stochastic reconfiguration. Therefore, at the VMC level, one can optimize not only the amplitude of the WF ({\it i.e.}, the Jastrow factor), but also the nodal surfaces ({\it e.g.}, 
the Slater determinant). This 
leads  to
a better variational energy  in general, and also improves the corresponding 
FN-DMC
energy.
$\rm(i\hspace{-.08em}i\hspace{-.08em}i)$
The energy derivatives
({\it e.g.}, atomic forces) are calculated very efficiently thanks to an 
implementation 
based on
the Adjoint Algorithmic Differentiation (AAD). As  a  consequence, one can perform structural optimizations and Langevin molecular dynamics.
$\rm(i\hspace{-.08em}v\hspace{-.06em})$
The code implements the newly developed 
Lattice Regularized Diffusion Monte Carlo (LRDMC), a 
stable DMC
algorithm
that, very recently, has shown  to have  
a better scaling with 
the atomic number $Z$, compared with standard DMC.{~\cite{2019NAK2}}

\vspace{1mm}
This review is organized as follows: in Sec.{{~\ref{methods}}}, we briefly explain the 
fundamental
QMC algorithms implemented in \tvb, namely, 
the VMC and LRDMC;
in Sec.{{~\ref{wavefunction}}, we describe the 
RVB
WF;
in Sec.{{~\ref{periodic_system}}, we 
extend the
WF 
to treat
periodic systems; in Sec.{{~\ref{sec_dft}} we introduce the   DFT algorithm  efficiently implemented in  \tvb;
in Sec.{{~\ref{derivatives}}, we describe the way  
energy derivatives are computed
({\it e.g.}, atomic forces) by means of AAD;
in Sec.{{~\ref{sec_opt_wf}}, we 
list the \tvb, steps 
to optimize a many-body WF; in Sec.{{~\ref{molecular_dynamics}}, we introduce the first and the second-order Langevin molecular dynamics implemented in \tvb; 
in Sec.{{~\ref{implementation_operation}}, we 
summarize
the typical QMC workflow from the WF generation to the final LRDMC calculation;
in Sec.{{~\ref{benchmark}}, we show weak and strong scaling results of \tvb, measured on the Marconi/CINECA supercomputer;
in Sec {{~\ref{postprocessing}}, we list the physical properties that can be calculated by \tvb;
in Sec.{{~\ref{applications}}, 
we 
review
the major \tvb\, applications done so far; in Sec.{{~\ref{gturbo}}, we introduce 
a
python-based workflow system, named \tvbg.

\begin{widetext}
\begin{figure*}[htbp]
\includegraphics[width=7in]{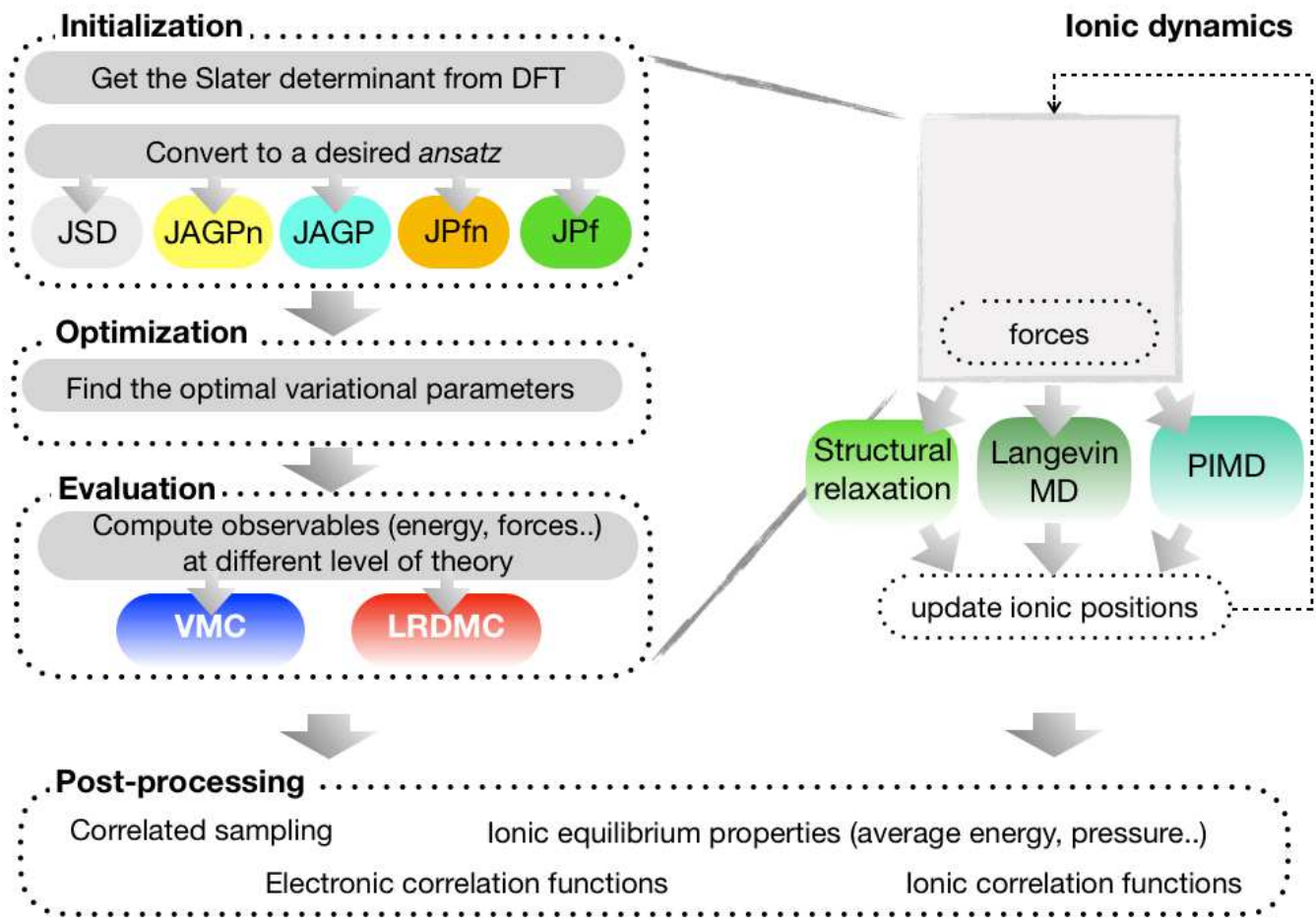}
 \caption{Schematic picture of some  of the possible tools available  in TurboRVB. The arrows indicate  typical workflow calculations.}
 \label{fig:workflow}
\end{figure*}
\end{widetext}

%
%


\begin{center}
\begin{table}[htbp]
\caption{\label{notations} Summary of the notations used in this review.}
\vspace{1mm}
\begin{tabular}{c|p{6.8cm}}
\Hline
 Notation & Description \\ 
\Hline
%
 ${\mathbf{r}_i}$ & $i$-th electron coordination. \\
 ${\mathbf{R}_a}$ & $a$-th ion coordination. \\
 ${\sigma_i}$ & spin of $i$-th electron. \\
 ${\mathbf{x}_i}$ & $N$ set of electron coordinations including spins $\equiv$ $\left( {{{\mathbf{r}}_1}{\sigma _1},{{\mathbf{r}}_2}{\sigma _2}, \ldots {{\mathbf{r}}_N}{\sigma _N}} \right)_{i}$. The index $i$ referes Monte Carlo sampling index. \\
${\mathbf{x}}$ & $N$ set of electron coordinations including spins, the same as ${\mathbf{x}_i}$. \\
\hline
%
 $\mathbf{i}$, $\mathbf{j}$ & compact notation for  (${\mathbf{r}_i}, {\sigma_i}$) and (${\mathbf{r}_j}, {\sigma_j}$), respectively. \\
$\int d\mathbf{i}$ & compact notation for $\sum_{\sigma_i} \int d\mathbf{r}_i.$ \\ 
 $a,b,I,J$ & indices of atoms. \\
 $i,j$ & indices of electrons / orbitals / Monte Carlo sampling. \\
 $\mu$, $\nu$ & indices of orbitals. \\
 $l,m$ & indices of orbitals. \\
\hline
%
 $M$ & the number of configurations of electrons (the number of Monte Carlo sampling). \\
 $N_\textrm{at}$ & the number of atoms. \\
 $N$ & the number of electrons. \\
 $N_u$ & the number of electrons with spin up. \\
 $N_d$ & the number of electrons with spin down. \\
\hline
%
 ${{\alpha _1}, \cdots ,{\alpha _p}}$ & variational parameters of the trial WF \\
 $\alpha_{k}$ & $k$-th variational parameter. \\
\hline
%
 ${\Psi}$ & many-body WF ansatz. $\Psi  =  \Phi _\text{AS} \times \exp J$. \\
 ${\cal A}$ & antisymmetrization operator. \\
 $\Phi_\text{AS}$ & antisymmetric (AS) part. \\
 $\Phi _\text{SD}$ & single Slater determinant (SD). \\
 $\Phi _\text{AGP}$ & 
antisymmetrized Geminal Power (AGP). \\
 $\Phi _\text{Pf}$  & pfaffian (Pf). \\
 $J$ & Jastrow factor, composed of $J_1$, $J_2$, and $J_{3/4}$. \\
 ${\Psi _{\text{T}}}$ & many-body trial WF. \\
 ${\Psi _{\text{G}}}$ & many-body guiding function. \\
\hline
%
 $G$ & $N\times N$ matrix with elements $G_{i,j}=g(\bvec{i},\bvec{j})$. \\
 $\Phi$ & $N$ $\times$ $k$ matrix with elements $\Phi_{i,k}={{\phi }_k}\left( \bvec{i} \right)$. \\
 $g \left( \bvec{i},\bvec{j} \right)$ & pairing function between electrons $\bvec{i}\leftrightarrow ({\bf r}_i, \sigma_i)$ and $\bvec{j}\leftrightarrow ({\bf r}_j, \sigma_j)$. \\
 $f\left( {\mathbf{r}}_{i},{\mathbf{r}}_{j} \right)$ & spatial part of the pairing function. \\
${\psi _i}\left( {\mathbf{r}} \right)$ & primitive atomic orbital for the antisymmetric part. \\
${{\phi }_k}\left( {\mathbf{r}} \right)$ & $k$-th molecular orbital for the antisymmetric part. \\
${\chi _i}\left( {\mathbf{r}} \right)$ & primitive atomic orbital for the Jastrow part. \\
$\mathbf{S}$ & overlap matrix with elements $S_{\mu,\nu} \equiv \langle \psi_\mu | \psi_\nu \rangle$. \\
\hline
%
 ${f_i}$ & general force for $i$-th variational parameter ${f_i}$ $\equiv$ - $\frac{{\partial E}}{{\partial {\alpha _i}}}$. \\
 ${{\mathbf{F}}_a}$ & atomic force for an atom $a$, ${{\mathbf{F}}_a}$ $\equiv$ $- \frac{{dE}}{{d{{\mathbf{R}}_a}}}$. \\
 $ {O_k}\left( {{{\mathbf{x}}}} \right) $ & the logarithmic derivative of a many-body WF ${O_k}\left( {{{\mathbf{x}}}} \right)$ $\equiv$ $\frac{{\partial \ln \Psi \left( {{{\mathbf{x}}}} \right)}}{{\partial {\alpha _k}}}$. \\
${\boldsymbol{\mathcal{S}}}$  & variance-covariance matrix of the logarithmic derivative of a many-body WF. \\
${\boldsymbol{\mathcal{S}}}^{{{\text{QMC}}}}\left( {\mathbf{R}} \right)$  & variance-covariance matrix of QMC forces. \\
\hline
\Hline
\end{tabular}
\end{table}
\end{center}

\section{Methods}
\label{methods}
\tvb\, implements two types of well established Quantum Monte Carlo methods: Variational Monte Carlo (VMC) and Lattice Regularized Diffusion Monte Carlo (LRDMC). We summarize these methods in this section. The interested readers should also refer to the comprehensive review of QMC~{\cite{2001FOU}} for details. \tvb\, also implements an original DFT engine to generate trial WFs, which is more suitable for the VMC and LRDMC calculations, as shown in Sec.{~\ref {sec_dft}}.

\subsection{Variational Monte Carlo}
Starting from the variational principle, the expectation value of the energy 
evaluated for  a given WF $\Psi$ can be written as:
\begin{equation}
\left\langle E \right\rangle  = \frac{{\int {d{\mathbf{x}}{\Psi ^2}\left( {\mathbf{x}} \right) \cdot \hat {\mathcal{H}} \Psi \left( {\mathbf{x}} \right)/\Psi \left( {\mathbf{x}} \right)} }}{{\int {d{\mathbf{x}}{\Psi ^2}\left( {\mathbf{x}} \right)} }} = \int {d{\mathbf{x}}{e_L}\left( {\mathbf{x}} \right)\pi \left( {\mathbf{x}} \right)},
\end{equation}
where ${\mathbf{x}} = \left( {{{\mathbf{r}}_1}{\sigma _1},{{\mathbf{r}}_2}{\sigma _2}, \ldots {{\mathbf{r}}_N}{\sigma _N}} \right)$ here and henceforth is  a shorthand notation for the   $N$ electron coordinates and their spins, whereas 
$$
{e_L}\left( {\mathbf{x}} \right) \equiv {\hat {\mathcal{H}}\Psi \left( {\mathbf{x}} \right) \over \Psi \left( {\mathbf{x}} \right)} 
\quad , \text{and} \quad 
\pi \left( {\mathbf{x}} \right) \equiv { {\Psi ^2}\left( {\mathbf{x}} \right) \over  \int {d{\mathbf{x'}}{\Psi ^2}\left( {\mathbf{x'}} \right)} }, 
$$ 
are the so-called local energy and the probability of the configuration 
$\mathbf{x}$,  respectively.
This multidimensional integration can be evaluated stochastically by generating a set $\left\{ {{{\mathbf{x}}_i}} \right\}$ according to the distribution $\pi \left( {\mathbf{x}} \right)$ using the Markov chain Monte Carlo such as the (accelerated{~\cite{1993UMR, 1998STE}}) Metropolis method, and by averaging the obtained local energies ${e_L}\left( {{{\mathbf{x}}_i}} \right)$:
\begin{equation}
{E_{{\text{VMC}}}} = {\left\langle {{e_L}\left( {\mathbf{x}} \right)} \right\rangle _{\pi \left( {\mathbf{x}} \right)}} \approx \frac{1}{M}\sum\limits_{i = 1}^M {{e_L}\left( {{{\mathbf{x}}_i}} \right)},
\end{equation}
%
which has an associated statistical error of $\sqrt{{\text{Var}[e_L(\mathbf{x}_i)] / \tilde M}}$, 
where $\text{Var}[e_L(\mathbf{x}_i)]$ is the variance of the sampled local energies, and $\tilde M$ is the sampling size $M$ divided by the autocorrelation time. This indicates that the precision of the VMC evaluation is inversely proportional to the square root of the number of samplings ({\em i.e.}, of the computational cost). 
It worth to notice that, if $\Psi(\bvec{x})$ is an eigenfunction of $\hat {\mathcal{H}}$, say with eigenvalue $E_0$, then $e_L(\bvec{x}) = E_0$ for each $\bvec{x}$, implying  that  the variance of the local energy is zero and $E_\text{VMC} = E_0$ with no stochastic uncertainty. This feature is known as the zero-variance property.

\vspace{1mm}
The probability distribution used for the importance sampling can also differ from $\pi \left( {\mathbf{x}} \right)$.
Indeed, one can use an arbitrary probability distribution function $\pi '\left( {\mathbf{x}} \right) = \Psi _{\text{G}}^2\left( {\mathbf{x}} \right)/\int {d{\mathbf{x}}\Psi _{\text{G}}^2\left( {\mathbf{x}} \right)}$, and estimate a generic local observable $O\left( {\mathbf{x}} \right)$ either by using 
$
{{\bar O}_{{\text{VMC}}}} 
= {\left\langle {O\left( {\mathbf{x}} \right)} \right\rangle _{\pi \left( {\mathbf{x}} \right)}} 
$ 
or:
\begin{equation}
{{\bar O}_{{\text{VMC}}}} 
= \frac{{{{\left\langle {O\left( {{\mathbf{x'}}} \right)\mathcal{W}\left( {{\mathbf{x'}}} \right)} \right\rangle }_{\pi '\left( {{\mathbf{x'}}} \right)}}}}{{{{\left\langle {\mathcal{W}\left( {{\mathbf{x'}}} \right)} \right\rangle }_{\pi '\left( {\mathbf{x}} \right)}}}} 
\approx \frac{{\sum\nolimits_{i = 1}^{M'} {O\left( {{{{\mathbf{x'}}}_i}} \right)\mathcal{W}\left( {{{{\mathbf{x'}}}_i}} \right)} }}{{\sum\nolimits_{i = 1}^{M'} {\mathcal{W}\left( {{{{\mathbf{x'}}}_i}} \right)} }},
\end{equation}
where $\mathcal{W}\left( {{{{\mathbf{x}}}}} \right) = |{\Psi }\left( {\mathbf{x}} \right)/\Psi _{\text{G}}\left( {\mathbf{x}} \right)|^2$,
and the points $\bvec{x'}_i$ are distributed according to $\pi'$. This {\it reweighting} scheme is very important when evaluating atomic forces, as discussed in Sec.~\ref{ionic_forces}. Since the evaluations of the standard deviations is nontrivial in this case, \tvb\ employs the bootstrap and jackknife methods in order to estimate the mean value and the statistical error{~\cite{2017BEC}}, which are also used when evaluating those of the local energy, forces, and so on. Indeed, the code outputs the history of the local energies, forces, or other properties in appropriate  files (when the corresponding option is true), thus allowing the error bar estimates  by simple post-processing.  
The user can also use the reblocking (binning) technique to remove the autocorrelation bias.\cite{1989FLY}

\vspace{1mm}
One can optimize the WF based on the \emph{variational theorem} by introducing a set of parameters $\left( {{\alpha _1},{\alpha _2}, \cdots ,{\alpha _p}} \right)$ to the WF ${\Psi \left( {{\mathbf{x}},\mathbf{\alpha} } \right)}$:
\begin{equation}
{E_{{\text{VMC}}}}\left( \alpha  \right) = \int {d{\mathbf{x}}{e_L}\left( {{\mathbf{x}},\alpha } \right)\pi \left( {{\mathbf{x}},\alpha } \right)}  \geqslant {E_{{\text{exact}}}}.
\end{equation}
However, the  optimization of a many-body WF remains a difficult challenge 
not only because optimizing a cost function containing many parameters 
is a complex numerical task, due to the presence of 
several local minima  in the energy landscape,  
but also because  this difficult task is further complicated by the presence  of statistical errors in the  QMC evaluation of any quantity.
Nevertheless, a great improvement in this field has been achieved 
when the QMC optimization techniques have made use of the explicit evaluation 
of energy derivatives with finite statistical errors.
In particular, in  \tvb\, the adjoint algorithmic differentiation (AAD) has been implemented,  by allowing  the efficient calculation of  generalized forces (${f_i} = -\frac{{\partial E}}{{\partial {\alpha _i}}}$){~\cite{2010SOR}}, and very efficient optimization methods, the so-called ``stochastic reconfiguration"~{\cite{1998SOR, 2007SOR}} and ``the linear method"~{\cite{2005SOR, 2007UMR, 2007TOU}}. These methods are discussed in Sec.~\ref{sec:SR} and \ref{sec:linearmethod}.

\subsection{Lattice regularized Monte Carlo}

Lattice regularized diffusion Monte Carlo (LRDMC), which was initially proposed by M. Casula {\it et al.}~{\cite{2005CAS}}, is a projection technique that allows us to improve a variational ansatz systematically. This method is based on Green's function Monte Carlo (GFMC){~\cite{1995TEN, 1998BUO, 2000SOR}}, filtering out  the ground state WF ${\left| {{\Upsilon _0}} \right\rangle }$ from a given  trial WF $\left| {{\Psi _{\text{T}}}} \right\rangle$: since the eigenstates of the Hamiltonian have the completeness property, the trial WF can be expanded as:
\begin{equation}
\left| {{\Psi _{\text{T}}}} \right\rangle  = \sum\limits_n {{a_n}\left| {{\Upsilon _n}} \right\rangle },
\end{equation}
where ${a_n}$ is the coefficient for the $n$-th eigenvectors (${{\Upsilon _n}}$). Therefore, by applying ${{\mathcal{G}}^M} = {\left( {{\boldsymbol{\Lambda}}  - \hat {\boldsymbol{\mathcal{H}}}} \right)^M}$, one can obtain
\begin{equation}
\begin{split}
\left| {{\Upsilon _0}} \right\rangle &\propto \mathop {\lim }\limits_{M \to \infty } {\left( {{\boldsymbol{\Lambda}}  - \hat {\mathcal{H}}} \right)^M}\left| {{\Psi _{\text{T}}}} \right\rangle  \\
& = \mathop {\lim }\limits_{M \to \infty } {\left( {\lambda  - {E_0}} \right)^M}\left[ {{a_0}\left| {{\Upsilon _0}} \right\rangle  + \sum\limits_{n \ne 0} {{{\left( {\frac{{\lambda  - {E_n}}}{{\lambda  - {E_0}}}} \right)}^M}{a_n}\left| {{\Upsilon _n}} \right\rangle } } \right],
\end{split}
\end{equation}
where ${\boldsymbol \Lambda}$ is a diagonal matrix with ${{\Lambda _{x,x^\prime}}} = \lambda \delta_{x,x^\prime}$ ($\lambda$ should be sufficiently large to obtain the ground state), and ${{E_n}}$ is $n$-th eigenvalue of $\hat {\mathcal{H}}$. Since $\frac{{\lambda  - {E_n}}}{{\lambda  - {E_0}}} < 1$, the projection filters out the ground state WF ${{\Upsilon _0}}$ from a given trial WF $\left| {{\Psi _{\text{T}}}} \right\rangle$, as long as the trial WF is not orthogonal to the true ground state ({\it i.e.}, ${a_0} \equiv \left< \Upsilon_0 | \Psi _{\text{T}} \right> \ne 0$).
To apply the GFMC for {\it ab initio} electron calculations, the original continuous Hamiltonian is regularized by allowing electron hopping with step size $a$, in order to mimic the electronic  kinetic energy.
The corresponding Hamiltonian ${{\hat{\mathcal{H}}}^a}$  is then defined  such that ${{{\hat {\mathcal{H}}}^a}} \to {\hat {\mathcal{H}}}$ for $a \to 0$. Namely, the kinetic part is approximated by a finite difference form:
\begin{equation}
\begin{split}
{\Delta _i}f\left( {{x_i},{y_i},{z_i}} \right) &\approx \Delta _i^af\left( {{x_i},{y_i},{z_i}} \right) \\
&\equiv \frac{1}{{{a^2}}}\left\{ {\left[ {f\left( {{x_i} + a} \right) - f\left( {{x_i}} \right)} \right] + \left[ {f\left( {{x_i} - a} \right) - f\left( {{x_i}} \right)} \right]} \right\} \\ 
& \leftrightarrow {y_i} \leftrightarrow {z_i},
\end{split}
\end{equation}
and the potential term is modified as:
\begin{equation}
{V^a}\left( {\mathbf{x}} \right) = V\left( {\mathbf{x}} \right) + \frac{1}{2}\left[ {\frac{{\sum\nolimits_i {\left( {\Delta _i^a - {\Delta _i}} \right) \Psi_\text{G}\left( {\mathbf{x}} \right)} }}{{{\Psi _{\text{G}}}\left( {\mathbf{x}} \right)}}} \right].
\label{lrdmc_potential}
\end{equation}
The corresponding Green's function matrix elements are:
\begin{equation}
{{\mathcal{G}}_{{x',x}}} = 
\braket{{x'}|( {\boldsymbol{\Lambda}}  - {\boldsymbol{{\hat{\mathcal{H}}}}}^a )|{x}} \equiv
\left( {\boldsymbol{\Lambda}}  - {\boldsymbol{{{\mathcal{H}}}}}^a \right)_{{x',x}}
\end{equation}
and the single LRDMC iteration step is given by the following equation:
\begin{equation}
{\Psi _{n + 1}}\left( {x'} \right) = \sum\limits_x {{{\mathcal{G}}_{x',x}}{\Psi _n}\left( x \right)}.
\label{single-step-LRMDC}
\end{equation}
The sketch  of the  LRDMC algorithm, a Markov chain that evolves the many-body WF according to the Eq.{~\ref{single-step-LRMDC}}, is as follows{~\cite{2005CAS}}: (STEP 1) Prepare a walker with configuration $x$ and weight $w$ ($w_{0}$ = 1). (STEP 2) A new configuration $x'$ is generated by the transition probability:

\begin{equation}
{p_{x',x}} = {{\mathcal{G}}_{x',x}}{\rm{ /}}{b_x},
\label{lrdmc_transition_p}
\end{equation}
where
\begin{equation}
{b_x} = \sum\nolimits_{x'} {{{\mathcal{G}}_{x',x}}}
\label{lrdmc_normalization_b}
\end{equation}
is a normalization factor.
By applying the discretized Hamiltonian to a given configuration (${{\hat {\mathcal{H}}}^a} \ket{x}$),  $(6N+1)$ configurations $\ket{x'}$ are determined  according to the probability ${p_{x',x}}$ in Eq.~{\ref{lrdmc_transition_p}}, where $N$ is the number of electrons in the  system~{\cite{2005CAS_phdthesis}}. This allows the evaluation of the normalization factor ${b_x}$ in Eq.{~\ref{lrdmc_normalization_b}} even in a continuous model. Notice that $6N$ comes from the diffusion of each electron in two directions ($\pm a$) and the remaining $1$ stands for the starting configuration $x$ before the possible hoppings (all $N$ electrons) ({\it i.e.}, $x'=x$). (STEP 3) Finally, update the weight with ${w_{n + 1}} = {w_n}{b_x}$, and return to the STEP I.
After a sufficiently large number of iterations (the Markov process is equilibrated), one can calculate the ground state energy ${E_0}$:
\begin{equation}
{E_0} \approx \frac{{\left\langle {{w_n}{e_L}\left( {{x_n}} \right)} \right\rangle }}{{\left\langle {{w_n}} \right\rangle }},
\label{independent-walkers-energy}
\end{equation}
where ${\left\langle {\cdots} \right\rangle}$ denotes the statistical average over many independent samples generated by the Markov chain, and ${e_L}\left( x \right)$ is called the (bare) local energy that reads:
\begin{equation}
{e_L}\left( x \right) = \sum\limits_{x'} {{{\mathcal{H}}_{x',x}} = \lambda  - {b_x}}.
\label{local-energy}
\end{equation}
Indeed, the ground state energy can be calculated after many independent $n$-step calculations. A more efficient computation can be realized by using the so-called "correcting factor" technique: after a single simulation that is much larger than the equilibration time, one can imagine starting a projection of length $p$ from each  $(n-p)^{th}$  iteration. The accumulated weight for each projection is:
\begin{equation}
{\mathcal{G}}_n^p = \prod\limits_{j = 1}^p {{b_{n - j}}}.
\label{single-walker-weight}
\end{equation}
Then, the ground state energy ${E_0}$ can be estimated by:
\begin{equation}
{E_0} \approx \frac{{\sum\nolimits_n {{\mathcal{G}}_n^p{e_L}\left( {{x_n}} \right)} }}{{\sum\nolimits_n {{\mathcal{G}}_n^p} }}.
\label{single-walker-energy}
\end{equation}
This straightforward implementation of the above simple method is not suitable for realistic simulations due to fluctuations of weights, large correlation times, the sign problems, and so on. \tvb\, implements the following state-of-art techniques for real electronic structure calculations.

%
\vspace{1mm}
If the potential term (Eq.~{\ref{lrdmc_potential}}) is unbounded (it is the case in {\it ab initio} calculations), the bare weight ${b_x}$ (Eq.~{\ref{lrdmc_normalization_b}}) and the local energy ${e_L}\left( x \right)$ (Eq.~{\ref{local-energy}}) significantly fluctuate, making the numerical simulation very unstable and inefficient. To overcome this difficulty, the code employs the importance sampling scheme~{\cite{2017BEC}}, in which the original Green's function is modified using the so-called guiding function ${\Psi _{\text{G}}}$ as:
\begin{equation}
{{\tilde {\mathcal{G}}}_{x',x}} = {{\mathcal{G}}_{x',x}}\frac{{{\Psi _{\text{G}}}\left( {x'} \right)}}{{{\Psi _{\text{G}}}\left( x \right)}},
\label{is-green-function}
\end{equation}
and the projection is modified as:
\begin{equation}
{\Psi _{\text{G}}}\left( {x'} \right){\Psi _{n + 1}}\left( {x'} \right) = \sum\limits_x {{{\tilde {\mathcal{G}}}_{x',x}}{\Psi _{\text{G}}}\left( x \right){\Psi _n}\left( x \right)}.
\label{is-projection}
\end{equation}
In practice, the guiding function is prepared by a VMC calculation. The modified Green's function for importance sampling ${{\tilde {\mathcal{G}}}_{x',x}}$ has the same
eigenvalues as the original one, and this transformation does not change the formalism of LRDMC. The weight is updated by:
\begin{equation}
{{\tilde b}_x} = \sum\limits_{x'} {{{\tilde {\mathcal{G}}}_{x',x}}},
\label{lrdmc_mod_weight}
\end{equation}
and the local energy with importance sampling is:
\begin{equation}
{{\tilde e}_L}\left( x \right) = \frac{\braket{{\Psi _{\text{G}} | \hat {\mathcal{H}} | x}}} {{\braket{\Psi _{\text{G}} | x}}} = \sum\limits_{x'} {{{\mathcal{H}}_{x',x}}\frac{{\Psi _{\text{G}}^{}\left( {x'} \right)}}{{\Psi _{\text{G}}^{}\left( x \right)}}}.
\label{lrdmc_mod_local_energy}
\end{equation}
Eq.~{\ref{lrdmc_mod_local_energy}} implies that if the guiding function ${\Psi _{\text{G}}}$ is an exact eigenstate of the Hamiltonian, there are no statistical fluctuations, implying the zero variance property, namely the computational 
efficiency to obtain  a given statistical error on the energy
improves with the quality of the variational WF. 
In this respect, it is also important to emphasize that a meaningful reduction 
of the statistical fluctuations is obtained by satisfying the so-called 
cusp conditions.
As long as  they are satisfied, 
the resulting local energy  does not diverge
at the coalescence points where two particles are overlapped,
 despite the singularity of the Coulomb potential term ($V(x)$ in Eq.~{\ref{lrdmc_potential}})~{\cite{2005CAS_phdthesis}}. 
In addition, the importance sampling maintains the  electrons in a region  away from the nodal surface, since the guiding function vanishes there ({\it i.e.}, the RHS of  Eq.{\ref{is-green-function}} $\to$ 0).
This clearly enhances the efficiency of the sampling 
because the local energy diverges at the nodal surface. 

\vspace{1mm}
The Green's function cannot be made strictly positive for fermions; therefore, the fixed-node (FN) approximation 
has to  be introduced{~\cite{2017BEC}} in order to avoid the sign problem. Indeed, the Hamiltonian is modified using the spin-flip term ${\mathcal{V}_{{\rm{sf}}}}\left( {x} \right) = \sum\limits_{x':{s_{x',x}} > 0}^{} {{{\mathcal{H}}_{x',x}}{\Psi _\text{G}}\left( {x'} \right)} /{\Psi _\text{G}}\left( {x} \right)$:
\begin{equation}
{\mathcal{H}}_{x',x}^{\text{FN},\gamma } = 
 \begin{cases}
  {{\mathcal{H}}_{x,x}} + \left( {1 + \gamma } \right){\mathcal{V}_{{\rm{SF}}}}\left( {x} \right)\,\,\,\,{\rm{for}}\,\,\,\,x' = x,\\
  {{\mathcal{H}}_{x',x}}\,\,\,\,\,\,\,\,\,\,\,\,\,\,\,\,\,\,\,\,\,\,\,\,\,\,\,\,\,\,\,\,\,\,\,\,\,\,\,\,{\rm{for}}\,\,\,\,x' \ne x,{s_{x',x}} < 0,\\
  - \gamma {{\mathcal{H}}_{x',x}}\,\,\,\,\,\,\,\,\,\,\,\,\,\,\,\,\,\,\,\,\,\,\,\,\,\,\,\,\,\,\,\,\,{\rm{for}}\,\,\,\,x' \ne x,{s_{x',x}} > 0,
\end{cases}
\end{equation}
where ${s_{x',x}} = {\Psi _\text{G}}\left( {x'} \right){{\mathcal{H}}_{x',x}} / {\Psi _\text{G}}\left( {x} \right)$ and $\gamma \ge 0$ is a real parameter. The use of the fixed-node Green's function:
\begin{equation}
\tilde {\mathcal{G}}_{{x',x}}^{{\text{FN}}} = \left( {\boldsymbol{\Lambda}  - {\mathcal{H}}^{{\text{FN}}}} \right)_{x',x}\frac{{\Psi _{\text{G}}^{}\left( {x'} \right)}}{{\Psi _{\text{G}}^{}\left( x \right)}}
\end{equation}
can prevent the crossing of regions where the configuration space yields a sign flip of the Green's function; therefore, the walkers are constrained in the same nodal pockets, and avoid the sign problem. 

%
\vspace{1mm}
\tvb\, also implements the many-walker technique and the branching (denoted as reconfiguration~{\cite{1998BUO}} in \tvb) scheme for a more efficient computation~{\cite{2017BEC}}. 
The code performs the branching as follows:
(1) Set the new weights equal to the average of the old ones:
\begin{equation}
{{w'}_{\alpha ,n}} = \bar w \equiv \frac{1}{{{N_w}}}\sum\limits_\beta  {{w_{\beta ,n}}}.
\label{branching-1}
\end{equation}
(2) Select the new walkers from the old ones with a probability that is proportional to the old walkers' weights:
\begin{equation}
{p_{\alpha ,n}} = \frac{{{w_{\alpha ,n}}}}{{\sum\limits_\beta  {{w_{\beta ,n}}} }},
\label{branching-2}
\end{equation}
which does not change the statistical average of weights, but suppresses the fluctuations by dropping walkers having small weights. The code performs branching (reconfiguration) after a projection time $\tau$, that can be chosen as a user input  parameter. In practice, within the many-walker and branching schemes, the average weights are stored and are set to one for all walkers after each branching.
The user can retrieve the accumulated weights at the end of  the simulation:
\begin{equation}
{\mathcal{G}}_n^p = \prod\limits_{j = 1}^p {{{\bar w}_{n - j}}},
\label{many-walker-weight}
\end{equation}
and calculate the ground state energy:
\begin{equation}
{E_0} \approx \frac{{\sum\nolimits_n {{\mathcal{G}}_n^p{e_L}\left( {{x_n}} \right)} }}{{\sum\nolimits_n {{\mathcal{G}}_n^p} }},
\end{equation}
where ${e_L}\left( {{x_n}} \right)$ is the mean local energy averaged over the walkers, which reads:
\begin{equation}
{e_L}\left( {{x_n}} \right) = \frac{{\sum\nolimits_\alpha  {{w_{\alpha ,n}}{e_L}\left( {{x_{\alpha ,n}}} \right)} }}{{\sum\nolimits_\alpha  {{w_{\alpha ,n}}} }}
\end{equation}
and  is evaluated just before each  reconfiguration. Notice that $p$ is also an input parameter, that has to be carefully chosen by the user 
to allow energy convergence.

%
\vspace{1mm}
When $\lambda$ is sufficiently large, the correlation time also becomes large because the diagonal terms of the Green's function become very close to one ({\it i.e.}, a walker remains in the same configuration), which causes a very large correlation time. In \tvb,  this difficulty is solved by  considering in a different way the diagonal and non-diagonal moves.
In a given interval of $M$ iterations, The non-diagonal updates are efficiently calculated using a random number according to the probability that the configuration remains in the same one (diagonal updates). This technique can be generalized to the continuous-time limit, namely, $M \to \infty$, at ${M\over \Lambda}=\tau$ fixed. In the $M \to \infty$ limit, the projection ${\left( {\Lambda  - \hat {\mathcal{H}}} \right)^M}$ is equal to the imaginary time evolution $\exp \left( { - \tau \hat {\mathcal{H}}} \right)$, apart for an irrelevant constant $\Lambda^M$.
Thus the user should specify only  $\tau$ as  input parameter. Indeed, the walker weight is updated by $w \to w\exp \left( { - {\tau _\xi }{e_L}\left( x \right)} \right)$ and the imaginary time is updated ${\tau _{{\text{left}}}} \to {\tau _{{\text{left}}}} - {\tau _\xi }$ at each non-diagonal update until ${\tau _{{\text{left}}}}$ becomes 0, where ${\tau _\xi} = - \log \left( 1 - {\xi} \right)/{b_x}$ is a diagonal move time step determined by a uniform random number $0 \le \xi < 1$. The branching (reconfiguration) is performed after each projection time of length $\tau$ that a user puts in the input file within the many walker and the branching implementation.

%
\vspace{1mm}
In practice, there are three important features in LRDMC.
First, there is not a time-step error in LRDMC because the Suzuki-Trotter decomposition is not necessary, unlike the standard DMC algorithm{~\cite{2017BEC}}.
Instead, there is a finite-size lattice error due to the discretization of the Hamiltonian ($a$). Therefore, in order to obtain an unbiased FN energy, 
it is important to  extrapolate the LRDMC  energy to the  $a \to 0$ limit by  using several results corresponding to different lattice spaces~{\cite{2005CAS}}. This is then
consistent with the 
standard DMC energy estimate (Fig.{~\ref{2005CAS_lrdmc_ext}}) obtained in the limit of an infinitely small time step.
Probably one of the most important advantages of the LRDMC method 
is that the extrapolation  to the $a\to 0$ limit is very smooth and reliable,
so that unbiased FN energies are easily obtained with low order polynomial fits.
Secondly, LRDMC can straightforwardly handle
different length scales of a WF by introducing different mesh sizes ($a$ and $a'$), so that electrons in the vicinity of the nuclei and those in the valence region can be appropriately diffused{\cite{2005CAS, 2019NAK2}}, which defines  the so-called double-grid LRDMC.
This scheme saves a substantial computational cost in all-electron calculations, especially for a system including  large atomic number atoms{\cite{2019NAK2}}, with a typical computational cost scaling  with $Z^{\sim 6}$ where $Z$ is the maximum atomic number. 
\tvb\, makes use of an appropriate ratio of the mesh sizes ({\it i.e.}, $a$/$a'$) the smaller one $a$ used when electrons are close to the nuclei and the 
larger one $a^\prime \gg a$ adopted in  the valence region. 
By choosing a proper Thomas-Fermi characteristic length around the nuclei,
 where  short hops of lengths $a$ mostly occur, 
 a significant improvement of  the scaling ({\it i.e.}, $Z^{\sim 5.6}$ $\to$ $Z^{\sim 5}$) has been recently reported.{\cite{2019NAK2}}
%
Finally, the inclusion of non-local pseudopotentials in this framework is straightforward by means of an additional spherical grid defined in an appropriate mesh{~\cite{2005CAS_phdthesis}}. As described in Ref.~{\citenum{2005CAS, 2010CAS}}, LRDMC provides an upper bound for the true ground-state energy and allows the estimation of $E_{\text{FN}}$, even in the presence of non-local pseudopotentials. 
Notice that this variational property has also been extended to the standard DMC framework{~{\cite{2006CAS}}, with the introduction of the so-called T-moves.
Moreover, the recently introduced determinant locality approximation
(DLA)\cite{2019ZEN} to deal with non-local pseudopotentials is also
implemented in \tvb\ and can be optionally used in LRDMC. 

\begin{figure}[htbp]
 \centering
 \includegraphics[width=7.6cm]{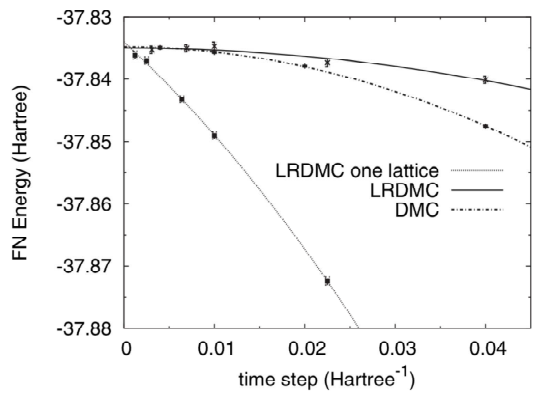}
 \caption{Fixed-node energies for the all-electron carbon atom computed within DMC, 
 single-grid LRDMC (one lattice), and double-grid LRDMC. The lattice discretization 
 parameter $a$ is mapped to the time-step $\tau$ as $a = \sqrt \tau$.
 Reprinted with permission from {\citet{2005CAS}}, {\it Phys. Rev. Lett.}, published by APS in 2005.
 }
 \label{2005CAS_lrdmc_ext}
\end{figure}

%
%

\section{Wave functions}
\label{wavefunction}


Both the accuracy and the computational efficiency of QMC approaches crucially depend on the WF ansatz. 
The optimal ansatz is typically a tradeoff between accuracy and efficiency.
On the one side, a very accurate ansatz can be involved and cumbersome, having many parameters and being expensive to evaluate.
On the other hand, an efficient ansatz is described  only by the most relevant  parameters and can be quickly and easily evaluated.
In particular, in sec.{\ref{methods}}, we have seen that QMC algorithms, both at the variational and fixed-node level, imply several calculations of the local energy $e_L(\mathbf{x})$ and the ratio $\Psi(\mathbf{x})/\Psi(\mathbf{x}')$ for different electronic configurations $\mathbf{x}$ and $\mathbf{x}'$. The computational cost of these operations determines the overall efficiency of QMC and its scaling with systems size.

\vspace{1mm}
\tvb\, employs a many-body WF ansatz $\Psi$ which can be written as the product of two terms:
\begin{equation} 
\Psi  =  \Phi _\text{AS} \times \exp J \,,
\end{equation}
where the term $\exp J$, conventionally dubbed Jastrow factor, is symmetric under electron exchange, and the term $\Phi _\text{AS}$, also referred to as the determinant part of the WF, is antisymmetric.
The resulting WF $\Psi$ is antisymmetric, thus fermionic.

\vspace{1mm}
In the majority of QMC applications, the chosen $\Phi _\text{AS}$ is a single Slater determinant (SD) $\Phi _\text{SD}$, {\it i.e.}, an antisymmetrized product of single-electron WFs. 
Clearly, SD alone does not include any correlation other than the exchange. However, when a Jastrow factor, explicitly depending on the inter-electronic distances, is applied to $\Phi _\text{SD}$ the resulting ansatz $\Psi_\text{JSD} = \Phi _\text{SD} * \exp J$  
often provides over 70\% of the correlation energy\footnote{The correlation energy is typically defined as the difference between the exact energy and the Hartree-Fock energy, which is the variational minimum for a SD ansatz.} at the variational level.
Thus, the Jastrow factor proves very effective in describing the correlation, employing only a relatively small number of parameters, and therefore providing a very efficient way to improve the ansatz.\footnote{
However, the Jastrow factor makes $\Psi$ not factorizable when expectation values of quantum operators are evaluated. For this reason it is not a feasible route to traditional quantum chemistry approaches, as it requires  stochastic approaches to evaluate efficiently the corresponding multidimensional integrals.}
A Jastrow correlated SD (JSD) function yields a computational cost for QMC simulations  -- both VMC and FN -- about $\propto N^3$, namely  the same scaling of  most DFT codes. Therefore, although QMC has a much larger prefactor,  it represents an approach much cheaper than traditional quantum chemistry ones, 
at  least for large enough systems.

\vspace{1mm}
While the JSD ansatz is quite satisfactory in several applications, there are situations where very high accuracy is required, and a more advanced 
ansatz is necessary.
The route to improve JSD is not unique, and different approaches have been attempted within the QMC community.
First, it should be mentioned that improving the Jastrow factors is not an effective approach to achieve higher accuracy at the FN level, as the Jastrow is positive and cannot change the nodal surface.  
A popular approach is through the employment of backflow{~\cite{2006ROP}}, which is a remapping of the electronic configurations that enters into $\Phi_\text{AS}$ (SD as a special case) where each electron position is appropriately changed depending on nearby electrons and nuclei.
Backflow is an effective way to recover correlation energy, both at the variational and FN level. However, it can  be used at a price to increase significantly an already large computational cost.
{\footnote{Indeed, with backflow each time an electron is moved, all the entries in $\Phi_\text{AS}$ (or several of them, if cut-offs are used) can be changed, resulting in a 
much more expensive algorithm.}
Another possibility is to improve $\Psi_\text{SD}$ similarly to conventional quantum chemistry approaches, namely by  considering  $\Phi_\text{AS}$ as a linear expansion of several Slater determinants.
While this second approach can provide very high accuracy, it may be  extremely expensive, as the number of determinants necessary to remain with  a pre-defined accuracy grows combinatorially with the system size.

\vspace{1mm}
The vision embraced in \tvb, is that the route toward an improved ansatz should not compromise  the efficiency and good scaling of QMC.
For this reason, neither backflow nor explicit multideterminant expansions are implemented in the code.
Within the \tvb\ project, the main goal is instead to consider an ansatz that  can be implicitly equivalent to a multideterminant expansion, but remains 
in practice as efficient as a single determinant.
There are five alternatives for the choice of $\Phi_\text{AS}$, which correspond to 
$\rm(\hspace{.18em}i\hspace{.18em})$ 
the Pfaffian (Pf), 
$\rm(\hspace{.08em}ii\hspace{.08em})$ 
the Pfaffian with constrained number of molecular orbitals (Pfn)
$\rm(i\hspace{-.08em}i\hspace{-.08em}i)$ 
the Antisymmetrized Geminal Power (AGP),
$\rm(i\hspace{-.08em}v\hspace{-.06em})$ 
the Antisymmetrized Geminal Power 
with constrained number of molecular orbitals (AGPn),
and
$\rm(\hspace{.06em}v\hspace{.06em})$
the single Slater determinant.
\begin{figure}[htbp]
\includegraphics[width=3.3in]{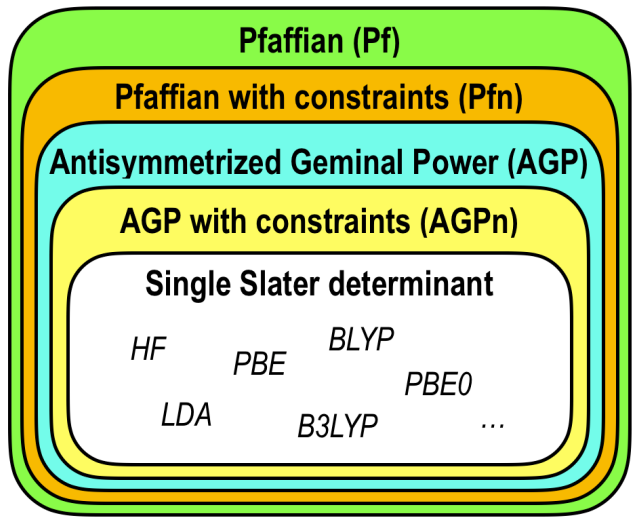}
 \caption{Ansatz hierarchy. The output of Hartree-Fock (HF) or DFT simulations with different exchange-correlation functionals are special instances of the SD ansatz. }
 \label{fig:ansatz}
\end{figure}
It is interesting to observe that
all the latter four WFs are obtained by introducing specific constraints on the most general Pf ansatz.
The hierarchy of the five ans\"atze is represented in the Venn diagram of Fig.~\ref{fig:ansatz}.
Clearly, a more general ansatz is more accurate in the total energy 
but not necessarily in the energy differences. Moreover, it is described by 
more variational parameters,  that could imply a more challenging optimization and a slightly higher cost.
\tvb\, includes several tools to go from one ansatz to another, as represented in Fig.~\ref{fig:ansatz-conv}.
Typically the starting point is SD, which can be obtained from a Hartree-Fock (HF) or a DFT calculation with different exchange-correlation functionals. Both 
methods are not expected to provide the optimal parameters when the Jastrow factor is included in the WF.
\begin{figure}[htbp]
\includegraphics[width=2.5in]{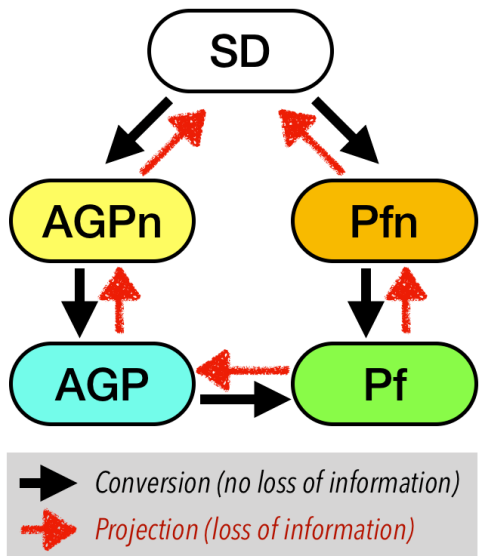}
 \caption{Ansatz conversion.}
 \label{fig:ansatz-conv}
\end{figure}
%
Indeed, in QMC, the WFs are always meant to include the Jastrow factor, which proves fundamental to improve the properties of the overall WF. For instance, AGP carries more correlation than SD. However, it is not size-consistent unless it is multiplied by a Jastrow factor.
Thus, a fundamental step to take advantage of the WF ansatz  is the possibility  to perform reliable optimizations of the parameters. Optimization will be discussed in section~\ref{sec_opt_wf}.
In this section, we will describe the functional form of the Jastrow factor implemented in \tvb\ (sec.~\ref{sec:Jastrow}), the Pfaffian (sec.~\ref{sec:Pf}), the AGP (sec.~\ref{sec:AGP}), the AGPn and Pfn (sec.~\ref{sec:AGPn}), the SD (sec.~\ref{sec:SD}), the multiconfigurational character of the AGP (sec.~\ref{sec:multiconf}), the basis set (sec.~\ref{sec:atomic_basis}), the pseudopotentials (sec.~\ref{sec:PP}), the contractions of the orbitals (sec.~\ref{sec:contraction}), and the conversion tool (sec.~\ref{sec:conv}).

\subsection{Jastrow factor (J)}\label{sec:Jastrow}


The Jastrow factor ($\exp J$) plays an important role
in improving the correlation of the WF and in fulfilling Kato's cusp conditions~\cite{1957KAT}.
\tvb\, implements the Jastrow term composed of one-body, two-body, and three/four-body factors ($J = {J_1}+{J_2}+{J_{3/4}}$).
The one-body and two-body factors are used to fulfill the electron-ion and electron-electron cusp conditions, respectively, and the three/four-body factors are employed to consider a systematic expansion, in principle converging 
to the most general electron pairs contribution.
The one-body Jastrow factor $J_1$ is the sum of two parts, the homogeneous part (enforcing the electron-ion cusp condition):
\begin{equation}
J_1^h \left( \mathbf{r}_1,\ldots,\mathbf{r}_N \right) = \sum_{i=1}^N \sum_{a=1}^{N_\text{at}} \left( { { - {{\left( {2{Z_a}} \right)}^{3/4}}u_a\left( {(2{Z_a})^{1/4}\left| {{\mathbf{r}_i} - {{\mathbf{R}}_a}} \right|} \right)} } \right),
\label{onebody_J_hom}
\end{equation}
and the corresponding inhomogeneous part:
\begin{equation}
{J_1^{inh}}\left( {{{\mathbf{r}}_1}{\sigma _1}, \ldots, {{\mathbf{r}}_N}{\sigma _N}} \right) =  \sum_{i=1}^N \sum_{a=1}^{N_\text{at}} \left( {\sum\limits_{l} {M_{a,l}^{{\sigma _i}} \chi_{a,l}\left( {{{\mathbf{r}}_i}} \right)} } \right) ,
\label{onebody_J_inhom}
\end{equation}
where ${{{\mathbf{r}}_i}}$ are the electron positions, ${{{\mathbf{R}}_a}}$ are the atomic positions with corresponding atomic number $Z_a$, $l$ runs over atomic orbitals $\chi _{a,l}$ ({\it e.g.}, GTO) centered on the atom $a$, ${{\sigma _i}}$ represents the electron spin ($\uparrow$ or $\downarrow$), 
$\{ M_{a,l}^{\sigma _i} \}$ are variational parameters,
and ${u_a\left( \mathbf{r} \right)}$ is a simple bounded function. In \tvb, the most common choice for $u_a$ is:
\begin{equation}
u_a\left( r \right) = \frac{ 1 }{2 b_{\text{e}a}} \left( {1 - {e^{ - r b_{\text{e}a}}}} \right) \,,
\label{onebody_u}
\end{equation}
depending on a single variational parameter $ b_{\text{e}a}$, that may be optimized 
independently for each atomic species.

\vspace{1mm}
The two-body Jastrow factor is defined as:
\begin{equation}
{J_2}\left( {{{\mathbf{r}}_1}{\sigma _1}, \ldots, {{\mathbf{r}}_N}{\sigma _N}} \right) =  {\sum\limits_{i < j} {{v_{{\sigma _i},{\sigma _j}}}\left( {\left| {{{\mathbf{r}}_i} - {{\mathbf{r}}_j}} \right|} \right)} },
\label{twobody_jastrow}
\end{equation}
where
$v_{{\sigma _i},{\sigma _j}}$
is another  simple bounded  function. There are several possible choices for $v_{{\sigma _i},{\sigma _j}}$ implemented in \tvb\ (all listed in the file input.tex in the doc folder), and one of them is, for instance, the following spin-dependent form:
\begin{equation}
  {v_{{\sigma _i},{\sigma _j}}}\left( {{r_{i,j}}} \right) = 
  \begin{cases}
    \cfrac{{{r_{i,j}}}}{4} \cdot {\left( {1 + b_{\rm{ee}}^{\rm{para}} \cdot {{r_{i,j}}}} \right)^{ - 1}} & ({\sigma _i} = {\sigma _j}) \\
    \cfrac{{{r_{i,j}}}}{2} \cdot {\left( {1 + b_{\rm{ee}}^{\rm{anti}} \cdot {{r_{i,j}}}} \right)^{ - 1}} & ({\sigma _i} \neq {\sigma _j}) 
  \end{cases}
\label{twobody_v}
\end{equation}
where  ${r_{i,j}} = \left| {{{\mathbf{r}}_i} - {{\mathbf{r}}_j}} \right|$, 
and $b_{\rm{ee}}^{\rm{para}}$ and $b_{\rm{ee}}^{\rm{anti}}$ are variational parameters.

\vspace{1mm}
The three/four-body Jastrow factor reads:
\begin{equation}
J_{3/4}\left( {{{\mathbf{r}}_1}{\sigma _1}, \ldots, {{\mathbf{r}}_N}{\sigma _N}} \right) = 
\sum_{i < j}
\left(  
\sum_{a,l} \sum_{b,m}
M_{a,l,b,m}^{{\sigma _i},{\sigma _j}}
\chi _{a,l}( \mathbf{r}_i )
\chi _{b,m}( \mathbf{r}_j )
\right),
\label{threebody_jastrow}
\end{equation}
where the indices $l$ and $m$ again indicate different orbitals centered on
corresponding atoms $a$ and $b$,
and $\{ M_{a,l,b,m}^{\sigma _i,\sigma_j} \}$ are variational parameters. 
Sometimes it is convenient to set to zero part of the coefficients of the four-body Jastrow factor, namely those corresponding to $a \ne b$, as they increase the overall variational space significantly and make the optimization more challenging, without being much more effective in improving the variational WF.

\subsection{Pfaffian Wave function (Pf)}\label{sec:Pf}

The SD is an antisymmetrized product of single-electron WFs.
Thus, SD neglects almost entirely the correlation between electrons.
A natural way to improve this description is to include explicitly in the ansatz pairwise correlations among electrons. This is precisely what the Pfaffian WF does.\cite{2016BAJ, 2008BAJ, 2019GEN2}
The building block of the Pfaffian WF is the pairing function $g(\mathbf{i},\mathbf{j})$ between any pair of electrons $i$ and $j$. 
Henceforth we denote with  the generic bolded index $\mathbf{i}$ both the space ${\bf r}_i$ coordinates and the spin values $\sigma_i$: 
\begin{equation}
\mathbf{i} \leftrightarrow ({\bf r}_i, \sigma_i)
\end{equation}
corresponding to the $i^{th}$ electron.

\vspace{1mm}
For simplicity, let us first consider a system with an even number $N$ of electrons.
The WF, written in terms of pairing functions, is:
\begin{equation}\label{eq:wfPF0}
\Phi_\text{AS} (\mathbf{1},\ldots,\mathbf{N}) = {\cal A} \left\{ g(\mathbf{1},\mathbf{2}) g(\mathbf{3},\mathbf{4}) \ldots g(\mathbf{N-1},\mathbf{N}) \right\}
\end{equation}
where ${\cal A}$ is the antisymmetrization operator: 
\begin{equation}
{\cal A} \equiv {1\over N!} \sum_{P\in S_N} \epsilon_P \hat P,
\end{equation}
$S_N$ the permutation group of $N$ elements, $\hat P$ the operator corresponding to  the generic permutation $P$, and $\epsilon_P$  its sign.

\vspace{1mm}
Let us define $G$ the $N\times N$ matrix with elements $G_{i,j} = g(\mathbf{i},\mathbf{j})$.
Notice that
\begin{equation}
g(\mathbf{i},\mathbf{j}) = -g(\mathbf{j},\mathbf{i}) \; (\text{and} \; G_{i,j} = -G_{j,i}),
\end{equation}
as a consequence of the statistics of fermionic particles, thus $G$ is skew-symmetric ({\it i.e.}, $G^T = -G$, being $^T$ the transpose operator), so the diagonal is zero and the number of independent entries is $\sum_{n=1}^{N-1} n = N(N-1)/2$.

\vspace{1mm}
The Pfaffian\cite{1963KAS} of a $N\times N$ skew-symmetrix matrix $G$ is defined as:
\begin{equation}\label{eq:Pfdef}
\text{Pf}(G) \equiv {1\over 2^{N/2} (N/2)!} \sum_{P\in S_N} \epsilon_P G_{P(1),P(2)} \cdots G_{P(N-1),P(N)}
\end{equation}
if $N$ is even, and it is zero if $N$ is odd.\footnote{
Some properties of the Pfaffian operator and their proofs are given in \citet{2008BAJ}.
}
Therefore, the WF $\Phi_\text{AS}$ defined in the right-hand side (RHS) of Eq.~\ref{eq:wfPF0} equals  
$ {1\over N!!} \text{Pf}(G) $
where the semifactorial  
$N!! \equiv { N! \over 2^{N/2} (N/2)!}$
is irrelevant in QMC, as it affects only the normalization of the WF.
Thus, we can define our electronic WF as: 
\begin{equation}\label{eq:wfPFeven}
\Phi_\text{Pf} = \text{Pf}(G) \,.
\end{equation}

\vspace{1mm}
Notice that the $\Phi_\text{Pf}$ here defined allows the description of  any system with $N_u$ electrons with spin-up and $N_d$ electrons with spin-down, provided that $N=N_u+N_d$ is even.
Indeed, with no loss of generality, we can assume that electrons $i=1,\ldots,N_u$ have $\sigma_i=1/2$ and electrons with $i=N_u+1,\ldots,N$ have $\sigma_i=-1/2$. 
Thus, the $N\times N$ skew-symmetric matrix $G$ is written as:
\begin{equation}
G = \left[\begin{array}{c|c} G_{uu} & G_{ud} \\ \hline
G_{du} & G_{dd}\end{array}\right]
\end{equation}
where 
$G_{uu}$ is a $N_u\times N_u$ skew-symmetric matrix with elements $g_{uu}(\mathbf{i},\mathbf{j})$, 
$G_{dd}$ is a $N_d\times N_d$ skew-symmetric matrix with elements $g_{dd}(\mathbf{i},\mathbf{j})$, 
$G_{ud}$ is a $N_u\times N_d$ matrix with elements $g_{ud}(\mathbf{i},\mathbf{j})$, 
and $G_{du} = -{G_{ud}}^T$, {\it i.e.}, $g_{du}(\mathbf{i},\mathbf{j})=-g_{ud}(\mathbf{j},\mathbf{i})$. 
$g_{uu}$ describes the pairing between a pair of electrons with spin-up:
\begin{equation}
g_{uu}(\mathbf{i},\mathbf{j}) = f_{uu}({\bf r}_i,{\bf r}_j) \left| \uparrow  \uparrow \right\rangle 
\end{equation}
where the function $f_{uu}$ describes the spatial dependence on the coordinates ${\bf r}_i,{\bf r}_j$ for $i,j\le N_u$.
Notice that $f_{uu}({\bf r}_j,{\bf r}_i) = - f_{uu}({\bf r}_i,{\bf r}_j)$ as a consequence of the properties of $g$.
The spin part $\left| \uparrow  \uparrow \right\rangle$ describes a system with unit total spin  and spin projection along the $z$-axis,  and will be indicated by  $\left| 1, +1 \right\rangle$.
Similarly, $g_{dd}$ describes the pairing between pairs of electrons with spin-down for $i,j> N_u$:
\begin{equation}
g_{dd}(\mathbf{i},\mathbf{j}) = f_{dd}({\bf r}_i,{\bf r}_j) \left| \downarrow  \downarrow \right\rangle 
\end{equation}
with $f_{dd}({\bf r}_j,{\bf r}_i) = - f_{dd}({\bf r}_i,{\bf r}_j)$,
and the spin part $\left| \downarrow  \downarrow \right\rangle$ describes a system with total unit spin  and negative spin projection along the $z$-axis, indicated with $\left| 1, -1 \right\rangle$.
$g_{ud}$ describes the pairing between pairs of electrons with unlike spins. 
Since two electrons with unlike spins can form a singlet  
$\left| 0,0 \right\rangle = { {\left| \uparrow  \downarrow \right\rangle - \left| \downarrow  \uparrow \right\rangle}\over \sqrt{2}}$
or a triplet 
$\left| 1,0 \right\rangle = { {\left| \uparrow  \downarrow \right\rangle + \left| \downarrow  \uparrow \right\rangle}\over \sqrt{2}}$, in the general case
the pairing function $g_{ud}$ will be a linear combination of the the two components:
\begin{equation}
g_{ud}(\mathbf{i},\mathbf{j}) = 
f_{S}({\bf r}_i,{\bf r}_j) { {\left| \uparrow  \downarrow \right\rangle - \left| \downarrow  \uparrow \right\rangle}\over \sqrt{2}} 
+
f_{T}({\bf r}_i,{\bf r}_j) { {\left| \uparrow  \downarrow \right\rangle + \left| \downarrow  \uparrow \right\rangle}\over \sqrt{2}} 
\end{equation}
where $f_{S}({\bf r}_i,{\bf r}_j) = f_{S}({\bf r}_j,{\bf r}_i)$ describes the spatial dependence of the singlet part of $g_{ud}$, and 
$f_{T}({\bf r}_i,{\bf r}_j) = -f_{T}({\bf r}_j,{\bf r}_i)$ describes the spatial dependence of the triplet part.
Therefore, the generic pairing function $g(\mathbf{i},\mathbf{j})$ is the sum of  all the four components mentioned above, namely :
\begin{equation}
\begin{split}
g\left( \mathbf{i},\mathbf{j} \right) 
&= f_{S}({\bf r}_i,{\bf r}_j) \left| 0,0 \right\rangle + f_{T}({\bf r}_i,{\bf r}_j) \left| 1,0 \right\rangle \\
&+ f_{uu}({\bf r}_i,{\bf r}_j) \left| 1,+1 \right\rangle + f_{dd}({\bf r}_i,{\bf r}_j) \left| 1,-1 \right\rangle \,.
\end{split}
\label{eq:g}
\end{equation}
%
%

\vspace{1mm}
The pairing functions $f_{S}$,  $f_{T}$, $f_{uu}$, and $f_{dd}$  are expanded over atomic orbitals (see sec.~\ref{sec:atomic_basis}). Say, for a generic pairing function $f$ we have 
\begin{equation}\label{agp_expansion}
f\left( {{{\mathbf{r}}_i},{{\mathbf{r}}_j}} \right) = \sum\limits_{l,m,a,b} {{{A}_{\left\{ {a,l} \right\},\left\{ {b,m} \right\}}}{\psi _{a,l}}\left( {{{\mathbf{r}}_i}} \right){\psi _{b,m}}\left( {{{\mathbf{r}}_j}} \right)}
\end{equation}
where ${\psi_{a,l}}$ and ${\psi_{b,m}}$ are primitive or contracted atomic orbitals, their indices $l$ and $m$ indicate different orbitals centered on atoms $a$ and $b$, while $i$ and $j$ label the electron coordinates.
Symmetries on the system, or properties of the underlying pairing function $f$ imply constraints on the coefficients. For instance, the coefficients of  $f_{S}$ are such that $A_{\left\{ {a,l} \right\},\left\{ {b,m} \right\}} = A_{\left\{ {b,m} \right\},\left\{ {a,l} \right\}}$ because $f_{S}({\bf r}_i,{\bf r}_j) = f_{S}({\bf r}_j,{\bf r}_i)$, whereas $A_{\left\{ {a,l} \right\},\left\{ {b,m} \right\}} = -A_{\left\{ {b,m} \right\},\left\{ {a,l} \right\}}$ for $f_{T}$, $f_{uu}$, and $f_{dd}$.

\vspace{1mm}
Let us consider now the remaining case of a system with an odd
number of electrons $N$.
The simplest way to handle this case is to consider 
a system with an extra fictitious  
electron ${\bf N+1} \leftrightarrow  (\bvec{r}_{N+1}, \sigma_{N+1}) $
that we set at infinity $\bvec{r}_{N+1} \to \infty$, 
thus non interacting with all the $N$ physical electrons.
The extra matrix elements are easily computed:
\begin{eqnarray}\label{eq:gfictitious}
g(\mathbf{N+1},\mathbf{N+1})&=& 0,   \nonumber \\
g(\mathbf{i},\mathbf{N+1})=-g(\mathbf{N+1},\mathbf{i})&=&\phi({\bf i}) ~~~ {\rm for ~} i\le N,
\end{eqnarray}
where $\phi({\bf i}) \equiv \phi(\bvec {r}_i,\sigma_i)$ can be considered an extra spin dependent  unpaired orbital vanishing 
at infinity. 

\vspace{1mm}
The antisymmetric part of the WF  of the overall system is then: 
\begin{equation}\label{eq:wfPF0odd}
\Phi_\text{AS} (\mathbf{1},\ldots,\mathbf{N}) 
= {\cal A} \left\{ g(\mathbf{1},\mathbf{2}) g(\mathbf{3},\mathbf{4}) \ldots g(\mathbf{N},\mathbf{N+1}) \right\}
\end{equation}
where the antisymmetrization operator acts on the (now even) $N+1$ electrons.
This is  a perfectly allowed $N$ electron fermionic WF, because 
by definition: 
$\rm(\hspace{.18em}i\hspace{.18em})$ it is antisymmetric for  all permutations 
of the $N+1$ particles and in particular for the $N-$ physical ones; 
$\rm(\hspace{.08em}ii\hspace{.08em})$  it does not depend by definition 
on the extra $N+1^{th}$ coordinates and spin as the fictitious 
particle is at infinity. On the other hand, it  is easy to show that this is 
a nontrivial and nonvanishing WF,  because 
we can use the basic Pfaffian formula in Eq.~{\ref{eq:Pfdef}}
for the antisymmetrization 
of the $N+1$ particles and we readily obtain  that:
\begin{equation}\label{eq:wfPFodd}
\Phi_\text{Pf} = \text{Pf}(\tilde G) \,,
\end{equation}
where 
\begin{equation} \label{eq:pfodd}
\tilde G = \left[\begin{array}{c|c} G & \Phi \\ \hline
					-\Phi^T & 0 \end{array}\right]
\end{equation}
is a $(N+1)\times(N+1)$ skew-symmetric matrix, and $\Phi$ is a $N$-dimensional vector whose elements are $\phi(\textbf{i})$, $i=1,\ldots,N$.
Its Pfaffian is quite generally nonvanishing for at least some configuration 
provided the diagonalization of the Pfaffian (see later)  has at least  $(N+1)/2$ non-zero eigenvalues (likewise for the Slater determinant  it is enough that the $N-$molecular orbitals are linearly independent) and we obtain in this way that 
the  Pfaffian is perfectly defined even in  the odd number of electron case.

\vspace{1mm}
Finally, we would like to emphasize that the above argument can be generalized 
to arbitrary number $k$ of unpaired orbitals and arbitrary boundary conditions including calculations on periodic supercells. 
We can indeed assume at the beginning that, when $N$ is odd (even), we add an odd  (even) number $k$ of unpaired orbitals, such that there are $N$ electrons coupled by the matrix $G$ plus $k$ unpaired electrons.
The  $k$ unpaired electrons are paired with $k$ fictitious electrons (with a pairing function satisfying conditions analogous to those in Eq.~\ref{eq:gfictitious}).
This yields  a Pfaffian of even linear dimension $N+k$ of the 
same form as the one in Eq.~{\ref{eq:pfodd}} but with $\Phi$ being a block $N \times  k$ rectangular matrix, determined by  $k$ different spin-dependent orbitals. 
The same argument holds that the antisymmetrized WF 
is written as the Pfaffian of the corresponding $(N+k) \times (N+k)$ skew-symmetric matrix, 
and does not depend on the fictitious 
particle coordinates 
introduced and therefore is an allowed physical WF antisymmetric 
over the $N$ physical particles.

\vspace{1mm}
A special case that is of interest is when all the $N$ electrons are set to be unpaired ({\it i.e.}, when $k=N$ and $\Phi$ is a generic $N\times N$ matrix) and it is assumed that there is no pairing among them ({\it i.e.}, $G=0$ in the $N\times N$ block  matrix in Eq. \ref{eq:pfodd}).
Thus,
using a well-known relation of a Pfaffian 
\begin{equation} \label{eq:pfafftodet}
\Pf{\begin{array}{c|c} 0 & \Phi \\ \hline
-\Phi^T & 0\end{array}} 
= (-1)^{N \times (N-1)\over 2} \det(\Phi),
\end{equation}
we recover the standard Slater determinant.
In this way, we can clearly see that the Pfaffian represents a quite remarkable generalization of the single determinant picture, and that a larger variational freedom is exploited by allowing pairing correlations with a non zero matrix $G$.

\vspace{1mm}
Recently we have noticed that, following the above derivation,  it is possible to generalize further the Pfaffian WF ansatz by introducing fictitious particles and relaxing the constraints on the $\tilde G$ matrix, even in the lower block diagonal part, that  can be assumed to be an arbitrary non-zero skew-symmetric matrix. Work is in progress in order to understand whether this extended 
formulation can further improve the accuracy within a single Pfaffian 
WF.

\subsection{Antisymmetrized Geminal Power (AGP)}\label{sec:AGP}


If we consider only the case of a pairing function $g(\mathbf{i},\mathbf{j})$ that is a spin singlet (namely, $f_{uu}$, $f_{dd}$ $f_{T}$ in Eq.~\ref{eq:g} are set to zero, yielding $g(\mathbf{i},\mathbf{j})=f_{S}({\bf r}_i,{\bf r}_j) \left| 0,0 \right\rangle$) then we obtain the singlet Antisymmetrized Geminal Power (denoted as AGPs).

\vspace{1mm}
Let us consider first an unpolarized system, having an even number $N$ on electrons, and without loss of generality, we can assume that the electrons $i=1,\ldots,N/2$ have spin up and electrons $j=N/2+1,\ldots,N$ have spin down.
Then, the matrices $G_{uu}$ and $G_{dd}$ defined in Sec.~\ref{sec:Pf} are both zero matrices of size $N/2\times N/2$, and the matrix $G_{ud}$ has only the contribution coming from the singlet, that we dub $G_S$. 
The antisymmetrization operator implies the computation of 
\begin{equation}
\Pf{\begin{array}{c|c} 0 & G_{S} \\ \hline
				-G_{S}^T & 0\end{array}} 
= (-1)^{N/2\times (N/2-1)\over 2} \det(G_S)
\end{equation}
where the equality follows from a property of the Pfaffian (see Eq.~\ref{eq:pfafftodet}).
The overall sign is arbitrary for a WF; thus the antisymmetrized product of singlet pairs (geminals) is indeed equivalent to the computation of the determinant of the matrix $G_S$:
\begin{equation}\label{eq:wfAGPs}
\Phi_\text{AGPs} = \det(G_S) \,.
\end{equation}
It should be noticed that it is not necessary  that the 
matrix $G_{ud}$ is symmetric  to reduce the Pfaffian  to a single determinant evaluation. As long as the matrices $G_{uu}$ and $G_{dd}$ are zero, the Pfaffian is indeed equivalent to  $\det(G_{ud})$ and describes an antisymmetric WF. However, if $G_{ud}$ is not symmetric the function
\begin{equation}
\Phi_\text{AGP} = \det(G_{ud})
\end{equation}
is not an eigenstate of the spin. In other terms, there is a spin contamination, similarly to the case of unrestricted HF calculations.

\vspace{1mm}
The AGP ansatz can be generalized to describe polarized systems, {\it i.e.}, systems where the number $N_u$ of electrons of spin up is different from the number $N_d$ of electrons with spin down.
With no loss of generality, we can assume that $N_u>N_d$, thus the system is constituted by a number $p=N_d$ of electron pairs and a number $k=N_u-N_d$ of unpaired electrons (clearly, $N=N_u+N_d=2p+k$).
We aim at evaluating:
\begin{equation}\label{eq:wfPF_generic1}
{\cal A} \left\{ g(\mathbf{1},\mathbf{2}) \ldots g(\mathbf{2p-1},\mathbf{2p}) \phi_1(\mathbf{2p+1}) \ldots \phi_k(\mathbf{N}) \right\}.
\end{equation}
As above, we assume that the pairing function is zero for like-spin pairs.
With no loss of generality, we can assume that electrons $i=1,\ldots,N_u$ have spin up and electrons $j=N_u+1,\ldots,N$ have spin down.
Moreover, as it was done at the end of the previous subsection,  we can add $k$ fictitious entries to $g(\mathbf{i},\mathbf{j})$, such that $g(\mathbf{i},\mathbf{N+l})=\phi_l(\mathbf{i})$ for $l=1,\ldots,k$ and $i=1,\ldots,N_u$.
Thus, the antisymmetrization implies the use  of  Eq.{~\ref{eq:pfafftodet}}, 
{\it i.e.}, :
\begin{equation}
\Pf{\begin{array}{c|c} 0 & \tilde G  \\ \hline
				-\tilde G^T & 0\end{array}} 
= (-1)^{N_u*(N_u-1)\over 2} \det(\tilde G)
\end{equation}
with 
the $N_u\times N_u$ matrix  
$
\tilde G = \left[\begin{array}{c|c} G_{ud} & \Phi \end{array}\right]
$,
the $N_u\times N_d$ matrix  $G_{ud}$ describing the pairing between the $N_u$ spin up electrons and the $N_d$ spin down electrons,
and the $N_u\times k$ matrix  $\Phi$ describing the $k$ unpaired orbitals.
Thus, we need to evaluate:
\begin{equation}\label{eq:wfAGPunpaired}
\Phi_\text{AGP} = \det(\tilde G) \,.
\end{equation}

\vspace{1mm}
One of the most important advantages of the AGP ansatz is that it is equivalent to a linear combination of Slater determinants ({\it i.e.}, multi configurations), but the computational cost remains at the level of  a single-determinant one{~{\cite{2003CAS, 2014ZEN, 2015ZEN2}}}, see Section~\ref{sec:multiconf}.
This is especially important for large systems because the naive multireference approach requires an exponentially large number of Slater determinants, which drastically increases the computational cost.
The AGP ansatz was applied to the {\it ab initio} calculation by M. Casula and S. Sorella ~{\cite{2003CAS}} for the first time in 2003; then it has been  also implemented in other QMC codes.

\subsection{AGP and Pf with constrained number of molecular orbital (AGPn and Pfn)}\label{sec:AGPn}

A convenient way to impose constraints on the variational parameters defining the AGP  or  Pf  WF is obtained by rewriting the expansion of the geminal in terms of molecular orbitals (MOs).
As shown in Eq.~\ref{agp_expansion}, a geminal $g(\bvec{i},\bvec{j})$ is natively expressed in terms of the atomic orbitals $\{\psi_{a,l}({\bf r})\}$, by summing over all the atoms $a$, the corresponding orbitals $l$ and the spin 
index $\sigma$ (as the elements of the basis may in principle 
also depend on the spin, even  though the most common choice is to take the 
same orbital for each of the two possible values of the spins).
In order to simplify the notation here, let us merge the atomic  orbital and spin indices in a unique one that is indicated with a greek symbol ({\it e.g.}, $\mu \leftrightarrow (a,l)$) running from 1 to the total dimension of the $2L$ atomic orbitals used, $L$ for each spin component.
Therefore, 
\begin{equation}\label{eq:gparam}
g(\mathbf{i},\mathbf{j}) = \sum\limits_{\mu,\nu}^{2L} A_{\mu,\nu} \psi_\mu(\mathbf{i}) \psi_\nu(\mathbf{j})
\end{equation}
and clearly the symmetry of $g$ implies that $A_{\mu,\nu} = -A_{\nu,\mu}$.
The coefficients $A_{\mu,\nu}$ define a $2L\times 2L$ skew-symmetric matrix $A$.
If we define the $2L$ dimensional vector $\Psi_i=\left( \psi_1(\mathbf{i}), \ldots, \psi_{2L}(\mathbf{i}) \right)^T $, 
Eq.~\ref{eq:gparam} rewrites as 
$g(\mathbf{i},\mathbf{j}) = \Psi_i^T A \Psi_j $.
Moreover, 
the overlaps $S_{\mu,\nu} \equiv \langle \psi_\mu | \psi_\nu \rangle$ between atomic orbitals define the 
overlap matrix $S$, that in the case of  a spin-dependent basis is 
block diagonal:
\begin{equation}
S = \left[\begin{array}{c|c} S_{uu} & 0 \\ \hline
					0 & S_{dd}\end{array}\right]
\end{equation}
with $S_{uu}$ and $S_{dd}$ positive definite $L\times L$ square matrices ($S_{uu} = S_{dd}$ when orbitals are the same for spin up and spin down).
In \tvb, the overlap matrix $S$ is computed on a suitable uniform 
mesh with an efficient 
and general parallel algorithm. Then  an orthonormal basis is defined:
\begin{equation}
\tilde \psi_\mu(\mathbf{i}) = \sum\limits_\nu S^{-1/2}_{\mu,\nu} \, \psi_\nu (\mathbf{i}) 
\end{equation}
where $S^{-1/2}$ is well defined since $S$ is strictly positive definite.\footnote{  
$S^{-1/2}$ can be computed after a standard  block diagonalization  of the matrix $S= U D U^\dag$,
being $U$ an unitary matrix and $D=\text{diag}\{d_1,\ldots,d_{2L}\}$ is a diagonal matrix, such that $S^{-1/2} = U D^{-1/2} U^\dag$, where 
$D^{-1/2}$ is the diagonal matrix obtained by taking the inverse square root 
of each diagonal element $d_i$  of $D$. 
At  this stage we carefully remove from the basis the elements corresponding 
to the smallest  eigenvalues $d_i$ in order to work with a  sufficiently 
large  condition number that guarantees stable finite precision numerical 
calculations.
}
The matrix elements of $g$ can be recasted in the new orthonormal basis $\left\{ \tilde \psi_\mu(\mathbf{i}) \right\}$, yielding
$
g(\mathbf{i},\mathbf{j}) = \sum_{\mu,\nu}^{2L} \tilde A_{\mu,\nu} \tilde\psi_\mu(\mathbf{i}) \tilde\psi_\nu(\mathbf{j})
= \tilde\Psi_i^T \tilde A \tilde\Psi_j 
$,
with the matrix $\tilde A \equiv S^{1/2} A S^{1/2} $ that is skew-symmetric.

\vspace{1mm}
At this point, from the spectral theory of skew-symmetric matrices, it is 
possible to perform the Youla decomposition of $\tilde A$ (see Ref.~\onlinecite{2019GEN3}), 
which can be written in the form $\tilde A = Q \Sigma Q^T$, 
where $Q$ is unitary (also real  if $\tilde A$ is real), 
and the matrix $\Sigma$ is block diagonal with $\Sigma_{2k-1,2k}=\lambda_k=-\Sigma_{2k,2k-1}$ for $k=1,...,L$, and zero everywhere else, with $\lambda_k \ge 0$.\footnote{The nonzero eigenvalues of $\tilde A$ are $\pm i \lambda_k$.}
So, the pairing function $g(\bvec{i},\bvec{j})$ can be written as $\Phi_i^T \Sigma \Phi_j$, 
where $\Phi_i = Q^T \tilde\Psi_i$ for each $i$.
This defines a basis of $L$ MOs $\left\{\phi_k(i)\right\}$  and 
corresponding twinned ones  $\left\{ \bar \phi_k(i) \right\}$, forming 
together a basis of $2L$ mutually orthonormal elements 
for which the original geminal function reads:
\begin{equation}\label{eq:geminal_MOs}
g\left( \mathbf{i},\mathbf{j} \right) = \sum\limits_{k=1}^L \lambda_k \left[
\phi_k(\mathbf{i}) \bar \phi_k(\mathbf{j}) -
\bar\phi_k(\mathbf{i}) \phi_k( \mathbf{j}) 
\right]
\end{equation}
with $\lambda_k \ge 0$.
After these transformations these MOs can be finally written 
in the chosen (hybrid) atomic basis:
\begin{eqnarray} \label{eq:basisexp}
\phi_k(\mathbf{i}) &=& \sum\limits_{\nu=1}^{2L} P_{\mu,l} \psi_{\mu} (\mathbf{i}) \nonumber \\
\bar  \phi_k(\mathbf{i}) &=& \sum\limits_{\nu=1}^{2L}  \bar P_{\mu,l} \psi_{\mu} (\mathbf{i})
\end{eqnarray}
by appropriate $ p \times 2 L$ rectangular matrices $P$  and $\bar P$.
Then, with no loss of generality, we can assume that the molecular orbitals $\left\{ \phi_k(\mathbf{i}), \bar \phi_k(\mathbf{i}) \right\} $ are ranked such that 
$ \lambda_1  \ge \lambda_2  \ge \ldots \ge  \lambda_L \ge 0 $.
The above expression highlights that the most important MOs are those corresponding to the larger values of $ \lambda_k$.
Therefore, it is possible to constrain the variational freedom by neglecting all the orbitals with $k>n$, yielding the pairing function:
\begin{equation}\label{eq:geminal_MOs_cut}
g_n\left( \mathbf{i},\mathbf{j} \right) = \sum\limits_{k=1}^n \lambda_k \left[
\phi_k(\mathbf{i}) \bar \phi_k(\mathbf{j}) -
\bar\phi_k(\mathbf{i}) \phi_k( \mathbf{j}) 
\right] \,,
\end{equation}
where $n$ is conveniently chosen and is $\ll L$.
This yields the AGPn ansatz and the $\Phi_\text{AGPn}$ WF, which can be useful to improve the stability of the wave-function optimization. The corresponding algorithm (enabled by setting \verb|molopt=-1| in the optimization section input of \tvb), based on projection operators 
in the space of the $n$ molecular orbitals considered, has been described extensively in Ref.~\onlinecite{2017BEC}. 
Moreover, in the original paper\cite{2009MAR} introducing the AGPn,  
a precise recipe was given to 
improve the evaluation of the binding energies.
Indeed, despite  a constraint on the variational parameters 
necessarily increases the variational 
energy expectation value, 
 energy differences may actually improve by an appropriate choice of $n$. 
In the mentioned  work~{\cite{2009MAR}}, this promising approach was applied with an AGP containing only singlet correlations, but the binding energies were defined without using a rigorous  size consistent criterium. This drawback can be now removed, by exploiting the full variational freedom of the Pf WF  combined with a general spin-dependent Jastrow factor (see {\it e.g.}, Fig.~\ref{fig:dissociationO2}).
Work is in progress in this interesting research  direction.

\vspace{1mm}
The variational  optimization  of an AGP  with a fixed number $n$ of molecular orbitals can be 
easily generalized to the Pf case,  by exploiting that the constrained
Pf WF, dubbed Pfn, can be written either in the canonical form with MOs as  in Eq.~{\ref{eq:geminal_MOs_cut}}
or in the localized basis set expansion, as in Eq.{~\ref{eq:gparam}}, with a  
corresponding matrix $A^n_{\mu,\nu}= \sum\limits_{k=1}^n  \lambda_k \left[  P_{\mu,k} \bar P_{\nu,k} - P_{\nu,k} \bar  P_{\mu,k} \right]$.
According  to Eq.{~\ref{eq:geminal_MOs_cut}} an arbitrary small variation $\delta g_n$
of the constrained pairing function $g_n$ reads:
\begin{eqnarray} \label{eq:deltaproj}
\delta g_n\left( \mathbf{i},\mathbf{j} \right) &=& \sum\limits_{k=1}^n 
\delta \lambda_k \left[
\phi_k(\mathbf{i}) \bar \phi_k(\mathbf{j}) -
\bar\phi_k(\mathbf{i}) \phi_k( \mathbf{j}) \right] \nonumber \\ 
&+&
\sum\limits_{k=1}^n \lambda_k \left[
 \delta \phi_k(\mathbf{i}) \bar \phi_k(\mathbf{j}) -
 \delta \bar\phi_k(\mathbf{i}) \phi_k( \mathbf{j}) \right]  \nonumber \\
&+&
 \sum\limits_{k=1}^n \lambda_k \left[
 \phi_k(\mathbf{i}) \delta \bar \phi_k(\mathbf{j}) -
  \bar\phi_k(\mathbf{i}) \delta \phi_k( \mathbf{j}) \right]
\end{eqnarray}
and therefore satisfies  the following property, as it will be shown later:
\begin{equation} \label{eq:satn}
(\hat I -{\mathbf L}) \delta g_n ( \hat I -{\mathbf R}) =0
\end{equation}
where $\hat I$ is the identity operator, ${\mathbf L}$ and ${\mathbf R}$ are  projection operators over the occupied MOs,  
{\it i.e.}, $\mathbf{L}^2(\mathbf{i},\mathbf{j}) = 
\int d{\mathbf{k}} \mathbf{L}(\mathbf{i},\mathbf{k}) \mathbf{L}(\mathbf{k},\mathbf{j})=\mathbf{L}(\mathbf{i},\mathbf{j})$, 
and similarly  $\mathbf{R}^2=\mathbf{R}$, where here and henceforth  the shorthand integration symbol $\int d\mathbf{k} = \sum_{\sigma_k} \int d\mathbf{r}_k$ contains implicitly also the spin summation.
These operators are then  defined as follows:
\begin{eqnarray} \label{eq:deflr}
\mathbf{L} (\mathbf{i}, \mathbf{j}) &=& \sum\limits_{k=1}^n 
\left[ \phi_k(\mathbf{i}) \phi^*_k(\mathbf{j}) +
\bar \phi_k(\mathbf{i}) \bar \phi^*_k(\mathbf{j})\right] \nonumber \\
\mathbf{R} (\mathbf{i}, \mathbf{j}) &=& \sum\limits_{k=1}^n 
\left[ \phi^*_k(\mathbf{i}) \phi_k(\mathbf{j}) +
\bar \phi^*_k(\mathbf{i}) \bar \phi_k(\mathbf{j}) \right].
\end{eqnarray}
With the above definitions, Eq.{~\ref{eq:deltaproj}} is easily  verified  
because  each term of Eq.{~\ref{eq:deltaproj}} is annihilated either by the  
left ($\hat I  -\mathbf{L}$) or the right ($\hat I - \mathbf{R}$)  projection
over the unoccupied MOs. 
Notice that $\mathbf{L}=\mathbf{R}$ in the real case and $\mathbf{L}=\mathbf{R}^*$  in the most general complex case.
In this way,  in order to implement  a constrained variation $\delta g_n$ 
of the Pfn  WF, corresponding to an appropriate variation of its matrix $\delta A^n_{\mu,\nu}$,   
it is useful  to work with a small free  variation $\delta g$ (with corresponding 
$\delta A_{\mu,\nu}$). This  is then projected  
onto the chosen restricted ansatz by means of the following 
equation:
\begin{equation} 
\delta g_n =  \delta g - (\hat I-{\bf L}) \delta g (\hat I- {\bf R})  \label{proPf}.
\end{equation}
Indeed,
it is easy to show that the RHS of the above equation vanishes  
if we apply $\hat I-{\bf L} $ and $\hat I- {\bf R}$  to its left  and its  right, respectively, 
just because $\hat I -\mathbf{R}$ and $ \hat I-\mathbf{L}$ are projection operators, being 
such $\mathbf{R}$ and $\mathbf{L}$, yielding  
  $(\hat I- {\bf R})^2= (\hat I- {\bf R}) $ and 
$(\hat I- {\bf L})^2= (\hat I- {\bf L}) $, from which Eq.{~\ref{proPf}}
fulfills Eq.{~\ref{eq:satn}}.
Eq.{~\ref{proPf}} represents, therefore, a {\em linear relation} applied to the variational parameter matrix change $\delta A_{\mu,\nu}$ corresponding  to the unconstrained  geminal $g$ in Eq.{~\ref{eq:gparam}}, yielding the new  constrained variation $\delta  A^n_{\mu,\nu}$.
Indeed, by using the definitions of the projector operators in Eq.{~\ref{eq:deflr}} and the expansion of  the MOs in  the atomic (hybrid) basis
(see Eq.{~\ref{eq:basisexp}}), Eq.{~\ref{proPf}} turns to a number  of   matrix-matrix operations 
acting on $\delta A$, $P$, $\bar P$ and the overlap matrix $S$ 
that can be easily and  efficiently implemented\cite{2017BEC}.

\vspace{1mm}
This linear relation between  $A$ and $A^n$   can be therefore easily implemented together with the 
corresponding  derivatives necessary  to the optimization of the  energy 
\footnote{The output of AAD are matrices 
 $D_{\mu,\nu}={\partial F \over \partial A_{\mu,\nu}}$ where $F$ is either the log of the WF or the 
 corresponding local  energy computed on a given configuration. Then the projected derivatives 
 corresponding to $\delta A^n_{\mu,\nu}$ easily follows from Eq.{~\ref{proPf}}, by applying the chain rule} 
and allows the explicit  calculation of  the 
new  matrix $A^n_{\mu,\nu}$, yielding the new  constrained geminal $  g_n + \delta g_n$.
Then the new geminal can be recasted in the form of Eq.{~\ref{eq:geminal_MOs_cut}} by  the mentioned 
diagonalization of skew-symmetric matrices, in this way 
implicitly  neglecting nonlinear  contributions 
that are irrelevant close to convergence, when $\delta g_n\to 0$.
After employing several iterations of this type, 
the lowest energy ansatz of the JPfn type  can  be obtained in a relatively simple and very  efficient  way.

\vspace{1mm}
Finally, it is also important to emphasize that this constrained optimization algorithm 
allows the reduction 
of the number of  parameters, by efficiently exploiting locality, namely that variational parameters 
$A_{\mu,\nu}$ corresponding to atoms at a distance larger than a reasonable cutoff (an 
input named \verb|rmax| in \tvb) can be safely disregarded with negligible error\cite{2017BEC}.


\subsection{Single Slater determinant (SD)}\label{sec:SD}

An important special case of the AGPn and Pfn ans\"atze  discussed in sec.~\ref{sec:AGPn} is when we constrain the pairing function $g_n$ to use the minimum possible number of molecular orbitals providing  a non zero WF.
The minimum number for an unpolarized system with $N$ electrons is equal to the number of electron pairs, that is $n=N/2$.
The WF obtained in this way starting  from the AGP is indeed the single Slater determinant (SD)~\cite{2017BEC, 2009MAR}, and we dub it $\Phi_\text{SD}$.
In principle also the Pfn  WF with $n=N/2$  corresponds to a single Slater 
determinant with spin dependent molecular orbitals, a case that has never been  considered so far, but it represents an available option  within the most recent versions of \tvb.

\vspace{1mm}
Similarly, in a polarized system having $N_u$ spin up electrons and $N_d$ spin down electrons (we assume $N_u>N_d$), the SD ansatz is obtained by using $N_u - N_d$ unpaired electrons, and using a constrained pairing function $g_n$ with $n=N_d/2$.

\subsection{Implicit multiconfigurational character of the AGP}\label{sec:multiconf}
In secs.~{\ref{sec:Pf}} and ~{\ref{sec:AGP}}, it has been mentioned that the Pfaffian and the AGP ans\"atze 
have a multiconfigurational character, despite they can be evaluated at the cost of a single determinant.
In this section, we expand the AGP ansatz in terms of Slater determinants, to show this aspect explicitly.
In order to simplify the derivation, we will consider here a simplified case,
while the most general case could be studied with a similar approach but involving more cumbersome expressions. 

\vspace{1mm}
We consider the real AGPs ansatz (Eq.~\ref{eq:wfAGPs}) for an unpolarized system of $N=2 N_p$ electrons ({\it i.e.}, $N_u=N_d=N/2$).
The symmetry implies that the twin molecular orbitals $\left\{ \phi_k(\mathbf{i}), \bar \phi_k(\mathbf{i}) \right\} $
appearing in the RHS of Eq.~\ref{eq:geminal_MOs}
have the same spatial part, that we denote $\tilde\phi_k(\bvec{r})$, modulus a sign (because they are orthonormal), and they have opposite spin part.
Without loss of generality we can assume that $\phi_k$ is spin up and $\bar \phi_k$ is spin down.
Given this convention, the scalar product $l_k$ between the spatial parts of $\phi_k$ and $\bar \phi_k$ with either be +1 or -1. 
We define $\tilde\lambda_k = l_k \lambda_k$, 
where $\lambda_k$ is the same as the one in the RHS of Eq.~\ref{eq:geminal_MOs}.
Notice that $\left| \tilde\lambda_k \right| = \lambda_k$,
so $\left\{ \tilde\lambda \right\}$ are ranked in decreasing order of their absolute value ({\it i.e.}, $\left| \tilde\lambda_k \right| \ge \left| \tilde\lambda_{k+1} \right|$).
The pairing function then can be written as
\begin{equation}
g\left( \mathbf{i},\mathbf{j} \right) = \sum_{k=1}^L \tilde\lambda_k 
\tilde\phi_k(\mathbf{r}_i) \tilde\phi_k(\mathbf{r}_j) 
\left( {\left| \uparrow  \downarrow \right\rangle - \left| \downarrow  \uparrow \right\rangle} \right)
\,.
\end{equation}
This expression  is useful for comparing with the standard CI expansion.

\vspace{1mm}
It is convenient now to use the second quantization notations in order to simplify the derivation.
In particular, we indicate with ${\hat a_{k,\uparrow}}^\dag$  (${\hat a_{k,\downarrow}}^\dag$)
the operator that creates an electron of spin up (down) in the orbitals $\tilde\phi_k$, and satisfies the canonical anticommutation relations.
We can write the pairing functions 
$g(\bvec{i},\bvec{j}) \equiv \left< \bvec{i}, \bvec{j} \middle| \hat g \middle| 0 \right>$, 
where $\left| 0 \right>$ is the vacuum, $\left| \bvec{i}, \bvec{j}\right>$ is the WF of a system with one electron with coordinates  $({\bf r}_i, \sigma_i)$ and another with coordinates $({\bf r}_j, \sigma_j)$, and $\hat g$ is the operator:
\begin{equation}\label{eq:goperator}
\hat g = \sum_{k=1}^L \tilde\lambda_k  \hat a_{k,\uparrow}^\dag \hat a_{k,\downarrow}^\dag \,,
\end{equation}
where 
$\tilde\lambda_k$ is the one defined above.
Using this notation, it is easy to show (see Appendix of Ref.~\onlinecite{2015ZEN2}) that the pairing function $g$ is equivalent to the complete active space of 2 electrons on the $L$ molecular orbitals $\left\{ \tilde\phi_k \right\}$.

\vspace{1mm}
Within this notation, the AGPs WF is
\begin{equation}
|\Phi_\text{AGPs}\rangle
=
 {\hat g}^{p} | 0 \rangle \,,
\end{equation}
where $p\equiv N/2$ is the number of electron pairs.
If we substitute in ${\hat g}^{p}$ the expansion for $\hat g$ in Eq.~\ref{eq:goperator},
after having conveniently defined the operator $\hat b_k \equiv \hat a_{k,\uparrow}^\dag \hat a_{k,\downarrow}^\dag$ which created an electron pair on the orbital $\tilde\phi_k$,
and having noticed that $\hat b_k \hat b_l = \hat b_l \hat b_k$ and ${\hat b_k}^2=0$ (as following from the anticommutation relations of the $\hat a_k^\dag$ operators)
we obtain that:
\begin{equation}\label{eq:gtop_exp}
{\hat g}^{p} =  {p!} \sum_{1\le i_1 < i_2 < \ldots < i_{p} \le L} 
\tilde\lambda_{i_1} \tilde\lambda_{i_2} \ldots \tilde\lambda_{i_p}
\hat b_{i_1} \hat b_{i_2} \ldots \hat b_{i_p} \,.
\end{equation}
The chosen order of the $\tilde\lambda_k$ coefficients implies that the leading term in ${\hat g}^{p}$
is given by the term with 
$( i_1, \ldots, i_p ) = (1,\ldots,p)$.\footnote{
It shall be noticed that there is not necessarily one leading term in the expansion in Eq.~\ref{eq:gtop_exp}.
For instance, when $\left| \tilde\lambda_p \right| = \left| \tilde\lambda_{p+1} \right|$ the term 
with $( i_1, \ldots, i_p ) = (1,\ldots,p)$ have equal weight to the term 
$( i_1, \ldots, i_{p-1}, i_p ) = (1,\ldots,p-1,p+1)$.
}
Thus, we can rewrite ${\hat g}^{p}$ in terms of this leading term and excitations with respect to it:
\begin{eqnarray}\label{eq:expand-AGP}
{ {\hat g}^{p} \left| 0 \right> \over p! \prod_{i=1}^{p} \tilde\lambda_i } & = & 
\prod_{i=1}^{p} \hat b_i |0\rangle +  \\
&+& \sum_{j = 1}^p \sum_{q=p+1}^L {\tilde\lambda_q \over \tilde\lambda_j} 
\left( \mathop{\prod_{i=1}^{p}}_{i\ne j} 
\hat b_i \right)
\hat b_q |0\rangle + \nonumber \\
&+& \sum_{1\le j<k\le {p}} \sum_{{p}< q<r \le L} 
{\tilde\lambda_q \tilde\lambda_r \over \tilde\lambda_j \tilde\lambda_k} 
\left( \mathop{\prod_{i=1}^{p}}_{i\ne k \land i\ne j} 
\hat b_i \right)
\hat b_q \hat b_r |0\rangle + \nonumber \\
&+& \ldots . \nonumber
\end{eqnarray}
In the above expansion, we shall recognize that the first element is a single Slater determinant of the molecular orbitals $\left\{ \tilde\phi_i \right\}_{i=1,\ldots,N/2}$;
the second element is the summation of all the possible double excitations both going from an orbital $j\le N/2$ to an orbital $q>N/2$;
the third element is the summation of a subset of the possible quadruple excitations, and so on.
In other terms, AGPs can be written as the zero seniority\footnote{The seniority number is the number of unpaired electrons in a determinant.} subset of the CI expansion, having some constraints on the coefficients of the expansion.

\vspace{1mm}
It is worth mentioning that  AGPs is also reliable in cases where a single Slater determinant is not a good reference, as for instance, in the case of a broken covalent bond, where the highest occupied molecular orbital (HOMO) and the lowest unoccupied molecular orbital (LUMO) are degenerate  (see Fig.~\ref{fig:c2h4}).
In fact, we can have that $\left| \tilde\lambda_{p} \right| = \left| \tilde\lambda_{p+1} \right|$, so the two determinants $\prod\limits_{i=1}^{p} \hat b_i \left| 0 \right>$ and $\left(\prod\limits_{i=1}^{p-1} \hat b_i \right) \hat b_{p+1} \left| 0 \right>$ have the same weight and they both are the leading terms of the constrained zero-seniority expansion in Eq.~\ref{eq:expand-AGP}.

\vspace{1mm}
Notice that in the AGPn ansatz we can perform an  expansion similar to Eq.~\ref{eq:expand-AGP}, but the excitations stops at the orbital $n$ rather than $L$. Thus, it is straightforward to see that for $n=N/2$ there are no excitations and the only term is a single Slater determinant. 

\begin{figure}[htbp]
\centering
\includegraphics[width=2.5in]{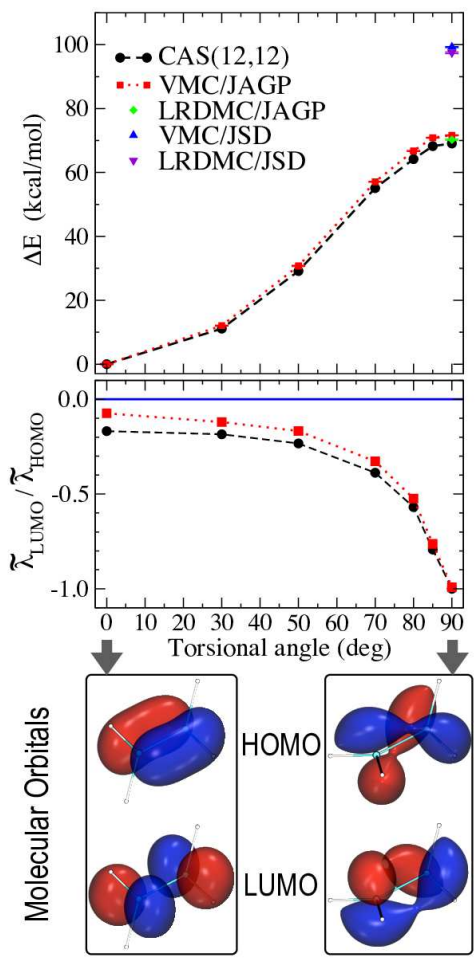}
\caption{
The twisted ethylene molecule (C$_2$H$_4$) is a prototypical system requiring a multireference method.
The ground state geometry is planar (torsional angle equal to zero); C atoms have a double bond and the HOMO and LUMO orbitals are separated. 
As the molecule is twisted, one bond is broken and the HOMO-LUMO orbitals become degenerate for a torsional angle of 90 degrees.
The upper panel shows evaluations of the torsional barrier as obtained using JSD and JAGP at the VMC and LRDMC level, and using a complete active space (CAS) of 12 orbitals in 12 electrons. JAGP provides reliable results, whereas JSD (which neglects the LUMO orbital) overestimate the barrier both at the variational and diffusion level. 
The middle panel shows the ratio $\tilde \lambda_\text{LUMO} / \tilde \lambda_\text{HOMO}$ in the JAGP ansatz, and the corresponding value in the CAS calculations.
The bottom panel shows the HOMO and LUMO orbitals for the planar and twisted geometries.
Readapted from {\citet{2014ZEN}}, {\it J. Chem. Theory Comput.}, published by ACS in 2014.
}\label{fig:c2h4}
\end{figure}

\subsection{Atomic basis set for the pairing function and the Jastrow factor}\label{sec:atomic_basis}

\tvb\, employs localized atomic orbitals such as the Gaussian type:
\begin{equation}
\psi _{l, \pm m ,I}^{{\text{Gaussian}}}\left( {{\mathbf{r}};\zeta } \right) = {\left| {{\mathbf{r}} - {{\mathbf{R}}_I}} \right|^l}{\operatorname{e} ^{ - \zeta {{\left| {{\mathbf{r}} - {{\mathbf{R}}_I}} \right|}^2}}} \cdot \Re[  (-i)^{1\pm 1 \over 2} {Y_{l,m,I}}\left( {\theta ,\varphi } \right)],
\label{gaussian}
\end{equation}
or the Slater type:
\begin{equation}
\psi _{l, \pm  m ,I}^{{\text{Slater}}}\left( {{\mathbf{r}};\zeta } \right) = {\left| {{\mathbf{r}} - {{\mathbf{R}}_I}} \right|^l}{\operatorname{e} ^{ - \zeta \left| {{\mathbf{r}} - {{\mathbf{R}}_I}} \right|}} \cdot \Re[ (-i)^{1\pm 1 \over 2} {Y_{l,m,I}}\left( {\theta ,\varphi } \right)],
\label{slater}
\end{equation}
where the real and the imaginary part ($m>0$)  of the 
 spherical harmonic function $ {Y_{l,m,I}}\left( {\theta ,\varphi } \right)$ 
 centered at ${{{\mathbf{R}}_I}}$ are (is) taken and rewritten in Cartesian coordinates in order to work  
with  real defined and easy to compute orbitals, 
$l$ is the corresponding angular momentum and $m\ge 0$ is its projection number along the $z-$quantization axis.
The localized atomic orbitals are also present in the onebody and three/four body Jastrow parts ({\it i.e.}, denoted as $\chi \left( {\mathbf{r}} \right)$ in Eq.~{\ref{onebody_J_inhom}} and Eq.~{\ref{threebody_jastrow}}).
One can use standard basis sets with  exponents $\zeta$ (also coefficients for a contracted basis set) taken from the available 
standard database such as the Basis Set Exchange{~\cite{2019PRI}} or from  other 
more specific references when using  pseudo potential~{\cite{2015TRA, 2016KRO, 2017TRA, 2017BEN, 2018BEN, 2018ANN}}. A python wrapper, named \tvbg, makes the above procedure much easier, as shown later.

\subsection{Pseudopotential}\label{sec:PP}

\tvb\, supports pseudopotential calculations both in VMC and LRDMC calculations. 
Many ECPs have been generated and successfully used in quantum chemistry codes, but they are usually tuned to match Density Functional Theory (DFT) or Hartree-Fock (HF) all-electron (AE) calculations, which are not expected 
to be optimal for state of the art many-body techniques. Recently some progress has been made in this direction and  pseudopotentials determined by correlated many-body techniques are also available~{\cite{2007BUR, 2008BUR, 2015TRA, 2016KRO, 2017TRA, 2017BEN, 2018BEN, 2018ANN}}. All the pseudopotentials used in QMC employ the standard semi-local form:
\begin{equation}
\hat V_{{\text{pp}}}^I\left( {{{\mathbf{r}}_i}} \right) = V_{{\text{loc}}}^I\left( {{r_{i,I}}} \right) + \sum\limits_{l = 0}^{{l_{\max }}} V_l^I\left( r_{i,I} \right) \sum\limits_{m =  - l}^l {\left| {{Y_{l,m}}} \right\rangle \left\langle {{Y_{l,m}}} \right|} 
\end{equation}
where $r_{i,I} = |{{\mathbf{r}}_i} - {{\mathbf{R}}_I}|$ is the distance between the $i$-th electron and the $I$-th ions, $l_{{\text max}}$ is the maximum angular momentum of the ion $I$, and $\sum\limits_{l = 0}^{{l_{\max }}} {\sum\limits_{m =  - l}^l {\left| {{Y_{l,m}}} \right\rangle \left\langle {{Y_{l,m}}} \right|} }$ is a projection operator on the spherical harmonics centered at the ion $I$. In \tvb, the angular momentum projector is calculated by using standard polyhedral quadrature formulas for the angular integrations{\cite{1990FAH}}.
As it is now becoming a  common  practice not only in QMC, 
both the local  $V_{{\text{loc}}}^I\left( {{r_{i,I}}} \right)$ and   the non-local $V_l^I\left( r_{i,I} \right)$ functions, are expanded over a simple 
Gaussian basis parametrized by coefficients ({\it e.g.}, effective charge ${Z_{{\text{eff}}}}$ and  other simple constants), multiplying simple powers of $r$, and  a 
 corresponding gaussian term:
\begin{equation}
{r^2}{V_l}\left( r \right) = \sum\limits_k {{\alpha _{k,l}}{r^{{\beta _{k,l}}}}\exp \left(- {{\gamma _{k,l}}{r^2}} \right)},
\end{equation}
where ${{\alpha _{k,l}}}$, ${{\beta _{k,l}}}$ (usually small positive integers), and ${{\gamma _{k,l}}}$ are the parameters obtained by appropriate fitting.
Several published pseudopotentials have already been tabulated in \tvb. Of course, one can also use any pseudopotential employing the semi-local form in  the mentioned Gaussian basis  with a straightforward little extra work for the input preparation.

\subsection{Contraction of the primitive atomic basis}\label{sec:contraction}

The contraction of atomic orbitals has been widely used in quantum chemistry and DFT codes, and  was originally introduced to define a pseudo-Slater orbital by combining several primitive Gaussian orbitals. This is also important in the QMC context because it decreases the number of variational parameters by a large factor, as it will be shown in this section. Although one can obtain a contracted basis set directly from a database, \tvb\, allows us to prepare a high-quality hybrid basis set (contraction) starting from a given  primitive basis, by the so-called ``geminal embedding scheme".~{\cite{2015SOR}} To this purpose,
let us decompose the genimal function in terms of atomic contribution, as far as the 
dependence over ${\mathbf i}$ (the left argument) is concerned:
\begin{equation}
g\left( {\mathbf{i},\mathbf{j}} \right) = \sum\limits_I {U_{{\text{proj}}}^I\left( {\mathbf{i},\mathbf{j}} \right)}  = \sum\limits_I {\sum\limits_{\mu ,\nu } {A _{\mu ,\nu }^I} } {\phi _\mu }\left( \mathbf{i} \right){\phi _\nu }\left( \mathbf{j} \right),
\end{equation}
where $I$ represents an atom in a system, $U_{{\text{proj}}}^I\left( {\mathbf{i},\mathbf{j}} \right)$ is the pairing function projected on the atom $I$, $A _{\mu ,\nu }^I$ is ${A _{\mu ,\nu }}$ if $\mu \in I$, otherwise ${A _{\mu ,\nu }^I}=0$, where $A_{\mu,\nu}$ is assumed to be given for the system under consideration, {\it e.g.}, obtained by a standard DFT calculation, where in this case, by Eq.{~\ref{eq:geminal_MOs_cut}}:
\begin{equation}
A_{\mu,\nu} = \sum\limits_{k=1}^n \left(c^k_\mu \bar c^k_\nu-  \bar c^k_\mu 
c^k_\nu \right) 
\end{equation}
where $c^k_\mu$ and $\bar c^k_\mu$ are the coefficients of the DFT molecular 
orbitals $\phi_k ({\mathbf i})=\sum_\mu c^k_\mu \phi_\mu({\mathbf i})$ and $\bar \phi_k ({\mathbf i})=\sum_\nu \bar c^k_\nu \phi_\nu ({\mathbf i})$ in the atomic basis expansion, respectively. 

\vspace{1mm}
Quite generally, the projected pairing function can be expanded in a truncated space spanned by $q$ terms:
\begin{equation}
\tilde U_{{\text{proj}}}^I\left( {\mathbf{i},\mathbf{j}} \right) = \sum\limits_{k = 1}^q {{\sigma _k}\psi _k^{{\text{GEO}}}\left( \mathbf{i} \right)\psi _k^{{\text{Env}}}\left( \mathbf{j} \right)}.
\end{equation}
In other words, the Schmidt decomposition is applied to 
the matrix $\tilde U_{{\text{proj}}}^I\left( {\mathbf{i},\mathbf{j}} \right)$
describing the coupling between a given atom $I$ and the 
enviroment, within the geminal ansatz.
 This procedure defines the so-called geminal embedded orbitals (GEOs), 
 that are determined 
 in terms of an expansion over all the atomic orbitals used for  the atom $I$:
\begin{equation}
\psi _k^{{\text{GEO}}}\left( \mathbf{i} \right) = \sum\limits_{\mu  \in I}^{} {\mu _{k,\mu }^{{\text{GEO}}}\psi _\mu ^{}\left( \mathbf{i} \right)} ,
\end{equation}
where $\psi _k^{{\text{GEO}}}\left( \mathbf{i} \right)$ are orthonormal.
Following the Schmidt decomposition, it is possible  to determine 
the best GEOs by minimizing the Euclidian distance between the original and the truncated geminal functions:
\begin{equation}
{d^2} = {\left| {U_{{\text{proj}}}^I - \tilde U_{{\text{proj}}}^I} \right|^2}. 
\end{equation}
Considering all possible unconstrained functions $\psi _k^{{\text{Env}}}\left( \mathbf{j} \right)$ and employing the steady condition $\frac{{\delta {d^2}}}{{\delta \psi _k^{{\text{Env}}}\left( \mathbf{j} \right)}} = 0$, ${{d^2}}$ reads:
\begin{equation}
{d^2} = {\left| {U_{{\text{proj}}}^I} \right|^2} - \sum\limits_{k = 1}^p {\int {d\mathbf{i}d\mathbf{j}D_{{\text{proj}}}^I\left( {\mathbf{i},\mathbf{j}} \right)\psi _k^{{\text{GEO}}}\left( \mathbf{i} \right)\psi _k^{{\text{GEO}}}\left( \mathbf{j} \right)} },
\label{d2}
\end{equation}
where ${\left| {U_{{\text{proj}}}^I} \right|^2} = \int {d\mathbf{i}d\mathbf{j}{{\left| {U_{{\text{proj}}}^I\left( {\mathbf{i},\mathbf{j}} \right)} \right|}^2}}$ and $D_{{\text{proj}}}^I\left( {\mathbf{i},\mathbf{j}} \right)$ is the density matrix that reads:
\begin{equation}
D_{{\text{proj}}}^I\left( {\mathbf{i},\mathbf{j}} \right) = \int {d\mathbf{k}U_{{\text{proj}}}^I\left( {\mathbf{i},\mathbf{k}} \right)^* U_{{\text{proj}}}^I\left( {\mathbf{j},\mathbf{k}} \right)} .
\label{density_matrix}
\end{equation}
Since the GEOs ${\psi _k^{{\text{GEO}}}\left( {\mathbf{r}} \right)}$ are orthonormal and the density matrix is hermitian, Eq.~{\ref{d2}} becomes minimum when the GEOs coincides with the $p$ eigenvectors of the density matrix with the maximum eigenvalues (denoted as ${w_i}$). The original atomic basis $\psi _\mu\left( \mathbf{i} \right)$ is usually non-orthogonal, so the problem turns into the generalized eigenvalue equation:
\begin{equation}
\sum\limits_{j \in J} {{{\left[ {\left( {{\boldsymbol {A}^{I}{}^{*}} {\boldsymbol S} {\boldsymbol {A}^{I}} } \right) {\boldsymbol S}} \right]}_{i,j}}\mu _{k,j}^{{\text{GEO}}}}  = {\omega _k}\mu _{k,i}^{{\text{GEO}}}.
\label{density_matrix_diag}
\end{equation}
The truncation error is readily estimated by the summation of the eigenvalues ${d^2} = {\left| {U_{{\text{proj}}}^I} \right|^2} - \sum\nolimits_{i = 1}^p {{w_p}} $. Since the eigenvalues are sorted in ascending order,  
a suitably chosen value of $p$ allows the user to neglect the most irrelevant 
vector components with small eigenvalues $w_i$ and work with enough accuracy 
even with a  few GEOs per atom, thus minimizing the number of variational 
parameters necessary to describe well the system, as it will be shown below.
With this construction  a new geminal is defined in  the GEO basis, namely:
\begin{equation}
\tilde g\left( {\mathbf{i},\mathbf{j}} \right) = \sum\limits_{\mu ,\nu }^{} {{{\tilde A }_{\mu ,\nu }}\psi _\mu ^{{\text{GEO}}}\left( \mathbf{i} \right)\psi _\nu ^{{\text{GEO}}}\left( \mathbf{j} \right)}
\end{equation}
where the matrix coefficients ${{\tilde A }_{\mu ,\nu}}$ are given by maximizing the normalized overlap ($Q$) between the original and the new geminals:
\begin{equation}
Q = \frac{\braket{ {\tilde g} | {g} }^{2}}{\braket{ {\tilde g} | {\tilde g} } \braket{ {g} | {g} } },
\end{equation}
where $\braket{{\tilde g} | {g}} = \int {d\mathbf{i}d\mathbf{j}\tilde g\left( {\mathbf{i},\mathbf{j}} \right)^* g\left( {\mathbf{i},\mathbf{j}} \right)} $.
It turns out that the overlap remains large even for small GEO basis set size $p$; implying that, by using this scheme, one can decrease the number of variational parameters corresponding to the matrix $A$, {\it i.e.}, from $ 4 L^2$ to $4 p^2 \ll 4 L^2.$

\subsection{Conversion of the WF}\label{sec:conv}

As described in Sec.{~\ref{wavefunction}}, \tvb\ implements different types of ansatz: 
$\rm(\hspace{.18em}i\hspace{.18em})$ 
the Pfaffian (Pf), 
$\rm(\hspace{.08em}ii\hspace{.08em})$ 
the Pfaffian with constrained number of molecular orbitals (Pfn)
$\rm(i\hspace{-.08em}i\hspace{-.08em}i)$ 
the Antisymmetrized Geminal Power (AGP),
$\rm(i\hspace{-.08em}v\hspace{-.06em})$ 
the Antisymmetrized Geminal Power 
with constrained number of molecular orbitals (AGPn),
and
$\rm(\hspace{.06em}v\hspace{.06em})$
the single Slater determinant (SD).
One can choose a proper ansatz depending on a target system, considering the computational cost of a chosen ansatz and the relevant physical and chemical properties of a target material. During the simulation, a user can go back and forth between the ansatz using modules implemented in \tvb, with/without losing the information of an optimized ansatz (see Fig.{~\ref{fig:ansatz-conv}}):
The first case is to add molecular orbitals to an ansatz, {\it i.e.}, JAGP $\Rightarrow$ JSD, JAGP $\Rightarrow$ JAGPn, or JPf $\Rightarrow$ JPfn. In \tvb, this is obtained by rewriting the expansion of the geminal in terms of molecular orbitals  (see sec.{~\ref{sec:AGPn} and sec.{~\ref{sec:SD}}}). The 
corresponding tool is convertfort10mol.x.
The second important case is to convert an ansatz among the available ones, {\it i.e.}, JSD, JAGP, or JAGPn $\Rightarrow$ JAGP. In \tvb, this is the purpose of the convertfort10.x tool and 
is achieved by maximizing the overlap between the two WFs, one of them  being the input 
(fort.10\_in) and the other  being the type (fort.10\_out) to be filled by  new geminal matrix coefficients (result written in fort.10\_new).
In more  details, in \tvb, the following overlap between two geminals is maximized:
\begin{equation}
\max Q = \frac{\braket{ {g} ^{{\text{new}}} | {g} ^{{\text{ori}}} }^{2} } { \braket{ {g} ^{{\text{new}}} | {g} ^{{\text{new}}} } \braket{ {g} ^{{\text{ori}}} | {g} ^{{\text{ori}}} } },
\end{equation}
in order to obtain  new geminal matrix coefficients ${A _{\mu,\nu}^{{\text{new}}}}$,  defining the new pairing function as:
\begin{equation}
{g^{{\text{new}}}}\left( {\mathbf{i},\mathbf{j}} \right) = \sum\limits_{\mu .\nu }^{} {A _{\mu .\nu }^{{\text{new}}}\psi _\mu ^{{\text{new}}}\left( \mathbf{i} \right)\psi _\nu ^{{\text{new}}}\left( \mathbf{j} \right)},
\end{equation}
while the original geminal was given in terms of the parameter 
matrix $A _{\mu ,\nu }^{\text{ori}}$:
\begin{equation}
{g^{{\text{ori}}}}\left( {\mathbf{i},\mathbf{j}} \right) = \sum\limits_{\mu ,\nu }^{} {A _{\mu ,\nu }^{{\text{ori}}}\psi _\mu ^{{\text{ori}}}\left( \mathbf{i} \right)\psi _\nu ^{{\text{ori}}}\left( \mathbf{j} \right)}.
\end{equation}
Notice that $0 \leqslant Q \leqslant 1$; therefore the larger is  $Q$, the better is the 
conversion, and $Q$ approaches the unit value if the conversion is perfect. For this type of conversion, one can also apply the geminal embedding scheme to construct a hybrid basis set, as described in the sec.{\ref{sec:contraction}}.

\vspace{1mm}
The final case is to convert a JAGP ansatz to JPf.
Since the JAGP ansatz is a special case of the JPf one,  where only $G_{ud}$ and $G_{du}$ terms are defined as described in the section \ref{sec:AGP}, the conversion can be realized just by direct substitution. Therefore, the main challenge is to find a reasonable initialization for the two spin-triplet sectors $G_{uu}$ and $G_{dd}$ that are not described in the JAGP and that otherwise have to be set to $0$. 
There are two possible approaches{~\cite{2019GEN2,2019GEN3}}: 
$\rm(\hspace{.18em}i\hspace{.18em})$
for polarized systems, we can build the $G_{uu}$ block of the matrix by using an even number of 
unpaired orbitals $\{ \phi_i\}$ and build an antisymmetric $g_{uu}$ by means 
of Eq.{~\ref{eq:geminal_MOs_cut}}, where the eigenvalues $\lambda_k$ are chosen to be 
large enough 
to occupy certainly  these unpaired states, as in  the standard Slater determinant used 
for our initialization.
 Again, we emphasize that  this works only for polarized systems.
$\rm(\hspace{.08em}ii\hspace{.08em})$
The second approach that also works in a spin-unpolarized case is to determine 
a standard broken symmetry single determinant ansatz ({\it e.g.}, by the \tvb\ built-in DFT within the LSDA)  and modify it with a global  spin rotation. Indeed, in the presence of finite local magnetic moments, it is often convenient to rotate the spin moments of the WF in a direction perpendicular to  the spin quantization axis chosen for  our spin-dependent Jastrow factor, {\it i.e.}, the $z$ quantization axis. In this way one can obtain reasonable initializations for  $G_{uu}$ and $G_{dd}$. \tvb\ allows every possible rotation, including an arbitrary small one close to the identity.
A particularly important case is when  a rotation of $\pi/2$ is applied around the $y$ direction. This operation maps
\begin{equation}
|\uparrow \rangle \rightarrow \frac{1} {\sqrt{2}} \left( |  \uparrow \rangle + |\downarrow \rangle \right)   \mbox{ and }  |\downarrow  \rangle  \rightarrow  \frac 1 {\sqrt{2}} \left( | \uparrow  \rangle - |\downarrow \rangle \right).	
\label{rot-ud}
\end{equation}
One can convert from a AGP the pairing function that is obtained from a VMC optimization:
\begin{equation}
{g_{ud}}(\mathbf{i},\mathbf{j}) = {f_S}({{\mathbf{r}}_i},{{\mathbf{r}}_j})\frac{{\left| { \uparrow  \downarrow } \right\rangle  - \left| { \downarrow  \uparrow } \right\rangle }}{{\sqrt 2 }} + {f_T}({{\mathbf{r}}_i},{{\mathbf{r}}_j})\frac{{\left| { \uparrow  \downarrow } \right\rangle  + \left| { \downarrow  \uparrow } \right\rangle }}{{\sqrt 2 }}
\end{equation}
to a Pf one:
\begin{equation}
{g_{ud}}(\mathbf{i},\mathbf{j}) \to g\left( {\mathbf{i},\mathbf{j}} \right){\text{ }} = {f_S}({{\mathbf{r}}_i},{{\mathbf{r}}_j})\frac{{\left| { \uparrow  \downarrow } \right\rangle  - \left| { \downarrow  \uparrow } \right\rangle }}{{\sqrt 2 }} + {f_T}({{\mathbf{r}}_i},{{\mathbf{r}}_j})\left( {\left| { \uparrow  \uparrow } \right\rangle  - \left| { \downarrow  \downarrow } \right\rangle } \right).
\end{equation}
This transformation provides a meaningful initialization to the Pfaffian WF that can be  then optimized for  reaching the best possible description of the ground state within this ansatz.

\section{Bulk systems}
\label{periodic_system}
The application of \tvb\, is not limited to open systems such as atoms
and molecules.
\tvb\, can also simulate bulk systems 
in large supercells with arbitrary 
twisted boundary conditions. These are 
used to minimize finite-size effects, and represent quite 
an important approach\cite{2001LIN,2016DAG} 
in order to reach a meaningful and accurate thermodynamic limit. 

\subsection{CRYSTAL basis set}

For periodic system calculations, the many-body WF should satisfy the many-body Bloch condition{~\cite{1994RAJ, 1995RAJ}}:
\begin{equation}
{\Psi _{{k_s}}}\left( {{{\mathbf{r}}_1}, \ldots ,{{\mathbf{r}}_i} + {{\mathbf{T}}_s} \ldots ,{{\mathbf{r}}_N}} \right) = {e^{i{{\mathbf{k}}_s} \cdot {{\mathbf{T}}_s}}}{\Psi _{{k_s}}}\left( {{{\mathbf{r}}_1}, \ldots ,{{\mathbf{r}}_i}, \ldots ,{{\mathbf{r}}_N}} \right),
\end{equation}
which follows  from the property that  the many-body Hamiltonian is invariant under the translation of {\it any electron coordinate} by a simulation-cell vector ${\mathbf{T}_s}$,  
where ${{\mathbf{T}}_{s}}= l {\mathbf a}+m {\mathbf b}+n {\mathbf c} $ 
is determined by arbitrary integers $l,n,m$ and the  three vectors 
${\mathbf  a},{\mathbf b}$ and ${\mathbf  c}$  define the supercell.

\vspace{1mm}
In \tvb, a single-particle basis set satisfies the following condition:
\begin{equation}
\psi _{l,m,I}^{{\text{PBC}}}\left( {{\mathbf{r}} + {{\mathbf{T}}_s};\zeta } \right) = {e^{ i{{\mathbf{k}}_{s}} \cdot {{\mathbf{T}}_s}}}\psi _{l,m,I}^{{\text{PBC}}}\left( {{\mathbf{r}};\zeta } \right)
\end{equation}
where ${{{\mathbf{k}}_{{s}}}}$ is a twist vector (${{\mathbf{k}}_{{s}}} = \left( {k_s^x,k_s^y,k_s^z} \right)$), and ${{\mathbf{T}}_s}$ represents an arbitrary simulation cell vector. 
Notice that the use   of  a non vanishing twist vector generally makes a many-body WF {\it complex}.
\tvb\, implements the CRYSTAL periodic basis{~\cite{1988PIS,2017BEC,2018DOV}}:
\begin{equation}
\psi _{l,m,I}^{{\text{PBC}}}\left( {{\mathbf{r}};\zeta } \right) = \sum\limits_{{{\mathbf{T}}_s}} {\psi _{l,m,I}^{}\left( {{\mathbf{r}} + {{\mathbf{T}}_s};\zeta } \right){e^{-i{{\mathbf{k}}_s} \cdot {{\mathbf{T}}_s}}}}
\end{equation}
where ${\psi _{l,m,I}^{}}$ is a non-periodic real atomic orbital such as Gaussian (Eq.~{\ref{gaussian}}) and Slater (Eq.~{\ref{slater}}). The use of Gaussian or Slater orbitals that rapidly decay far from nuclei guarantees that the above summation converges fast with  a finite small number of  ${\mathbf{T}_s}$. 
Notice that, in \tvb, ${\mathbf{T}_s}$ are not limited to orthorhombic ones but has been recently extended to include all possible crystal translation groups ({\it e.g.}, rhombohedral, hexagonal, triclinic). 

\vspace{1mm}
The same procedure is applied to the basis set for the Jastrow part, 
although
using  
simple periodic  boundary conditions{~\cite{2017BEC}}, because the twists do not 
affect the Jastrow part of the WF,  namely:
\begin{equation}
\chi _{l,m,I}^{{\text{PBC}}}\left( {{\mathbf{r}};\zeta } \right) = \sum\limits_{\mathbf{T}_s} {\chi _{l,m,I}^{}\left( {{\mathbf{r}} + {{\mathbf{T}}_s};\zeta } \right)},
\end{equation}
which satisfies $\chi _{l,m,I}^{{\text{PBC}}}\left( {{\mathbf{r}} + {{\mathbf{T}}_s};\zeta } \right) = \chi _{l,m,I}^{{\text{PBC}}}\left( {{\mathbf{r}};\zeta } \right)$.

\vspace{1mm}
Moreover,
  the homogeneous one-body and two-body Jastrow 
factors
have to be 
appropriately
periodized 
  because they are not defined in terms of  the above periodic basis. Namely, the homogeneous one-body Jastrow part (Eq.~{\ref{onebody_J_inhom}}) should satisfy:
\begin{equation}
{{\tilde J}_1}\left( {{\mathbf{r}} + {{\mathbf{T}}_s}} \right) = {{\tilde J}_1}\left( {\mathbf{r}} \right),
\end{equation}
and the two-body Jastrow part (Eq.~{\ref{twobody_jastrow}}) should fulfill:
\begin{equation}
{J_2}\left( {{{\mathbf{r}}_1}{\sigma _1}, \ldots ,\left( {{{\mathbf{r}}_i} + {{\mathbf{T}}_s}} \right){\sigma _i}, \ldots {{\mathbf{r}}_N}{\sigma _N}} \right) = {J_2}\left( {{{\mathbf{r}}_1}{\sigma _1}, \ldots ,{{\mathbf{r}}_i}{\sigma _i}, \ldots {{\mathbf{r}}_N}{\sigma _N}} \right).
\end{equation}
In order to satisfy the above constraints, we consider  the relative electron-nuclei or electron-electron coordinate differences ${\mathbf r}_d$, necessary to evaluate $J_1$ and $J_2$, respectively, and expand them as: 
\begin{equation}
{\mathbf r}_d  =  r_a  {\mathbf  a}+  r_b  {\mathbf b} 
+ r_c  {\mathbf  c},
\end{equation}
where $r_a,r_b$ and $r_c$ are appropriate transformed coordinates, that are conveniently defined 
within a cube of unit length, because of  the assumed periodicity of the supercell, 
namely $| r_a|,|r_b|,|r_c|< 1/2$. As a consequence, this mapping makes  the physical 
electron-electron and electron-ion distance periodic by definition ({\it i.e.}, they refer to the minimum distance image of the supercell). However, there may be 
divergences or singularity at the boundaries of this unit cube.
Therefore, before  computing the distance corresponding to ${\mathbf r}_d$, these coordinates 
are transformed $\left(r_a,r_b,r_c\right) \to \left( \bar r_a,\bar r_b,\bar r_c \right) 
=\left( p(r_a),p(r_b),p(r_c) \right)$ 
by use of an appropriate function $p(x)$, with at least continuous first derivative 
for $|x|<1/2$. This function is  
 chosen to preserve the physical meaning at short distances, {\it i.e.}, $p(x)=x$ in these cases, 
and being nonlinear elsewhere, in order to satisfy not only  the  periodicity 
but also the requirement of continuous first derivatives of the many-body 
WF $\Psi_{k_s}$.
We have, therefore, defined  $p(x)$ as   follows: 
\begin{equation}
  p\left( x \right) = 
  \begin{cases}
    x  & (- \frac{1}{4} < x  < \frac{1}{4}) \\
 -{1 \over 8 (1+2 x )}  & (- \frac{1}{2} \leqslant x  \leqslant  - \frac{1}{4}) \\
 {1 \over 8 (1-2 x )}   & (\frac{1}{4} \leqslant x  \leqslant \frac{1}{2}).
  \end{cases}
\end{equation}
Indeed, though the modified relative distance diverges ({\it i.e.}, $\left| {{\mathbf r}_d} \right| \to \infty$) at the edges of the Wigner-Seitz cell ({\it e.g.}, ${\mathbf{r}} =  \pm \frac{1}{2}{\mathbf{a}},\pm \frac{1}{2}{\mathbf{b}} , \pm\frac{1}{2}{\mathbf{c}}$), the exponential (Eq.{~\ref{onebody_u}}) and the Pad\'e (Eq.{~\ref{twobody_v}}) functions remain finite, $u\left( r \right) \to 1/2b_{\text{e}a}$ and ${v_{{\sigma _i},{\sigma _j}}}\left( {{r_{i,j}}} \right) \to 1/4b_{\rm{ee}}^{\rm{para}}$ or $ 1/2b_{\rm{ee}}^{\rm{anti}}$, respectively,  thus  preserving with  continuity the periodicity of 
the one-/two- body homogeneous parts of the Jastrow factor.
However, for the Pad\'e form, one has to change the expression for $p(x)$ in order to satisfy 
also the continuity in the WF derivatives,  
 even when the modified relative distances diverge, {\it i.e.},:
\begin{equation}
  p\left( x \right) = 
  \begin{cases}
    x  & (- \frac{1}{6} < x  < \frac{1}{6}) \\
 -{1 \over 54 (1/2+ x )^2}  & (- \frac{1}{2} \leqslant x  \leqslant  - \frac{1}{6}) \\
 {1 \over 54 (1/2- x )^2}   & (\frac{1}{6} \leqslant x  \leqslant \frac{1}{2}).
  \end{cases}
\end{equation}
In this case the region where $p(x)=x$ further shrinks. Thus, it 
is often more convenient to 
use  the exponential form to obtain a more accurate variational WF, because the long-range part is implicitly corrected by the inhomogeneous terms 
in Eq.{~\ref{threebody_jastrow}}.

\vspace{1mm}
Finally, we remark that the many-body WF also obeys the second Bloch condition{~\cite{1994RAJ, 1995RAJ}}, namely:
\begin{equation}
{\Psi _{{k_p}}}\left( {\left\{ {{{\mathbf{r}}_i} + {{\mathbf{T}}_p}} \right\}} \right) = {e^{i{{\mathbf{k}}_p} \cdot {{\mathbf{T}}_p}}}{\Psi _{{k_p}}}\left( {\left\{ {{{\mathbf{r}}_i}} \right\}} \right),
\end{equation}
where ${\mathbf{T}_p}$ represents a {\it unit-cell} (not supercell) vector, and ${{{\mathbf{k}}_p}}$ is the crystal momentum. This comes from the property that the 
many-body Hamiltonian is invariant under the {\it simultaneous translation} of all-electron coordinates by a unit-cell vector ${\mathbf{T}_p}$. Within \tvb, this condition can be employed 
by imposing the intra-unit cell translational 
symmetries
on the Jastrow and the pairing function, as simple  linear constraints in the variational parameters. However, this option  is restricted to the case ${{{\mathbf{k}}_p}}=0$. On the other hand, it is possible to use different twists on each spin 
component, that has proven very effective for implementing the  mentioned translation symmetries within pairing WFs.\cite{karakuzu}

\subsection{Finite-size effects}
The systematic error induced by a finite simulation cell is a long-standing issue in the {\it ab initio} QMC calculation.
There are two types of finite-size errors in QMC calculations; one is the so-called one-body effect that arises from the kinetic energy term of the Hamiltonian, and the other one is the so-called two-body effect that arises from the periodic Ewald contribution resulting from  the electron-electron interaction. Notice that, in the independent-particle calculations ({\it i.e.}, DFT), only 
the former is present, which can be readily evaluated by $k$ summation
in the first Brillouin zone, but the two-body finite-size effects are
not present because the exchange-correlation energy used in DFT
usually derives from DMC results extrapolated to the infinite
simulation cell size. To correct the one-body finite-size error, one
can use the twisted averaged boundary conditions~{\cite{2001LIN}},
special $k$-points methods~{\cite{1994RAJ, 1995RAJ}}, or the exact
special twist 
(EST) 
method~{\cite{2016DAG}}, all of which have been implemented in
\tvb. Within 
the 
\tvb\ implementation, the Jastrow part is independent of twists ({\it i.e.}, \tvb\ uses a common Jastrow for all twists),
\begin{equation}
{\Psi _{{k_s}}}\left(  {{{\mathbf{r}}_1}, \ldots ,{{\mathbf{r}}_N}} \right) = J\left( {{{\mathbf{r}}_1}, \ldots ,{{\mathbf{r}}_N}} \right)\Phi _{AS}^{_{{k_s}}}\left( {{{\mathbf{r}}_1}, \ldots ,{{\mathbf{r}}_N}} \right).
\label{wf-twist}
\end{equation}
As emphasized in Ref.~\onlinecite{2001LIN}, at variance with DFT,  the QMC computational effort is independent of the number of $k$-points used.
For the two-body finite-size effects, which cannot be alleviated by the above remedy, one can employ the model periodic Coulomb interaction (MPC)~{\cite{1995FRA, 1997WIL, 1999KEN}}. This method has not been implemented in \tvb\ yet.
Nevertheless, simpler alternatives exist. For instance one can alleviate the two-body finite size effects by directly increasing the supercell size, or one can estimate these effects by employing the KZK exchange-correlation function~{\cite{2008KWE}} at the DFT level. Moreover, it is also possible 
to employ systematic finite-size corrections based on the knowledge 
of the density structure factor in momentum space\cite{chiesa_fse_sk}, that can be readily computed within \tvb, with a short postprocessing computation.

\section{Built-in Density Functional theory (DFT) code}
\label{sec_dft}
Although most QMC codes load their trial WFs from available  DFT/quantum chemistry codes such as Gaussian{~{\cite{2016FRI}}}, CRYSTAL{~{\cite{2018DOV}}}, and Quantum Espresso{~{\cite{2009GIA}}}, \tvb\ 
is a self-consistent many-body package and does not require input from other codes. Indeed, in \tvb, an original DFT code 
is implemented that has the important feature to work in exactly  
the same basis and pseudopotentials (if any)  
used for  the QMC  many-body WF, and for instance can 
work with mixed Slater-Gaussian basis or with arbitrary contraction, {\it i.e.}, mixing angular momenta.  
On the other hand, the drawback is that only  very  few functionals 
are available so far because, within the \tvb\ approach, the DFT has to be used only to 
have  a reasonable initialization of the antisymmetric part of the many-body WF. This is then optimized at best by direct minimization 
of the energy in the presence of the Jastrow factor and should very weakly depend on the initialization, or for instance, on the DFT functional used. 
At present, \tvb\, supports Perdew-Zunger (PZ81) Local-Spin Density Approximation (LSDA){\cite{1981PER}}, and those combined with KZK exchange-correlation energy~{\cite{2008KWE}}, but, for instance, does not allow LDA + $U$ calculations. This will be the goal of a possible future work.

\subsection{Stabilization algorithm for large localized basis sets}
An important issue in any DFT algorithm based on localized basis sets, such as Gaussian, is how to reach the so-called complete basis set limit. In the plane-wave method, this is systematically achieved by increasing the number of plane waves ({\it e.g.}, kinetic cutoff energy). However, with a localized basis set,  by   increasing  its dimension, numerical instabilities may  occur 
as a consequence of the redundancy and non-orthonormality of the basis.
 On the other hand, the use of a small basis set implies too much biased results. In \tvb, an original and systematic algorithm to remove this redundancy{\cite{2010AZA}} has
 been implemented.
When the basis set is non-orthogonal, the key  self consistent step 
for solving the DFT   Kohn-Sham equations,  turns to 
a generalized eigenvalue problem:
\begin{equation}
{\boldsymbol {\mathcal{H}}} {\mathbf c} = E {\mathbf S} {\mathbf c},
\label{Kohn-Sham-original}
\end{equation}
where ${\boldsymbol{\mathcal{H}}}$ is the single-particle Hamiltonian matrix, ${\mathbf c}$ is the vector of the coefficients of the orbitals, $E$ is the single-particle energy, and the ${\mathbf S}$ is the $N_b$ $\times$ $N_b$ overlap matrix, where $N_b$ is the dimension  of the basis set:
\begin{equation}
{S_{i,j}} = \braket{{{{\phi _i}}} | {{{\phi _j}}}}= \int {d{\mathbf{r}}{\phi _i ^ *}\left( {\mathbf{r}} \right){\phi _j}\left( {\mathbf{r}} \right)}.
\label{overlap_matrix}
\end{equation}
%
The overlap matrix is strictly positive definite, namely all its eigenvalues 
are positive.
The problem mentioned above shows up when 
the overlap matrix ${\mathbf S}$ becomes  ill-conditioned.
If the condition number, namely the ratio between the largest (denoted as $s_{N_b}$) and the smallest eigenvalues (denoted as $s_1$), becomes very large, the CBS limit cannot be achieved by the standard procedure due to  numerical instabilities. Therefore, in \tvb, the small eigenvalues and the corresponding eigenvectors are disregarded.
Indeed, the original basis is modified as follows:
\begin{equation}
e_j^i = \frac{1}{{\sqrt {{s_i}} }}\upsilon _j^i\,\,{\text{for}}\,\,{s_i}/{s_N} \geqslant {\varepsilon _{{\text{mach}}}},
\end{equation}
where ${{{s}_i}}$ and $\upsilon _j^i$ are the $i$-th eigenvalues and eigenvectors of the overlap matrix, ${\varepsilon _{{\text{mach}}}}$ is usually set to the machine precision, $j$ runs from $1$ to $N_b$, $i$ runs from $1$ to $M_b$, and $M_b$ represents the number of eigenvalues satisfying the above inequality. 
If  infinite numerical accuracy were available, the matrix $S$ in  the  new 
basis would  be just the identity and the generalized eigenvalue 
problem in Eq.{~\ref{Kohn-Sham-original}} would turn to a standard one in this basis.
Within finite precision arithmetic it is better to iterate further the 
stabilization procedure and define a new basis:
\begin{equation}
\tilde e_j^i = \frac{1}{{\sqrt {{{\tilde s}_i}} }}\tilde \upsilon _j^i\,
\end{equation}
where ${{{\tilde s}_i}}$ and $\tilde \upsilon _j^i$ are the $i$-th eigenvalues and eigenvectors of the recomputed overlap matrix ${{\tilde s}_{i,j}} = \bra{e_i}{\mathbf{S}}\ket{e_j}$ (now well conditioned since very close to the 
identity), respectively.
For using the latter  basis set, the $N_b$ $\times$ $M_b$ global transformation matrix is stored, as it takes into account the projection from the original basis to the final modified one:
\begin{equation}
{U_{i,j}} = \sum\limits_k {e_j^k\tilde e_k^i}.
\end{equation}
On this basis set, the generalized eigenvalue problem becomes a well-conditioned one with a corresponding  $M_b$ $\times$ $M_b$ Hamiltonian matrix $\tilde{\boldsymbol{\mathcal{H}}} = {\mathbf U}{\boldsymbol {\mathcal{H}}} \tilde{\mathbf U}$. In this way, one can avoid the numerical instabilities induced by a too  redundant large basis set. This  stabilization  introduces an error at most $\sqrt {{\varepsilon _{{\text{mach}}}} \cdot N_b}$, which is typically  negligible compared with the  target chemical accuracy.

\subsection{Electron-ion cusp condition}
\label{electron_ion_cusp_dft}
Another feature of the  DFT code implemented in \tvb, 
is that electron-nuclei cusp conditions are exactly fulfilled 
for any basis ({\it i.e.}, Gaussian orbital)
even within the DFT framework. 
This is achieved by an appropriate modification of the standard basis sets 
commonly used ({\it e.g.}, ccpVTZ){~\cite{1989DUN}} for WF based calculations: the new basis 
is obtained by multiplying each element of  the original basis by 
 a suitably chosen  one-body Jastrow factor introducing the 
correct cusps, namely:
\begin{equation}
\tilde \phi _j^b\left( {{\mathbf{r}} - {{\mathbf{R}}_b}} \right) = \phi _j^b\left( {{\mathbf{r}} - {{\mathbf{R}}_b}} \right){{\tilde J}_1}\left( {\mathbf{r}} \right),
\label{onebody_j_single_DFT}
\end{equation}
where ${{\tilde J}_1}\left( {\mathbf{r}} \right)$ is the same as in Eq.{~\ref{onebody_J_hom}} and the parameter $b$ in Eq.{~\ref{onebody_u}} is optimized by direct minimization of the chosen DFT energy functional.
In this way  each element of the reshaped basis set satisfies the so-called electron-ion cusp conditions, namely that, when $ {\mathbf{r}} $ is close to any atomic position ${\mathbf{R}}_b$, the following relation holds:
\begin{equation}
\mathop {\lim }\limits_{{\mathbf{r}} \to {{\mathbf{R}}_b}} \frac{{\nabla \tilde \phi _j^a}}{{\tilde \phi _j^a}} =  - {Z_b}\frac{{{\mathbf{r}} - {{\mathbf{R}}_b}}}{{|{\mathbf{r}} - {{\mathbf{R}}_b}|}}
\end{equation}
for all $a,b$. This formulation allows us to reach the CBS limit  extremely fast in DFT calculations, 
especially within all electrons. 
Indeed,  we have experienced that  a  given target accuracy can be obtained with a smaller 
basis, {\it e.g.}, our modified ccpVDZ basis is typically equivalent in accuracy to 
the much larger ccpVTZ  (and  the ccpVTZ  equivalent to the ccpVQZ, and so on 
and so forth).  

\vspace{1mm}
For QMC application, this tool is particularly useful because 
the Slater determinant obtained with this new basis  does not have divergences in the local energy, that  are instead  present  in the original basis, when electron positions are close to the nuclear ones. 
This feature emphasizes further the clear  advantage of \tvb\ to allow 
the  use of a special purpose DFT code, just devoted to the 
optimal initialization of a QMC many-body electronic WF.

\subsection{Double-grid remedy for all-electron calculations}
As described in sec.~\ref{electron_ion_cusp_dft},
\tvb\ employs standard atomic orbitals, modified 
by means of an appropriately chosen one-body 
factor.
Thus the electron-electron integrations cannot be evaluated analytically even when the Gaussian atomic basis set is employed. Therefore, \tvb\, calculates the electron-electron Hartree potential by solving the Poisson's equation with the fast Fourier transform (FFT) on a Cartesian grid both in open and periodic systems. The numerical solution becomes problematic in all-electron calculations when the atomic number $Z$ becomes large. Indeed, in this case, 
the Kohn-Sham Hamiltonian matrix elements have to be computed in the presence of a rapidly varying electron density in the vicinity of nuclei, and therefore require   an exceedingly large FFT mesh, with an almost prohibitive 
computational cost for its evaluation.
 \tvb\, alleviates this drawback by a  double-grid scheme{~\cite{2019NAK}}, in which the Hartree potential is calculated with standard FFT convolution on a coarse mesh. 
Then, these values are interpolated on a much finer mesh in the vicinity of the nuclei. 
Although the DFT energy obtained with the above approximation is not exactly consistent with the one corresponding to a very dense uniform mesh, the corresponding  VMC energies and the variances are almost indistinguishable each other, at a much cheaper computational cost.

%
%

\section{Derivatives of energy}
\label{derivatives}
Derivatives of total energies with respect to variational parameters represent an  essential ingredient  for  optimizing a many-body WF. Forces (derivative with respect to atomic positions) are also essential for performing  structural optimization or  molecular dynamics. However, in a complex code, 
and especially in QMC, the evaluation of the functional  
derivatives, necessary for the WF optimization,  are very complicated, mainly for the complexity
of the algorithm that, in turn, may lead to a very inefficient implementation, 
though recent progress has been done{~\cite{2017ASS}}.
 For instance
, a simple approach is to compute them with finite difference expressions, 
 leading to  a   too   large computational time, because obviously proportional to the number of targeted derivatives. 

\vspace{1mm}
Algorithmic differentiation (AD) is a method capable 
of solving all the above  problems, essentially 
by a smart application of the differentiation chain rule.
There are two types of algorithms, the forward algorithmic differentiation (FAD) that implements  the chain rule straightforwardly from the beginning to the 
end of the algorithm and, the adjoint algorithmic differentiation (AAD), that  uses the chain  rule starting from the end of the algorithm (also known as  
backward propagation).
When the number of input parameters is much larger than the corresponding 
output ones, AAD is much more efficient than FAD and indeed allows the calculation of all possible derivatives in a computational time proportional (with a small prefactor, see {\it e.g.}, Fig.~\ref{2010SOR_aad})  to the one 
for computing  the target function ({\it i.e.}, the energy or the  WF value). 
The former is therefore 
the  ideal method for  Quantum Monte Carlo when the variational 
WF contains  several variational parameters.
The AAD was  applied for the calculation of atomic forces  
 in Ref.\onlinecite{2010SOR}. To our knowledge,
 this was  the first time that  AAD was used within QMC. 
Now, \tvb\, implements all  derivatives such as those of the  local energy ($\frac{\partial }{{\partial {\alpha _k}}}{e_L}\left( {{{\mathbf{x}}_i}} \right)$) or those corresponding to 
the WF logarithm ($\frac{\partial }{{\partial {\alpha _k}}}\ln \Psi \left( {{{\mathbf{x}}_i}} \right)$) using the AAD, which drastically improves the efficiency (Fig.~{\ref{2010SOR_aad}}){~{\cite{2010SOR}}} and reliability 
of the calculation.
%
\begin{figure}[htbp]
 \centering
 \includegraphics[width=7.6cm]{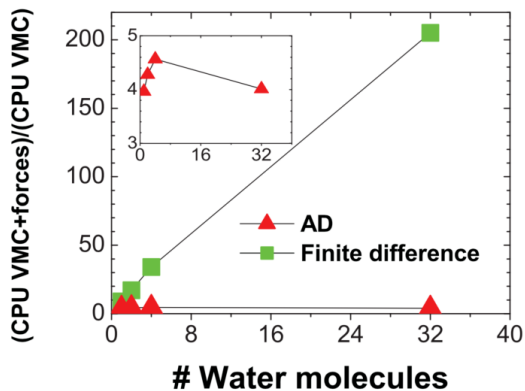}
 \caption{Ratio of CPU time required to compute energies and all force components referenced to the one required for the simple energy calculation within VMC. The calculations refer to 1, 2, 4, and 32 water molecules. The inset is an expansion of the lower part of the plot.
Reprinted with permission from {\citet{2010SOR}}, {\it J. Chem. Phys.}, published by AIP in 2010.
 }
 \label{2010SOR_aad}
\end{figure}

\vspace{1mm}
\subsection{Derivatives with respect to variational parameters}
The derivatives of the energy with respect to a given real variational parameter ${{\alpha _k}}$ (one complex parameter can be thought to be composed by two real ones, its real and imaginary part, respectively)  is represented as
a generalized \emph{force}:
\begin{equation}
{f_k} =- \frac{{\partial E\left( \alpha  \right)}}{{\partial {\alpha _k}}} =  - \frac{\partial }{{\partial {\alpha _k}}}\frac{{\braket{{\Psi _\alpha }|\hat {\mathcal{H}}|{\Psi _\alpha }}}}{{\braket{{\Psi _\alpha }|{\Psi _\alpha }}}}.
 \label{eq_general_force}
\end{equation}
In variational Monte Carlo, the derivative can be evaluated using $M$ configurations of electron coordinates{~\cite{2017BEC}}:
\begin{equation}
\begin{split}
{f_k}  
&=  - 2 \Re\left[ {\sum\limits_{\mathbf{x}} {\frac{{e_L ^ * \left( {\mathbf{x}} \right)\left( {{O_k}\left( {\mathbf{x}} \right) - {{\bar O}_k}} \right){{\left| {{\Psi _\alpha }\left( {\mathbf{x}} \right)} \right|}^2}}}{{\sum\nolimits_{\mathbf{x}} {{{\left| {{\Psi _\alpha }\left( {\mathbf{x}} \right)} \right|}^2}} }}} } \right] \\
&\approx  - 2 \Re\left[ {\frac{1}{M}\sum\limits_{i = 1}^M {e_L^ * \left( {{{\mathbf{x}}_i}} \right)\left( {{O_k}\left( {{{\mathbf{x}}_i}} \right) - {{\bar O}_k}} \right)} } \right],
\end{split}
\end{equation}
where ${e_L \left( {{{\mathbf{x}}}} \right)}$ is the local energy, 
${O_k}\left( {\mathbf{x}} \right)$ is the logarithmic derivative of the WF ({\it i.e.}, ${O_k}\left( {\mathbf{x}} \right) = \frac{{\partial \ln {\Psi _\alpha }\left( {\mathbf{x}} \right)}}{{\partial {\alpha _k}}}$), and ${{\bar O}_k}$ is its average over $M$ samples  ({\it {i.e.}}, ${{\bar O}_k} = \frac{1}{M}\sum\limits_{i = 1}^M {{O_k}\left( {{{\mathbf{x}}_i}} \right)}$). In \tvb, the logarithmic derivative (${O_k}\left( {\mathbf{x}} \right)$) is computed  very efficiently by using the AAD algoritm. Notice that the derivatives of the local energy are not needed here because the Hamiltonian does not depend on any variational parameter. Instead, these terms  are  necessary in order to calculate ionic forces ({\it i.e.}, derivatives of the  total energies with respect to atomic postions).
If the WF is an exact eigenstate of the hamiltonian, the generalized forces  ${f_k}$ exactly vanish without statistical errors because the local energy is no longer dependent on ${\mathbf{x}}$. In other words, the derivatives have the zero-variance property and  represent  therefore 
the fundamental ingredients for  
an efficient WF optimization, as described sec.{~\ref{sec_opt_wf}}.

\vspace{1mm}
In practice, during an optimization, the code monitors the variational energy ($E\left( \boldsymbol{\alpha} \right)$) and the maximum value of the signal to noise ratio among all  the force components, which is denoted as \verb| devmax |  in the code:
%
\begin{equation} \label{eq:devmax}
\verb| devmax | \equiv \max_k \left( {\left| {\frac{{{f_k}}}{{{\sigma _{{f_k}}}}}} \right|} \right)
\end{equation}
where  ${\sigma _{{f_k}}}$ represents the estimated error bar of the force ${f_k}$. 
This value  is used in \tvb\ as one of the convergence criteria  of the optimization. To our experience, after a reasonable number of  iterations devmax 
stabilizes to a value $\le$ 4.0, 
when a correct (possibly local) minumum is being approached.

\subsection{Atomic forces}
\label{ionic_forces}
In \tvb, a very useful  space warp coordinate transformation (SWCT){~\cite{2010SOR}} is employed for the calculation of atomic forces. This scheme was  introduced several decades ago in Ref.~\onlinecite{1989UMR},  in order to decrease the statistical errors of forces.
The finite-difference expression for  ionic forces calculations is:
\begin{equation}
{{\mathbf{F}}_a} =  - \frac{\Delta }{{\Delta {{\mathbf{R}}_a}}}E =  - \frac{{E\left( {{{\mathbf{R}}_a} + \Delta {{\mathbf{R}}_a}} \right) - E\left( {{{\mathbf{R}}_a}} \right)}}{{\Delta {{\mathbf{R}}_a}}} + O\left( {\Delta {{\mathbf{R}}_a}} \right).
\end{equation}
The SWCT is  used to mimic the displacement of charges around the nucleus,  
\begin{equation}
{{\mathbf{\bar r}}_i} = {{\mathbf{r}}_i} + \Delta {{\mathbf{R}}_a}{\omega _a}\left( {{{\mathbf{r}}_i}} \right), 
\end{equation}
\begin{equation}
{\omega _a}\left( {\mathbf{r}} \right) = \frac{{ \kappa \left( {\left| {{\mathbf{r}} - {{\mathbf{R}}_a}} \right|} \right)}}{{\sum\nolimits_{b = 1}^M {\kappa\left( {\left| {{\mathbf{r}} - {{\mathbf{R}}_b}} \right|} \right)} }},
\end{equation}
where ${\kappa\left( r \right)}$ is a function that decays sufficiently fast
for large $r$ because the charges far from the nuclei should not be 
affected by the  SWCT, and $\omega\to 0$, as a consequence of this requirement.  It turns out  that the original\cite{1989UMR} simple choice $\kappa\left( r \right) = 1/{r^4}$ works very well and is indeed adopted in \tvb.

\vspace{1mm}
Starting from the finite-difference expression, one can straightforwardly derive the corresponding differential expression{~\cite{2010SOR}}:
\begin{equation}
{{\mathbf{F}}_a} =  - \left\langle {\frac{d}{{d{{\mathbf{R}}_a}}}{e_L}} \right\rangle  + 2\left( {\left\langle {{e_L}} \right\rangle \left\langle {\frac{d}{{d{{\mathbf{R}}_a}}}\ln \left( {{{\mathcal{J}}^{\frac{1}{2}}}\Psi } \right)} \right\rangle  - \left\langle {{e_L}\frac{d}{{d{{\mathbf{R}}_a}}}\ln \left( {{{\mathcal{J}}^{\frac{1}{2}}}\Psi } \right)} \right\rangle } \right)
\label{forces:aad}
\end{equation}
where ${\mathcal{J}}$ is the Jacobian of the above transformation, and the brackets indicates a Monte Carlo average over the trial WF. All the terms above can be written by the partial derivatives of the local energy and those of the logarithm of the WF{~\cite{2010SOR}}:
\begin{equation}
\frac{d}{{d{{\mathbf{R}}_a}}}{e_L} = \frac{\partial }{{\partial {{\mathbf{R}}_a}}}{e_L} + \sum\limits_i^{} {{\omega _a}\left( {{{\mathbf{r}}_i}} \right)\frac{\partial }{{\partial {{\mathbf{r}}_i}}}{e_L}},
\end{equation}
\begin{equation}
\frac{d}{{d{{\mathbf{R}}_a}}}\ln \left( {{{\mathcal{J}}^{\frac{1}{2}}}\Psi } \right) = \frac{\partial }{{\partial {{\mathbf{R}}_a}}}\ln \left( \Psi  \right) + \sum\limits_i^{} {\left[ {{\omega _a}\left( {{{\mathbf{r}}_i}} \right)\frac{\partial }{{\partial {{\mathbf{r}}_i}}}\ln \left( \Psi  \right) + \frac{1}{2}\frac{\partial }{{\partial {{\mathbf{r}}_i}}}{\omega _a}\left( {{{\mathbf{r}}_i}} \right)} \right]}.
\end{equation}
In order to evaluate these differential expressions,  6$N$ + 6$N_{\rm at}$ components have to be evaluated, namely, $\left\langle {\frac{\partial }{{\partial {{\mathbf{r}}_i}}}{e_L}} \right\rangle$, $\left\langle {\frac{\partial }{{\partial {{\mathbf{r}}_i}}}\ln \Psi } \right\rangle$, $\left\langle {\frac{\partial }{{\partial {{\mathbf{R}}_a}}}\ln {e_L}} \right\rangle $, $\left\langle {\frac{\partial }{{\partial {{\mathbf{R}}_a}}}\ln \Psi } \right\rangle ${~{\cite{2010SOR}}}. These values are very efficiently computed in \tvb, by using the aforementioned AAD, that works even in presence of  pseudopotentials.

\vspace{1mm}
The SWCT significantly decreases the statistical errors, but the forces still have infinite variance properties because $\partial {e_L}$ and $\partial \ln \left( \Psi  \right)$ diverge in the vicinity of the nodal surfaces. Attaccalite and Sorella~{\cite{2008ATT}} developed a reweighting method 
to address this issue in an unbiased way.
Within this scheme, the QMC sampling is not driven by the chosen  trial WF ${\Pi _{\text{T}}}\left( {\mathbf{x}} \right) = \Psi _{\text{T}}^2\left( {\mathbf{x}} \right)$, but by a slightly different  guiding function ${{\Psi _{\text{G}}}\left( {\mathbf{x}} \right)}$ defined by
\begin{equation}
{\Psi _{\text{G}}}\left( {\mathbf{x}} \right) = \frac{{{R^\varepsilon }\left( {\mathbf{x}} \right)}}{{R\left( {\mathbf{x}} \right)}}{\Psi _{\text{T}}}\left( {\mathbf{x}} \right)
\label{guiding_function}
\end{equation}
where $R\left( {\mathbf{x}} \right)$ vanishes more weakly than the 
trial function does, when the configuration ${\mathbf  x}$ 
approaches the nodal surface, whereas ${R^\varepsilon }$ 
is an appropriately chosen regularization of $R$, depending  on  
$\varepsilon$, that never vanishes. 
In this way, the guiding function is larger when approaching 
the nodal surface, and this singular region can be  more accurately sampled in order to avoid  infinite variance problems in the calculation of forces, as it will be shown later.
For this purpose, $R\left( {\mathbf{x}} \right)$ is defined in a way 
to vanish as the antisymmetric part of the WF 
to some power $2 \theta_R \le 1$. However, in order 
to avoid too large fluctuations,
a simple relation is used  
that ${\rm Pf} (G) \simeq {1 \over  G^{-1}_{i,j}} $,
where ${G_{i,j}}$ is the pairing function (Eq.{\ref{eq:g}}), here referred to
the general AGP case (the other cases straightforwardly follows upon some 
restriction of the matrix $G$).
Indeed, if the Pfaffian goes to zero, most of the matrix elements  of the inverse of $G$ diverge inversely proportional to it. 
Thus the quantity
\begin{equation}
R\left( {\mathbf{x}} \right) = \left( S \sum\limits_{i,j} {{{\left| {G_{i,j}}^{ - 1} \right|}^2}}   \right)^{ - \theta_R}
\label{R_r}
\end{equation}
satisfies the  requirement without depending explicitly on the full Pfaffian 
that has fluctuations exponentially large in the number of particles and 
would lead to an inefficient regularization\cite{2008ATT}.
Here, we have introduced also the  scaling 
factor $S$, that  takes into account 
that  the WF may vanish also when a single particle is 
going to infinity, and not because it is approaching the nodal surface.
In this case, the scaling factor has to vanish in a proper 
way so as to allow a non-vanishing $R$.  In this way, 
the guiding function can decay sufficiently fast in this limit and 
can be  normalized in an open system. This represents  
a necessary condition to have  a stable 
simulation, otherwise all electrons will be kicked out at infinity, after 
a long  Markov  chain.
Therefore, the factor $S$ is chosen to vanish as long  as  one raw 
of the 
pairing function is vanishing, because it is corresponding to an 
electron going  to infinity (all matrix elements containing such 
electron coordinate have to vanish):
\begin{equation}
S={\rm min}_i \sum\limits_j |G_{ij}|^2.
\end{equation}
This definition is also useful, because the regularization proposed is scale-invariant, namely $R\left( {\mathbf{x}} \right)$ remains unchanged if $G$ is scaled by  an arbitrary constant and therefore is adopted as it is also for bulk systems, and it is particularly important when, in these cases, there  exist large regions of almost negligible electronic density.{\cite{2013ZEN}}

\vspace{1mm}
After that  ${R^\varepsilon }\left( {\mathbf{x}} \right)$ is defined as:
\begin{equation}
  {R^\varepsilon }\left( {\mathbf{x}} \right) = {\rm max}\left[R\left( {\mathbf{x}} \right), \varepsilon\right] 
\label{regularization}
\end{equation}
that is much simpler than the original proposal reported in Ref.~\onlinecite{2008ATT}.
By using the new probability, forces (the HF and the Pulay) can be evaluated {\em with finite variance} as:
\begin{equation}
 {\left\langle {\mathcal{W} \left( {\mathbf{x}} \right){\mathbf F} \left( {\mathbf{x}} \right)} \right\rangle _{{\Pi _{\text{G}}}\left( {\mathbf{x}} \right)}}/{\left\langle {\mathcal{W}\left( {\mathbf{x}} \right)} \right\rangle _{{\Pi _{\text{G}}}\left( {\mathbf{x}} \right)}},
\label{reweighting}
\end{equation}
where the $ \mathcal{W} \left( {\mathbf{x}} \right)$ is the new weight:
\begin{equation}
\mathcal{W}\left( {\mathbf{x}} \right) = {\left( {\Psi_\text{T} \left( {\mathbf{x}} \right)/{\Psi _{\text{G}}}\left( {\mathbf{x}} \right)} \right)^2} \equiv \left( {R\left( {\mathbf{x}} \right) \over {\rm max} \left[  R\left( {\mathbf{x}} \right) , \varepsilon\right] } \right)^2
\end{equation}
and ${\mathbf F} \left( {\mathbf{x}} \right)$, according to Eq.~\ref{forces:aad},  indicates  the average of an appropriate random variable containing 
either the derivative of the local energy (the HF contribution)  or the 
product of the local energy and the log derivative of the WF (the  Pulay one), 
both  contributions diverging as $\simeq {1 \over \delta^2}$, where $\delta$ is the distance from the nodal surface.
The point is that, since in the vicinity of the nodal surface ${\mathcal{W}\left( {\mathbf{x}} \right)}$ is proportional to ${\delta ^{4 \theta_R}}$,  whereas the probability vanishes much slower as ${\Pi _{\text{G}}}\left( {\mathbf{x}} \right) \simeq \delta^{2  -4 \theta_R}$, the reweighting scheme solves the infinite variance problem  for 
$1/4< \theta_R \le 1/2$  because the random variable $ {\mathcal{W}\left( {\mathbf{x}} \right)} {\mathbf F} \left( {\mathbf{x}} \right)$ remains  
with  {\em finite variance}. This follows  after simple inspection of its integrability in the small 
$\delta$ region, {\it i.e.}, $\int d\delta \delta^{4\theta_R-2} <\infty$. Thus, both  the HF and the Pulay terms can be evaluated with finite variances in the \tvb\ evaluation of atomic forces. In the most recent  versions of 
\tvb,  the default value of $\theta_R$ is taken to be at the middle  of  the  interval of stability, {\it i.e.}, $\theta_R=3/8$,  because 
the value $\theta=1/2$, previously chosen\cite{2008ATT}, leads to some instability, namely, during the QMC sampling,  the nodal surface is approached
too much closely, and the  determinants or Pfaffians become very ill-conditioned with  obvious  problems of numerical accuracy.
Moreover, the value of $\varepsilon$ is automatically selected during the first 
steps of the simulation in a way  that the average reweighting factor $\left\langle {\mathcal{W}\left( {\mathbf{x}} \right)} \right\rangle \simeq 0.8$, that represents empirically an almost  optimal setting.

\vspace{1mm}
Finally, we remark that this method to compute the energy derivatives 
with finite variance is applied by default  in \tvb\   even for the 
optimization of the WF variational parameters, and   
that a similar approach can be used when  computing  atomic forces within  the 
LRDMC algorithm. 

%
%

\section{Optimization of WF\MakeLowercase{s}}
\label{sec_opt_wf}

\subsection{Stochastic reconfiguration}\label{sec:SR}

Once the energy derivatives can be computed, the most straightforward strategy to optimize a WF is to employ the steepest descent method, where the WF parameters are iteratively updates as follows:
\begin{equation}
{\alpha_k} \to {\alpha'_k} = {\alpha _k} + \delta {\alpha _k}
\end{equation}
\begin{equation}
\delta {\alpha _k} =  \Delta {\mathbf{f}_k},
\end{equation}
where $\Delta$ is a small constant and ${f_k} \equiv -\frac{{\partial E}}{{\partial {\alpha _k}}}$ is the generalized force already defined in Eq.{~\ref{eq_general_force}}.
However, it does not work well when optimizing highly non-linear WF parameters because a small change of a given variational parameter may produce a very different WF, whereas another parameter change may weakly affect the WF. Of course, one can use  more sophisticated methods such as the Newton-Raphson method, the conjugate gradient, the quasi-Newton method, but the straightforward implementation of these optimizations do not work efficiently within a stochastic 
approach like QMC. In order to overcome this difficulty,  a more efficient change in the variational parameters  has been defined by means of a positive-definite {\it preconditioning} matrix ${\boldsymbol{\mathcal{S}}}$ and the generalized force vector $\mathbf{f}$:

\begin{equation}\label{eq:SRupdate}
{\alpha _k} \to {\alpha _k} + \Delta  \cdot {\left( {{{\boldsymbol{\mathcal{S}}}^{ - 1}}{\mathbf{f}}} \right)_k},
\end{equation}
where 
the matrix ${\boldsymbol{\mathcal{S}}}$ is stochastically evaluated by means of  $M$ configuration samples ${\mathbf{x}} = \left\{ {{{\mathbf{x}}_1},{{\mathbf{x}}_2}, \ldots {{\mathbf{x}}_M}} \right\}$:
\begin{equation}
{{\mathcal{S}}_{k,k'}} = \left[ {\frac{1}{M}\sum\limits_{i = 1}^M {\left( {{O_k}\left( {{{\mathbf{x}}_i}} \right) - {{\bar O}_k}} \right) ^ * \left( {{O_{k'}}\left( {{{\mathbf{x}}_i}} \right) - {{\bar O}_{k'}}} \right)} } \right],
\end{equation}
where ${O_k}\left( {{{\mathbf{x}}_i}} \right) = \frac{{\partial \ln \Psi \left( {{{\mathbf{x}}_i}} \right)}}{{\partial {\alpha _k}}}$ and ${{\bar O}_k} = \frac{1}{M}\sum\limits_{i = 1}^M {{O_k}\left( {{{\mathbf{x}}_i}} \right)}$.
The resulting approach is the so-called stochastic reconfiguration (SR) method{~\cite{1998SOR}}. 
Mazzola {\it et al.} {\cite{2012MAZ}} pointed out that the matrix ${\boldsymbol{\mathcal{S}}}$ is essentially a {\it metric} for the parameter space,  measuring  the distance of the underlying normalized WF. Therefore, Eq.~\ref{eq:SRupdate} is simply the steepest descent in this curved manifold. 
This observation connects the SR method with the so-called natural gradient method, widely used in the context of deep learning{~\cite{1998AMA}}.
In this context, for each parameter $\alpha$, 
$\left( {{{\hat O}_k} - {{\bar O}_k}} \right)$ and ${{\mathcal{S}}_{k,k'}}$ can be interpreted as the score function ({\it i.e.}, the gradient of the log-likelihood function) and the Fisher information matrix (FIM), respectively, while 
the WF square ${\left| {\Psi \left( {{{\mathbf{x}}}} \right)} \right|^2}$ plays the role of  the likelihood function.
In this sense, the stochastic reconfiguration method is essentially identical to the natural gradient optimization with the FIM that has been intensively used in the machine-learning community.
%
\begin{figure}[htbp]
 \centering
 \includegraphics[width=7.6cm]{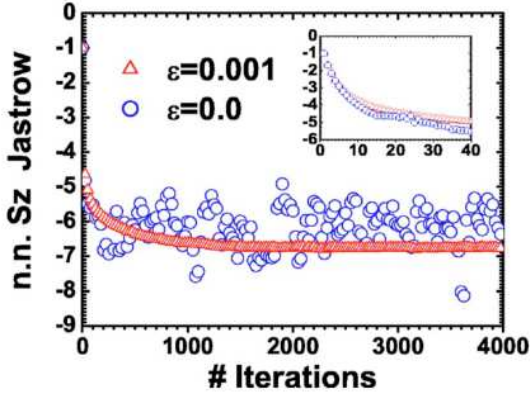}
 \caption{Optimization of the variational WF in the simple one-dimensional Heisenberg model ${\mathcal{H}} = J\sum\nolimits_i {{{\vec S}_i} \cdot {{\vec S}_{i + 1}}}$ with the standard SR ($\varepsilon$ = 0, open circles) and with the present regularization ($\varepsilon$ = 0.001, open triangles). In the figure, the evolution of the nearest neighbor spin-spin ($S_z$) Jastrow parameter is plotted. This figure clearly shows that the SR method with regularization is several orders of magnitude more efficient than the standard SR for determining the variational parameter with a given statistical accuracy. The inset shows the first few iterations.
Reprinted with permission from {\citet{2007SOR}}, {\it J. Chem. Phys.}, published by AIP in 2007.
 }
 \label{2007SOR_sr_eps}
\end{figure}
%

%
\vspace{1mm}
The straightforward implementation of the SR method is not stable mainly because the statistical noise sometimes makes the matrix ${\boldsymbol{\mathcal{S}}}$ ill-conditioned,
which deteriorates the efficiency of the optimization method{~\cite{2007SOR}}.
Therefore, in practice, the diagonal elements of the  preconditioning matrix ${\mathcal{S}}$ are shifted by a small positive parameter ($\varepsilon$) as:
\begin{equation}
{s'_{i,i}} = {s_{i.i}}(1 + \varepsilon ).
\label{regularization_s}
\end{equation}
This modification improves the efficiency of the optimization by  several orders of magnitude, as shown in Fig.{~\ref{2007SOR_sr_eps}}. Finally, the variational parameters are updated as:
\begin{equation}
{\alpha _k} \to {\alpha _k} + \Delta  \cdot {\left( {{{{\boldsymbol{\mathcal{S'}}}}^{ - 1}}{\mathbf{f}}} \right)_k}. \label{eq:sr}
\end{equation}
In practice, the user should provide  two input  parameters, namely, $\Delta$ (denoted as \verb|tpar| in the code input variables) and $\varepsilon$ (denoted as \verb|parr|).

\subsection{Linear method}\label{sec:linearmethod}
In \tvb, the state-of-the-art QMC optimization method is also implemented, namely the so-called linear method~{\cite{2005SOR, 2007UMR, 2007TOU}}. In this scheme, a many-body WF is 
expanded up the linear order ({\it i.e.}, considering only the first derivatives for the variational parameters):
\begin{equation}
\ket{{\Psi _{\alpha  + \delta \alpha }}} = \ket{{\Psi _\alpha }} + \sum\limits_{k = 1} {\delta {\alpha _k}\frac{{\partial}}{{\partial {\alpha _k}}} \ket{{\Psi _\alpha }} }  \equiv \ket{{\Psi _0}} + \sum\limits_{k = 1} {\delta {\alpha _k}\ket{{\Psi _k}}},
\end{equation}
where ${{\alpha _k}}$ is the $k$-th variational parameter, and ${{\Psi _k}\left( {\mathbf{x}} \right)}$ is the first derivative with respect to the $k$-th variational parameter. Within this expansion, the expectation value  of the  energy reads:
\begin{equation}
{E_{{\alpha}  + {z}}} = \frac{{\sum\nolimits_{k,k' = 0}^p {{z_k}^ *{z_{k'}} \braket{ {\Psi _k} | \hat {\mathcal{H}} | {\Psi _{k'}} } } }}{{\sum\nolimits_{k,k' = 0}^p {{z_k}^ *{z_{k'}} \braket{ {\Psi _k} | {\Psi _{k'}} } } }}.
\end{equation}
In order to obtain the  vector 
${\mathbf{z}}$ that minimizes this expectation value, one should solve the generalized eigenvalue problem:
\begin{equation}
{\boldsymbol{\mathcal{H}}\mathbf{z}} = E{\boldsymbol{\mathcal{S}}\mathbf{z}},
\label{gen_eigen_prob_linear}
\end{equation}
where $ {{\mathcal{H}}_{k,k'}} = \braket{ {\Psi _k} | \hat {\mathcal{H}} | {\Psi _{k'}}} $ and ${{\mathcal{S}}_{k,k'}} = \braket{{\Psi _k} | {\Psi _{k'}}}$.

\vspace{1mm}
This algorithm is more conveniently implemented in the so-called semi-orthogonal basis{~\cite{2017BEC}}:
\begin{equation}
\ket{{\Psi _{\alpha  + \delta \alpha }}} = {z_0} \ket{{{\tilde \Psi }_\alpha }} + \sum\limits_{k = 1} {{z_k}\left( {{{\hat O}_k} - {{\bar O}_k}} \right) \ket{{{\tilde \Psi }_\alpha }}},
\end{equation}
where $\ket{{{\tilde \Psi }_\alpha }} = \ket{{{\Psi _\alpha }}} / {{\left\| {{\Psi _\alpha }} \right\|}}$, $\braket{x|{\hat{O_k}}|x'} = {\delta _{x,x'}}{O_k}\left( {\mathbf{x}} \right)$, ${O_k}\left( {\mathbf{x}} \right) = \frac{\partial }{{\partial {\alpha _k}}}\ln \left| {{\Psi _\alpha }\left( {\mathbf{x}} \right)} \right|$, and ${{\bar O}_k} = \braket{{\Psi _\alpha }|{\hat{O_k}}|{\Psi _\alpha }}/\braket{{\Psi _\alpha }|{\Psi _\alpha }}$. \tvb\, calculates ${\mathbf{z}}$ by solving the generalized equation (Eq.~{\ref{gen_eigen_prob_linear}}) with the matrices:
\begin{equation}
{{\mathcal{H}}_{k,k'}} \approx \frac{1}{M}\sum\limits_{i = 1}^M {\left( {{O_k}\left( {{x_i}} \right) - {{\bar O}_k}} \right) ^ * \frac{{\braket{{x_i}|\hat {\mathcal{H}}\left( {\hat{O_{k'}} - {{\bar O}_{k'}}} \right)|{\Psi _\alpha }}}}{\braket{{x_i}|{\Psi _\alpha }}}}
\end{equation}
\begin{equation}
{{\mathcal{S}}_{k,k'}} \approx \frac{1}{M}\sum\limits_{i = 1}^M {\left( {{O_k}\left( {{x_i}} \right) - {{\bar O}_k}} \right) ^ * \left( {{O_{k'}}\left( {{x_i}} \right) - {{\bar O}_{k'}}} \right)}
\end{equation}
which can be readily evaluated by a Monte Carlo sampling, using not only the 
random variables $O_k(x_i)$ necessary  for the simpler SR technique, but also 
the parameter derivatives of the local energy  $\partial_{\alpha_k} e_L( x_i)$, that can be directly computed by AAD, 
and thus allow the calculation of the above Hamiltonian matrix elements, by straightforward algebra:
\begin{equation}
\partial_{\alpha_k} e_L(x_i) = \partial_{\alpha_k} { \langle x_i | \hat {\mathcal{H}}  |{\Psi _\alpha } \rangle \over   \langle  x_i  |{\Psi _\alpha } \rangle }= \frac{{\braket{{x_i}|\hat {\mathcal{H}}\left( {\hat{O_{k'}} - {{\bar O}_{k'}}} \right)|{\Psi _\alpha }}}}{\braket{{x_i}|{\Psi _\alpha }}}.
\end{equation}
 Notice that the code does not always take the eigenvector corresponding to the lowest eigenvalue (that in turn may be also complex, as in the linear method the matrix $ {\mathcal{H}}_{k,k^\prime}$ is not Hermitian), but takes 
 the  one that maximizes ${\left| {{z_0}} \right|^2}$ ({\it i.e.}, the coefficient of the zeroth-order term) in order to have the most stable 
 parameter change, namely the updated WF being  as close as possible to the old one. Finally, the code updates the variational parameters by using the obtained ${\mathbf{z}}$ and an input parameter $\Delta$ (denoted as \verb|tpar|) as:
\begin{equation}
{\alpha _k} \to {\alpha _k} + \Delta \cdot {z_k}/{z_0}
\end{equation}
where $\Delta \simeq 1/3$ is the default \tvb\ choice that is much more stable than the  original algorithm ($\Delta=1$),  but it is  
approximately $1/3$ slower.
The linear  method is usually  rather unstable  for large systems 
and many parameters, and also, in  such cases, 
each iteration requires a diagonalization of a huge matrix (a task  that can be  often prohibitive). In \tvb, a practical  
compromise has been devised by using 
the linear method for a restricted variational space containing up to  
a maximum number (\verb|npbra|) of  variational  parameters with largest 
signal to noise ratio (optionally larger than \verb|parcutpar|, see Eq.~\ref{eq:devmax})  and/or a number (\verb|ncg|)  of 
global line parameter directions spanned by the 
natural gradient ones ({\it e.g.}, $\Delta$ is one variational parameter of this form  
in Eq.~\ref{eq:sr}) calculated at the given optimization 
iteration or the previous closest ones{~\cite{2007SOR}}, optionally neglecting 
the ones with smaller signal-noise ratio ($<$ \verb|parcutmin|). 

\subsection{Practical Rule}

In most optimizations, the number of samplings in VMC is much  larger than the number of variational parameters and the optimization is stable and 
efficient. In the opposite case, the preconditioning matrix ${\mathbf S}$ becomes rank-deficient singular; therefore the optimization does not work properly ({\it c.f.} the Cram\'{e}r-Rao inequality). A practical rule is to set the number of samples $M$ such that{~\cite{2017BEC}}:
\begin{equation}
M \geqslant 5 \sim 10 \times p,
\end{equation}
where $p$ is the number of variational parameters.
However, one can also employ very small number of samplings with sufficiently large $\varepsilon$ in the stochastic reconfiguration method because the regularization of the matrix ${\mathbf S}$ in Eq.{~\ref{regularization_s}} can remove all instabilities  as long as $\varepsilon$ is sufficiently large, as in  the infinite $\varepsilon$  limit one recovers the very powerful stochastic gradient 
technique well-known within machine learning.
This method seems promising for optimizing even a huge amount of variational parameters and could certainly represent an important 
development in QMC.

\vspace{1mm}
At present, the inversion in Eq.{~\ref{eq:sr}} can be done in \tvb\ 
without storing explicitly the matrix{\cite{2012NEU}} as the associated 
linear problem is solved iteratively using conjugate gradients and implicit 
small matrix-vector operations, optimally distributed within the MPI 
protocol. Thus the cost of this inversion becomes negligible in the large $\varepsilon$ limit, and linear with the number of variational parameters.

\section{Molecular dynamics}
\label{molecular_dynamics}
In \tvb,  several types of  {\it ab initio} molecular dynamics (MD)
have been implemented
for both
classical and quantum nuclei. 
The MD is driven by
ionic forces $\mathbf{F} =
-\boldsymbol{\nabla}_{\mathbf{R}}V(\mathbf{R})$, where
$\mathbf{F} \equiv \{\mathbf{F}_1, \ldots, \mathbf{F}_N\}$,
$\mathbf{R} \equiv \{\mathbf{R}_1, \ldots, \mathbf{R}_N\}$, and the potential
energy landscape $V(\mathbf{R})$ is evaluated by VMC, namely
\begin{equation}
V(\mathbf{R}) = \frac{\langle \Psi_{\mathbf{R}}|{\hat{\mathcal{H}}}(\mathbf{R})|\Psi_{\mathbf{R}}\rangle}{\langle \Psi_{\mathbf{R}}|\Psi_{\mathbf{R}}\rangle}.
\label{V_wave_function}
\end{equation}
In Eq.~\ref{V_wave_function}, $|\Psi_{\mathbf{R}}\rangle$ is the QMC
WF that, according to the Born-Oppenheimer (BO) approximation,
minimizes the expectation value of ${\mathcal{H}}(\mathbf{R})$ at each ionic
position $\mathbf{R}$.
The expression for the ionic forces is quite
complex. Computing these forces with finite variance and in a fast
way, as done in \tvb,\ (Sec.~\ref{ionic_forces}),
is of paramount importance to make a QMC-based MD possible. 
Moreover, in the QMC framework the evaluation of $V(\bvec{R})$ and its
corresponding force estimators are intrinsically noisy. The
statistical noise must be kept under control, if one wants to have an unbiased sampling of the phase space
during the propagation of the trajectory. This issue has been solved
by resorting to a Langevin type of molecular dynamics in both
classical and quantum formulations, where the QMC noise becomes a
controlled source of thermalization at the target temperature $T$.


%
\vspace{1mm}
Two types of Langevin dynamics (LD) have been implemented in \tvb: 
the second-order LD, in its classical
(Sec.~\ref{second_order_langevin}) and quantum (Sec.~\ref{pioud}) variants, 
and 
the first-order LD accelerated by the covariance matrix
of QMC forces (Sec.~\ref{first_order_langevin}).

\subsection{Second-order Langevin dynamics}
\label{second_order_langevin}

The equations of motion of the second-order LD read{~\cite{2008ATT}}:
\begin{equation}
\dot {\boldsymbol \upsilon}  =  - \gamma \left( {\mathbf{R}} \right) \cdot {\boldsymbol \upsilon}  + {\mathbf{F}}\left( {\mathbf{R}} \right) + {\boldsymbol \eta} \left( t \right),
\label{second-order-stochastic-main}
\end{equation}
\begin{equation}
{\mathbf{\dot R}} = {\boldsymbol \upsilon} ,
\label{second-order-stochastic-velocity}
\end{equation}
\begin{equation}
\left\langle {{\eta _i}\left( t \right){\eta _j}\left( {t'} \right)} \right\rangle  = {\mathcal{S}_{i,j}}\left( {\mathbf{R}} \right)\delta \left( {t - t'} \right),
\label{second-order-stochastic-force}
\end{equation}
where ${\mathbf{R}}$, ${\boldsymbol \upsilon}$, ${\mathbf{f}}$,
${\boldsymbol \eta}$ are the 3$N_\textrm{at}$-dimensional vectors representing
atomic positions, velocities, deterministic and stochastic forces
of $N_\textrm{at}$ atoms, written in mass-scaled coordinates:
\begin{eqnarray}
 \bvec{R}_i &=& \bvec{R}_i^0  \sqrt{m_i},  \nonumber \\
 \bvec{F}_i &=& \bvec{F}_i^0 /\sqrt{m_i}, \nonumber \\
 \bsym{\eta}_i &=& \bsym{\eta}_i^0 \sqrt{m_i},
\label{mass_scaling}
\end{eqnarray}
for $i=1,\ldots,N_\textrm{at}$. The stochastic forces 
are related to the friction matrix $\gamma$ 
through the fluctuation-dissipation theorem, namely: 
\begin{equation}
\mathcal{S} \left( {\mathbf{R}} \right) = 2T\gamma \left( {\mathbf{R}} \right),
\label{cont-fluc-dis}
\end{equation}
with the temperature $T$ expressed in atomic units.
\tvb\, exploits the freedom in Eq.~\ref{cont-fluc-dis}, by assuming:
%
\begin{equation}
\mathcal{S} \left( {\mathbf{R}} \right) = {s_0} \mathcal{I} + {\Delta _0}{\mathcal{S}^{{\text{QMC}}}}\left( {\mathbf{R}} \right)
\label{stochastic_force}
\end{equation}
where ${s_0}$ and ${\Delta _0}$ are constants, 
$\mathcal{I}$ is the identity,
and ${\mathcal{S}^{{\text{QMC}}}}\left( {\mathbf{R}} \right)$ the covariance matrix of QMC forces:
\begin{equation}
\mathcal{S}_{i,j}^{^{{\text{QMC}}}}\left( {\mathbf{R}} \right) =
\Bigl\langle {\Bigl( {{\mathbf{F}_i}\left( {\mathbf{R}} \right) -
        \left\langle {{\mathbf{F}_i}\left( {\mathbf{R}} \right)}
        \right\rangle } \Bigr)} \Bigr\rangle 
\left\langle {\left( {{\mathbf{F}_j}\left( {\mathbf{R}} \right) 
     - \left\langle {{\mathbf{F}_j}\left( {\mathbf{R}} \right)} 
       \right\rangle } \right)} \right\rangle.
\label{qmc_force_cov}
\end{equation}
%
In the above Equation, $\left\langle  \cdots  \right\rangle$ refers to the average over
the QMC sampling at the given MD step.
The friction, a quantity that controls the nuclear sampling efficiency, is, therefore, position-dependent now.
Luo {\it et al.} have
shown
that the force covariance matrix 
$\mathcal{S}_{i,j}^{^{{\text{QMC}}}}\left( {\mathbf{R}} \right)$ is,
within a good approximation,
proportional to the dynamical matrix (Fig.~{\ref{2014LUO_force_cov}}).
%
\begin{figure}[htbp]
 \centering
 \includegraphics[width=6.7cm]{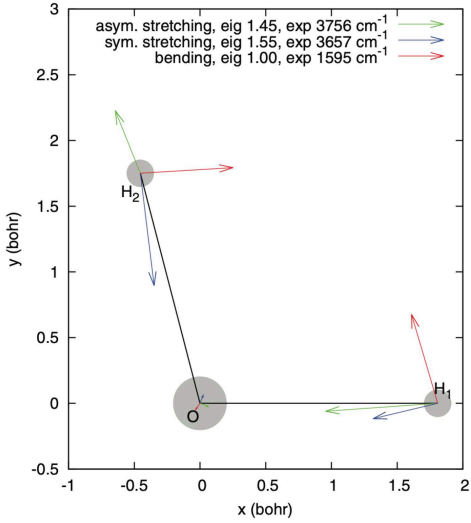}
 \caption{Eigenvectors of the 3 $\times$ 3 correlation matrix of QMC
   forces $\mathcal{S}^{\text{QMC}}$ in the water monomer. The
   obtained eigenvectors correspond to the well-known three
   vibrational modes: bending (red), symmetrical (blue), and
   asymmetrical (green) stretching. The smaller eigenvalue corresponds
   to the lowest-frequency vibrational mode. The eigenvalues in the
   plot are all rescaled by the lowest 
one.
Reprinted with permission from {\citet{2014LUO}}, {\it J. Chem. Phys.}, published by AIP in 2014.
 }
 \label{2014LUO_force_cov}
\end{figure}
%
Therefore, the choice in Eq.~\ref{qmc_force_cov} realizes 
the nearly optimal damping 
of the 
high-frequency
modes. 
After discretizing Eqs.~{\ref{second-order-stochastic-main}} and
{\ref{second-order-stochastic-velocity}} in the time interval ${t_n} -
\tau /2 < t < {t_n} + \tau /2$, where the index $n$ denotes the time 
slice ${t_n} = n\tau$, with time step $\tau$, and by integrating the
above Equations 
one obtains:
\begin{equation}
{{\boldsymbol \upsilon} _{n + 1}}\left( t \right) = {e^{ - {\gamma _n}\tau }} \cdot {{\boldsymbol \upsilon} _n} + {\Gamma _n}\left( {{{\mathbf{F}}_n} + \tilde {\boldsymbol \eta} } \right)
\label{disc-second-order-stochastic-main}
\end{equation}
\begin{equation}
{{\mathbf{R}}_{n + 1}} = {{\mathbf{R}}_n} + \tau {{\boldsymbol \upsilon} _{n + 1}}
\label{disc-second-order-stochastic-velocity}, 
\end{equation}
with ${{\mathbf{F}}_n} \equiv {\mathbf{F}}\left( {{{\mathbf{R}}_n}} \right)$, ${\gamma _n} \equiv \gamma \left( {{{\mathbf{R}}_n}} \right)$, 
\begin{equation}
{\Gamma _n} = \gamma _n^{ - 1}\left( {{{I}} - {e^{ - {\gamma _n}\tau }}} \right),
\end{equation}
and
\begin{equation}
\tilde {\boldsymbol \eta}  = \frac{{{\gamma _n}}}{{2\sinh \left(
      {{\gamma _n}\tau /2} \right)}}\int_{{t_n} - \tau /2}^{{t_n} +
  \tau /2} {{e^{ - {\gamma _n}\left( {t - {t_n}} \right)}} \cdot
  {\boldsymbol \eta} \left( t \right)dt}.
\label{eta_noise}
\end{equation}
By using 
Eq.~\ref{eta_noise},
it is easy to show that the correlator defining the discrete noise is
given by the following $ 3 N_\textrm{at}$ $\times$ $3 N_\textrm{at}$
matrix: 
\begin{equation}
\left\langle {{{\tilde \eta }_i}{{\tilde \eta }_j}} \right \rangle  \equiv \mathcal{S}^*  = T \gamma _n^2\coth \left( {{\gamma _n}\tau /2} \right).
\label{disc-fluctuation-dissipation}
\end{equation}
%
In this way,  $\tilde {\boldsymbol \eta}$ fulfill the
fluctuation-dissipation theorem required by the Langevin
thermostat. However, part of this noise is already present in the QMC force
evaluation (${{\tilde{\boldsymbol \eta} }^{{\text{QMC}}}}$), as quantified by
Eq.~\ref{qmc_force_cov}. 
Hence, \tvb\ adds the
external random force ${{\tilde {\boldsymbol \eta} }^{{\text{ext}}}}$, such
that ${\tilde {\boldsymbol
    \eta}} = {{\tilde{\boldsymbol \eta} }^{{\text{ext}}}} + {{\tilde{\boldsymbol \eta} }^{{\text{QMC}}}}$.
This is generated according to the following correlator: 
\begin{equation}
\left\langle {\tilde \eta _i^{{\text{ext}}}\tilde \eta _j^{{\text{ext}}}} \right\rangle  =  \mathcal{S}^* - {\mathcal{S}_{\text{QMC}}},
\label{correlator_ext_force}
\end{equation}
in order to fulfill the original fluctuation-dissipation theorem.
Since the correlation matrix in Eq.~\ref{correlator_ext_force} is
positive definite as long as $\tau  < {\Delta _0}$, 
the correlated external force can be readily generated by standard
algorithms ({\it e.g.}, the Box-Muller's method{~{\cite{1958BOX}}
combined with the matrix diagonalization). This noise correction
plays an important role in targeting a stable temperature
even in the presence of statistical errors in the QMC forces. The
original numerical integration 
detailed here
has been 
further
improved by
G. Mazzola and S. Sorella (see the supplementary notes of
Ref.~{\onlinecite{2014MAZ}}). 

\subsection{Path integral Ornstein-Uhlenbeck Dynamics}
\label{pioud}

F. Mouhat {\it et. al} 
have
recently introduced the nuclear quantum
effects into the second-order LD based on a path integral molecular
dynamics (PIMD) approach~{\cite{2017MOU}}, driven by QMC nuclear forces.
As a starting point of the Feynman path integral (PI) theory,
the quantum partition function $Z =
\text{Tr}\left[e^{-\beta {\mathcal{H}}}\right]$ (with $\beta = 1/k_BT$) 
can be written as\cite{1959TRO}
\begin{equation}
Z \simeq \frac{1}{(2\pi \hbar)^ {3N_{\textrm{at}}L} }\int \textrm{d}^{N_{\textrm{at}}L}  \mathbf{p}
\,\textrm{d}^{N_{\textrm{at}}L}  \mathbf{q} \, e^{-\tau_\beta {\mathcal{H}}_L(\mathbf{p},\mathbf{q})}, 
\label{Z} 
\end{equation}
with  $\tau_\beta=\beta/L$ denotes  the imaginary time step. We have replaced here the true quantum
particles by fictitious classical ring polymers, consisting of $L$
replicas (beads) of the system. The beads in these necklaces are
connected to each other by harmonic springs, evolving in the  
imaginary time. Without loss of generality, we can extend the
definition of classical vectors to the quantum case, by including all
$L$ replicas, such that the resulting vector $\bvec{q} \equiv \left(
  \bvec{R}^{(1)},\ldots,\bvec{R}^{(i)},\ldots,\bvec{R}^{(L)}
\right)^T$ is $3N_\textrm{at}L$-dimensional. If interpreted
classically, the partition function $Z$ in Eq.{~\ref{Z}} describes a
system at the effective temperature $L T$. The Hamiltonian ${\mathcal{H}}_L$
corresponding to this quantum-to-classical isomorphism reads as 
\begin{eqnarray}
{\mathcal{H}}_L(\mathbf{p},\mathbf{q})&=& \sum \limits_{i=1}^{N_\textrm{at}}\sum
                              \limits_{j=1}^L\left( \frac{1}{2}
                              [\mathbf{p}_i^{(j)}]^2 +
                              \frac{1}{2}\tilde{\omega}_L^2\left(\bvec{R}_i^{(j)}-\bvec{R}_i^{(j-1)}\right)^2\right)
                              \nonumber \\ 
&+& \sum \limits_{j=1}^L V(\bvec{R}_1^{(j)},..,\bvec{R}_{N_\textrm{at}}^{(j)}),
\label{mapping}
\end{eqnarray}
written in mass-scaled coordinates Eq.{~\ref{mass_scaling}}, with
$\tilde{\omega}_L = L/\beta \hbar$, and subjected to the ring boundary
condition $\bvec{R}_i^{(0)} \equiv \bvec{R}_i^{(L)}$. 
The system is thermalized by a Langevin thermostat, such that the
related Liouville operator ${\cal L}$  can be written as:
\begin{eqnarray}
  {\cal L} & = &  \sum_{i=1}^{3N_\textrm{at}L} \Biggl( {\cal F}_i \partial_{p_i} + p_i \partial_{q_i}  \Biggr. \nonumber \\
 & - &  \Biggl. \sum_{j=1}^{3N_\textrm{at}L} \gamma_{ij}
       \biggl(\partial_{p_i} p_j + k_B T
       M \partial_{p_i}\partial_{p_j}\biggr) \Biggr) 
\label{liouvillian}
\end{eqnarray}
in mass-scaled coordinates, with $\partial_{q_i} \equiv
\frac{\partial}{\partial_{q_i}}$. $ {\cal L}$ is built upon the
Hamiltonian propagation, driven by the first two terms, and the
Langevin thermostat, represented by the last two, deriving from the
Fokker-Planck equation. Hereafter in this Section, 
$i$ and $j$ run over all particles and beads indexed together.
In Eq.{~\ref{liouvillian}}, $ {\cal F}_i \equiv F^\textrm{BO}_i +
F^\textrm{harm}_i$ is the total force acting on each replica, comprising the BO (intra-replica) and
harmonic contributions (inter-replicas), where $F^\textrm{BO}_i \equiv
-\partial_{q_i}{\tilde V}$ and ${\tilde V}=\sum\limits_{j=1}^L
V(\bvec{R}_1^{(j)},..,\bvec{R}_{N_\textrm{at}}^{(j)})$. 

\vspace{1mm}
The quantum-to-classical isomorphism Hamiltonian in
Eq.{~\ref{mapping}} includes very different energy scales.  
To cope with this, we split the Liouvillian in Eq.{~\ref{liouvillian}}
into two operators, one containing only the physical (BO) modes,
the other depending exclusively on the harmonic modes of the rings. To
do so, we first separate the friction matrix into two contributions: 
\begin{equation}
\boldsymbol{\gamma} = \boldsymbol{\gamma}^\textrm{BO} + \boldsymbol{\gamma}^\textrm{harm}.
\label{gamma_split}
\end{equation}
We can then rewrite the total Liouvillian as the sum of two terms,
${\cal L} =  {\cal L}^\textrm{BO} + {\cal L}^\textrm{harm}$, where 
\small
\begin{eqnarray}
{\cal L}^\textrm{harm} & = &  \sum_i \Biggl( F^\textrm{harm}_i \partial_{p_i} + p_i \partial_{q_i} \Biggr. \nonumber \\
                                 & - & \Biggl. \sum_j \gamma^\textrm{harm}_{ij} \biggl(\partial_{p_i} p_j + k_B T M \partial_{p_i}\partial_{p_j}\biggr) \Biggr), \label{L_harm} \\
{\cal L}^\textrm{BO} & = & \sum_i \Biggl( F^\textrm{BO}_i \partial_{p_i}\Biggr. \nonumber \\ 
                             & - & \Biggl. \sum_j \gamma^\textrm{BO}_{ij} \biggl(\partial_{p_i} p_j + k_B T M \partial_{p_i}\partial_{p_j}\biggr) \Biggr),
\label{L_BO}
\end{eqnarray}
\normalsize
in such a way that we can break up the evolution operator via a Trotter
factorization\cite{1959TRO} to get the following symmetric propagator
\cite{1992TUC,1992SEX}: 
\begin{equation}
e^{-{\cal L}\delta t} \simeq e^{-{\cal L}^\textrm{BO} \delta t/2}
e^{-{\cal L}^\textrm{harm} \delta t}  e^{-{\cal L}^\textrm{BO} \delta
  t/2}. 
\label{secondalgo}
\end{equation}

\vspace{1mm}
The equations of motion
corresponding to the propagator in Eq.~\ref{secondalgo} have been
implemented
in \tvb,
where now
$F^\textrm{BO}_i$ can be computed in a QMC framework, and therefore, it
can be affected by an intrisic noise, as it was in the classical scheme
(Sec.~\ref{second_order_langevin}). 

\vspace{1mm}
We note that the equations of
motion generated by ${\cal L}^\textrm{harm}$ are \emph{linear} in both $\bvec{p}$ and
$\bvec{q}$, thus exactly solvable in an analytic closed form. This is
because the non-linear BO components of the total force have been put in
the ${\cal L}^\textrm{BO}$ factor. The dynamics generated by
${\cal L}^\textrm{harm}$ is a quantum Ornstein-Uhlenbeck process, which
describes the motion of Brownian quantum particles under the influence of
friction.

\vspace{1mm}
The exact integration of ${\cal L}^\textrm{harm}$ without
further splitting the Langevin thermostat from the harmonic
modes of the rings is a major achievement, which gives the
algorithm an enhanced stability and better scaling with respect to the time step size $\tau$ and
the number of replica $L$. Moreover, the time step error is
significantly smaller than the one of other PILD methods\cite{2010CER}. Thus, this
is a key feature of the PI algorithm implemented in \tvb. This is also why
the implemented method is called path integral Ornstein-Uhlenbeck dynamics
(PIOUD). 

\vspace{1mm}
If we now look at the ${\cal L}^\textrm{BO}$ factor in
Eq.~\ref{secondalgo}, the corresponding equations of motion are the
ones in Eq.~(12) of Ref.~\onlinecite{2014LUO}. Indeed, only the momenta
are evolved in ${\cal L}^\textrm{BO}$, and the resulting equations are
equal to those of the simple 
classical Langevin algorithm introduced in
Ref.~\onlinecite{2008ATT}, restricted to $\bvec{p}$, which in
the present case is a $3N_\textrm{at}L$-dimensional vector.
As in the classical second-order QMC-driven LD, the QMC noise
affecting the BO forces is controlled thanks to the friction
$\bsym{\gamma}^\textrm{BO}$ and the fluctuation-dissipation theorem.
A noise reduction scheme
similar to the one described in
Eqs.~\ref{disc-fluctuation-dissipation} and~\ref{correlator_ext_force} is also applied in this case, yielding a thermalization towards a steady
temperature ensemble even for quantum particles.

\vspace{1mm}
We refer the interested reader to Ref.~\onlinecite{2017MOU},
where we report a detailed description of the PIOUD algorithm and the 
integrated equations of motion.

\subsection{First-order Langevin dynamics}
\label{first_order_langevin} 

\tvb\ also features an accelerated first-order dynamics to sample 
the (classical) ionic canonical distribution.
This molecular dynamics scheme has been introduced recently by G. Mazzola and S. Sorella{~\cite{2017MAZ}}. 
Here, the ions formally undergo the following dynamics
\begin{equation}
{\mathbf{\dot R}} = {\mathcal{S}^{ - 1}}\left( {\mathbf{R}} \right) \cdot {\mathbf{F}}\left( {\mathbf{R}} \right) + {\boldsymbol \eta}
\label{first-order-Langevin-dynamics}
\end{equation}
\begin{equation}
\left\langle {{\eta _i}{\eta _j}} \right\rangle  = 2T\delta \left( {t - t'} \right)\mathcal{S}_{i,j}^{ - 1}\left( {\mathbf{R}} \right)
\label{first-order-Langevin-dynamics-stochastic-force}
\end{equation}
where $T$ is the temperature, ${\mathbf{F}}\left( {\mathbf{R}} \right)$ and ${\boldsymbol \eta}$ are the deterministic and stochastic forces, $\mathcal{S}$ is the covariance matrix of the QMC forces as introduced in Eq.~{\ref{qmc_force_cov}}. 
This  dynamics generalizes the Newton method, which is valid for structural relaxation, to finite-temperature.

\vspace{1mm}
Indeed, the use of the matrix $\mathcal{S}$, which is empirically
proportional to the Hessian matrix, can drastically decrease the
autocorrelation time of the Markov chain. In other words, this {\it
  metric} plays an important role in treating the different time scales of 
a given system. 
In this way, one can still use a quite large integration time step
$\tau$, which is beneficial in reducing the autocorrelation time of
quantities 
that depend on slowly varying degrees of freedom ({\it e.g.}, molecular diffusion), without introducing any bias in the sampling.

\vspace{1mm}
Discretization of Eq.~{\ref{first-order-Langevin-dynamics}} for a
finite time step $\tau$ at 
finite temperature is rather involved and  has been described in the Ref.~{\onlinecite{2017MAZ}}, and reads

\begin{equation}
\begin{split}
{\mathbf{R}}\left( {t + \tau } \right) &= {\mathbf{R}}\left( t \right) + \sqrt {2T\tau }  \cdot {\mathbf{z}}\left( t \right) + \tau {{\mathcal{S}}^{ - 1}}\left( {\mathbf{R}} \right){\mathbf{F}}\left( {\mathbf{R}} \right) \\
&- {{\mathcal{S}}^{ - 1}}\left( {\mathbf{R}} \right)\left( {\frac{{{\mathcal{S}}\left( {{\mathbf{R}}\left( {t - \tau } \right)} \right) - {\mathcal{S}}\left( {\mathbf{R}} \right)}}{2}} \right)\left( {{\mathbf{R}}\left( {t - \tau } \right) - {\mathbf{R}}\left( t \right)} \right),
\end{split}
\label{disc-first-order-Langevin-dynamics}
\end{equation}
with
\begin{equation}
\left\langle {{z_i}\left( t \right){z_j}\left( t \right)} \right\rangle  = \mathcal{S}_{i,j}^{ -1}\left( {{\mathbf{R}}\left( t \right)} \right).
\label{disc-first-order-Langevin-dynamics-stochastic-force}
\end{equation}

As the effective temperature depends on the finite integration time, the 
convergence to the target temperature $ T_\text{target} $ 
for $\tau \to 0$ can be improved 
by a renormalization of the temperature $T$ used in the  simulation, namely by using  the above equations with an appropriate choice of $T$:
\begin{equation}
T=T_\text{target} ~ ( 1- a \tau),
\end{equation}
where $a=1/2$ in the case of the harmonic potential leads to the exact result 
for any discretization time $\tau$, as long as $S$ is the exact elastic matrix. In the general case, the parameter $a$  should be tuned to speed up convergence  to the  $\tau \to 0$ limit. Conversely, the target temperature of the simulation can be measured a posteriori, as it is often done in the case of second-order Langevin dynamics, by measuring the average squared velocities of the particles.

\vspace{1mm}
The accelerated first-order Langevin dynamics has been successfully
employed in the most recent studies of dense liquid hydrogen,
providing equilibrated simulations even close to 
phase
transitions{~\cite{2017MAZ, 2018MAZ}}. 
Moreover, nuclear quantum effects can also be computed within this first-order LD, in exactly the same way as it was implemented for the second-order case discussed in sec.~\ref{pioud}. However, so far, it is not clear which method is more efficient and 
further studies are also necessary  to clarify  what is the optimal choice of 
the acceleration matrix  $S(\mathbf{R})$ within  the QMC framework.

%
%

\begin{center}
\begin{table*}[hbtp!]
\caption{\label{main_module} Main modules in \tvb.}
\vspace{1mm}
\begin{tabular}{c|p{15cm}}
\Hline 
 Module & Description \\ 
\Hline 
 makefort10.x             &   Generates a JAGP WF from a given basis set and structure. \\
 assembling\_pseudo.x     &   Prepares pseudo potentials. \\
 convertfort10mol.x       &   Adds molecular orbitals. If the number of molecular orbitals is equal to (larger than) half the number of electrons in a system, the resultant WF is the JSD (JAGPn). \\
 convertfort10.x          &   Converts a wave function type (fort.10\_in), {\it i.e.}, JSD/JAGP/JAGPn/JPf/JPfn, to another one with a different type (fort.10\_out). The output (fort.10\_new) is as close as possible to the input and may also include an optimal contracted (hybrid) basis set. \\
 convertfortpfaff.x       &   Converts a JAGP WF to a JPf one. \\
 prep.x                   &   Performs a DFT calculation.                                        \\
 turborvb.x               &   Performs VMC optimization, VMC evaluation, LRDMC, structural optimization and molecular dynamics.  \\
 readforward.x            &   Performs correlated samplings, and calculates various physical properties.                               \\
\Hline 
\end{tabular}
\end{table*}
\end{center}

\section{Implementation and operations}
\label{implementation_operation}
Table {\ref{main_module}} shows the main modules implemented in \tvb\
and a brief description 
of
their functionalities. 
All 
programs 
reported 
in Table~\ref{main_module}
are coded in \textsc{Fortran90}.
The flow-chart of 
a typical QMC
calculation
(see Fig.{~\ref{fig:workflow}}) is detailed as follows: 
starting from the geometry and the chemical composition, the user
first chooses the basis set for both Jastrow and antisymmetric factors,
and generates a JAGP-type ansatz using makefort10.x 
(fort.10 is the wave function filename in \tvb). 

\vspace{1mm}
The antisymmetric part of the WF is initially determined at the
DFT mean-field level.
Since the DFT
calculation is based on 
a single Slater determinant,
the generated JAGP WF should be converted to a JSD one. This is
performed by the convertfort10mol.x module, which
adds
$M$
molecular orbitals to the WF, where $M$ = $N^\uparrow$.
\footnote{We assume,
 without loss of generality, that ${N^\uparrow  \ge N^\downarrow}$}
The DFT molecular orbitals are then used to construct the initial trial WF for 
subsequent 
QMC
calculations. In
QMC,
the user
can choose to work with five different ans\"atze, namely, JSD, JAGP, JAGPn, JPfn, and JPf. 
The initial antisymmetric part of the JSD and JAGPn wave functions can be
directly obtained from the DFT orbitals.
In the JSD case, only the
occupied orbitals are imported, while in the JAGPn WF also the unoccupied orbitals
will be used in the geminal expansion, up to the $n$-th orbital.
Another possibility is to employ a JAGP ansatz. In that case, 
one has to convert the initial JSD trial WF obtained by DFT to the JAGP one using
convertfort10.x.

\vspace{1mm}
JAGPn can also be obtained by applying convertfort10mol.x
to a 
previously 
determined
JAGP ansatz.
Analogously, a JPf WF can be converted from a JAGP WF using
convertpfaff.x.
JPfn can be obtained by applying convertfort10mol.x
to the JPf ansatz.
These conversions can be done 
either
before 
or
after a
QMC optimization. The possibilities for ansatz conversion are
schematically drawn in Fig.~\ref{fig:ansatz-conv}.

\vspace{1mm}
After the initial determination of the antisymmetric part of the WF,
one performs the WF optimization at the VMC level. This is an important step
used to determine the best variational parameters of the Jastrow
factor at fixed determinant, or to fully relax the WF by
optimizing both Jastrow and determinant parameters.

\vspace{1mm}
Finally, the user can proceed to VMC and LRDMC calculations
using the optimized WF. In LRDMC, in order to remove the lattice discretization
error, it is better to repeat the calculations with several lattice
spaces $a$, and to extrapolate the results with a quartic function
$E\left( a \right) = {E_0} + {k_1}{a^2} + {k_2}{a^4}$, where ${E_0}$
is the extrapolated ($a \to 0$) energy. 

\vspace{1mm}
The python tool \tvbg\ makes 
all the above steps
more
straightforward, as discussed later.

%
%

\begin{figure*}[htbp!]
 \centering 
 \includegraphics[width=16.7cm]{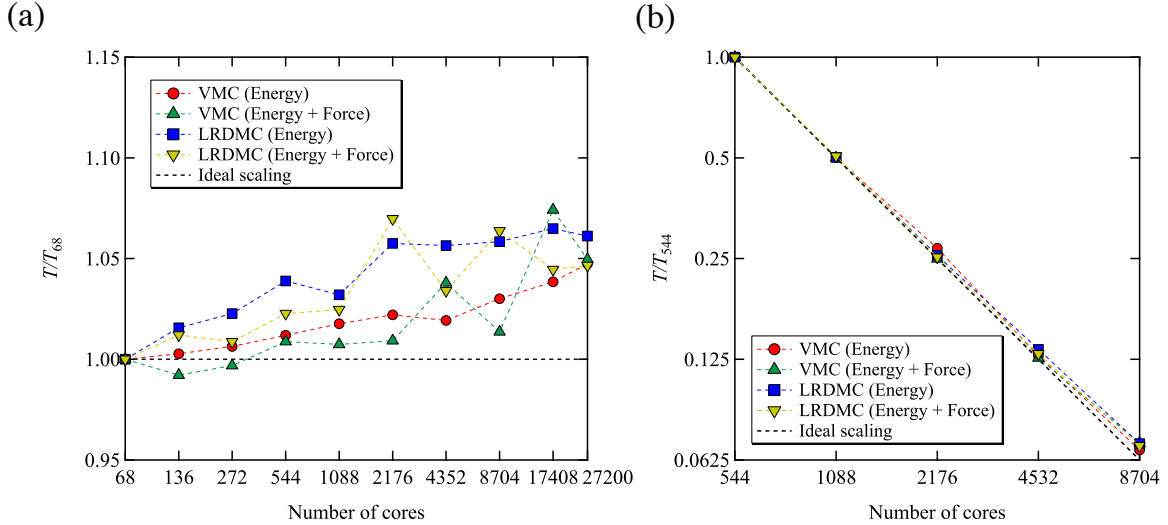}
 \caption{(a) The weak and (b) strong scalings of \tvb. The benchmarks were measured using the conventional 2 $\times$ 2 $\times$ 2 diamond with the ccECP pseudo potential (256 electrons in the simulation cell) on Marconi KNL nodes (Cineca supercomputer/SCAI). 
 }
 \label{bench-mark}
\end{figure*}

\section{Parallelization and benchmarks}
\label{benchmark}
\tvb\, has been parallelized using the 
MPI and OpenMP 
libraries, also supporting the hybrid MPI/OpenMP protocol. As well known, the QMC
algorithm is readily parallelized by employing many independent
random walks (called walkers), initialized with different seeds for
random number 
generators.
Fig.{~{\ref{bench-mark}}} shows the strong and the  weak scaling performances of \tvb, which are compared with the ideal scaling.
The benchmarks 
have been
measured using the conventional 2 $\times$ 2 $\times$ 2 diamond supercell containing 64 carbon atoms for a simulation with  256 electrons on Marconi KNL nodes (Cineca supercomputer/SCAI,  Intel Xeon Phi 7250 CPU, 1.40 GHz, 68 cores per node without hyperthreading).
For the weak scaling case, we set the number of walkers $M$ to the minimum possible one (one walker per core) and increased the number of cores from 68 (one node) to 27200 (400 nodes). On the other hand, for the strong scaling case, we fixed the number of walkers $M$ = 2176 and increased the number of cores from 544 (8 nodes) to 8704 (128 nodes) with the hybrid MPI/OpenMP protocol (the number of openMP threads was set to four).
For both cases, the performances are excellent, because all techniques
are off from the ideal scaling 
by $\sim$ 5\% 
at worst
in all cases performed here. At present, the code does not support the use of  GPU accelerators. This feature is currently under development.

%
%

\section{Postprocessing}
\label{postprocessing}

\subsection{Correlated Samplings}
A very efficient and easy-to-use  correlated sampling technique 
has been implemented 
in \tvb . This is very useful in evaluating 
total energy differences between two very similar WFs. 
One of its practical sides is 
to investigate whether a conversion of a WF ({\it {e.g.}}, JSD $\to$
JAGP}) has been successful. A user should run a short VMC calculation 
using one of the two mentioned WFs, 
during which 
the 
walkers' history is stored in an appropriate scratch area 
(turborvb.scratch). 
After that, a second run should be executed 
(by using the 
readforward.x module) 
to determine  at  the end  the correlated energy difference and the 
overlap between the two WFs.  Correlated samplings for  LRDMC  or for 
the difference in other quantities, {\it e.g.}, forces, have not been 
implemented 
yet. 
The latter could be very important for the frozen phonon 
calculation based on the ionic force difference method, which 
represents a possible subject of research.  
 
\subsection{Physical properties}
\tvb\, enables us to calculate various properties using the optimized 
many-body WF 
at both 
the VMC and LRDMC levels by averaging local observables and applying 
the forward walking technique{~\cite{2017BEC}}. The 
observables 
presently 
available in the readforward.x module can be categorized as follows:  
$\rm(\hspace{.18em}i\hspace{.18em})$ Charge density [$\rho \left(
  {\mathbf{r}} \right) = {\rho ^ \uparrow }\left( {\mathbf{r}} \right) 
+ {\rho ^ \downarrow }\left( {\mathbf{r}} \right)$]  and spin density 
[${\rho ^\sigma }\left( {\mathbf{r}} \right) = ({\rho ^ \uparrow }\left( {\mathbf{r}} \right) 
- {\rho ^ \downarrow }\left( {\mathbf{r}} \right))/2 $], 
fundamental 
properties obtained by a stochastic evaluation 
of 
multi-dimensional 
integrals,
involving 
the 
many-body WF 
and improved estimators, such as those proposed 
by Assaraf, Caffarel, and Scemama{~\cite{2007ASS}}. 
Fig.~{\ref{2019GEN_h4_ele}} shows the charge density of the square four hydrogen (H$_4$) depicted by C. Genovese {\it et al.}
$\rm(\hspace{.08em}ii\hspace{.08em})$
Electronic correlation functions, such as the 
charge-charge and spin-spin 
structure factors,
useful to 
study the critical behavior close to a phase transition, or the 
physical properties of a given phase. 
$\rm(i\hspace{-.08em}i\hspace{-.08em}i)$ 
The expectation values of the spin component along the quantization 
axis (${S_z}$) and the spin square (${S^2}$) operators 
inside a sphere centered on each atom{~\cite{2019GEN2, 2019GEN3}}, 
whose radius can be specified by the user. 
$\rm(i\hspace{-.08em}v\hspace{-.06em})$
The Berry phase:
${z_\alpha} = \braket{\Psi | {e^{i\left( {2\pi 
          /L^\alpha} \right)\sum\nolimits_{i = 1}^N {r_i^\alpha } }} |
  \Psi}$, 
where the complex polarization is computed for the $\alpha=\{x,y,z\}$
component ${r_i^\alpha }$ of the position vector $\mathbf{r}_i$,
and $L^\alpha$ is the supercell length in the same 
direction $\alpha${~\cite{2011STE}}. This property can be used to 
characterize 
the 
metal-insulator transition based on the many-body WF,
as it has been done in 
the one-dimensional hydrogen chain by L. Stella 
{\it et al.}
(Fig.~{\ref{2011STE_metal_insulator}}). 


\begin{figure}[htbp]
 \centering 
 \includegraphics[width=7.6cm]{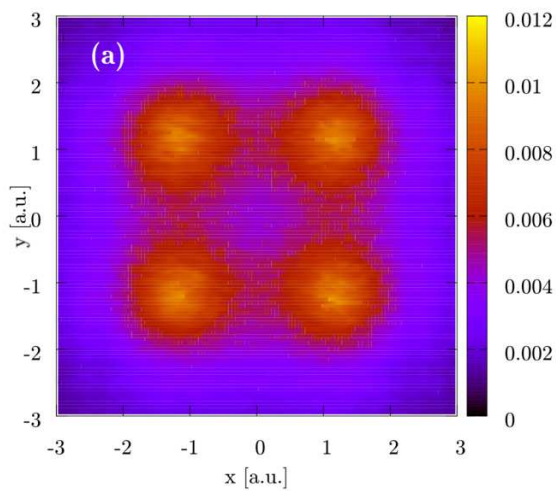}
 \caption{The $xy$-plane charge density of the square four hydrogen (H$_4$), calculated using a JAGP WF. Reprinted with permission from {\citet{2019GEN}}, {\it J. Chem. Phys.}, published by AIP in 2019. 
 }
 \label{2019GEN_h4_ele}
\end{figure}

\begin{figure}[htbp]
 \centering 
 \includegraphics[width=7.6cm]{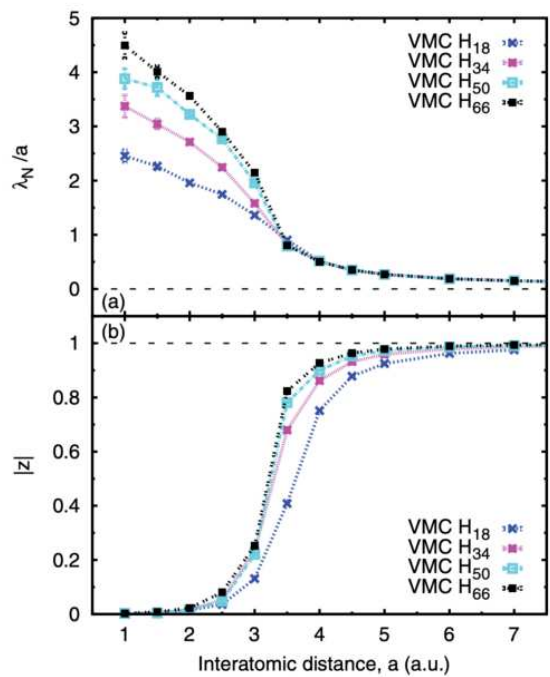}
 \caption{(a) Electronic localization length ${\lambda}_N$ divided by 
   the interatomic distance $a$ as a function of $a$, for the 
   one-dimensional hydrogen chain, where ${\lambda}_N$ is defined as 
   ${\lambda _N} = \left( {L/2\pi } \right)\sqrt { - \ln {{\left|
           {{z_N}} \right|}^2}/N}$, $N$ is the number  of electrons (hydrogen atoms), and $L$ is the length of the simulation cell along to the one dimensional chain. (b) Modulus of the complex polarization $z_N$ as a function of the interatomic distance of the hydrogen atoms. According to the previous works{~\cite{1999RES, 2005CAP}}, in the thermodynamic limit ($N \to 0$), a metal can be characterized by a vanishing modulus of the complex polarization ($\left| {{z_N}} \right| \to 0$, ${\lambda _N} \to + \infty$), while it becomes unity in the insulating case ($\left| {{z_N}} \right| \to 1$, ${\lambda _N} \to 0$). Thus, one can discuss the metal-insulator transition. Reprinted with permission from {\citet{2011STE}}, {\it Phys. Rev. B}, published by APS in 2011. 
 }
 \label{2011STE_metal_insulator}
\end{figure}

%
%

\section{Applications}
\label{applications}

Since the start of the 
\tvb\ project
in 2003 by M. Casula and S. Sorella~{\cite{2003CAS}}, 
the code has been applied to study atomic species, molecular systems, and various materials, including challenging systems such as large complexes, surfaces, liquids, and so on. Here, we provide a brief review of the applications done so far.


\subsection{Molecular systems}\label{sec:molecules}

\tvb\, has been 
employed
to study the properties of several molecular systems,
including small diatomic systems 
(such as
Li$_2$,{~\cite{2004CAS}} 
Be$_2$,{~\cite{2009MAR, 2014SOR}} 
B$_2$,{~\cite{2009MAR}} 
C$_2$,{~\cite{2009MAR, 2019GEN2}} 
N$_2$,{~\cite{2009MAR}}
O$_2$,{~\cite{2004CAS, 2009MAR, 2014ZEN2, 2019GEN3}},
F$_2$,{~\cite{2009MAR}}
Na$_2$,{~\cite{2019NAK}} 
LiF,{~\cite{2009MAR}} 
CN,{~\cite{2009MAR}}
Fe$_2${~\cite{2009CAS}}),
reactive oxygen species (singlet O$_2$, O$_2^-$, OH$^\bullet$, OH$^-$, NO$^\bullet$, NO$^-$, HOO$^\bullet$, HOO$^-$, cis and trans HOOO$^\bullet$),~\cite{2014ZEN2} 
aromatic molecules (benzene\cite{2004CAS, 2005CAS2}, oligoacene series\cite{2018DUP}, see section~\ref{sec:aromatic}),
the water molecule,\cite{2004CAS, 2013ZEN} dimer\cite{2008STE, 2015ZEN} and hexamer clusters,\cite{2015ZEN}
the Zundel ion (H$_5$O$_2^+$){~\cite{2014DAG, 2017MOU}},
H$_2$S, SO$_2$, NH$_3$, PH$_3$,\cite{2014LUO}
and others.
Most of the studies report 
both variational and FN diffusion Monte Carlo results, and several of the WF ans\"{a}tze available in \tvb\ are tested.
Often, it is shown that quite good results are already obtained at the
variational level, and the FN 
DMC
further improves the accuracy of the outcomes. For instance, see the case of the sodium dimer in  Fig.~\ref{2019NAK_Na2}. 

\begin{figure}[htbp]
 \centering
 \includegraphics[width=7.6cm]{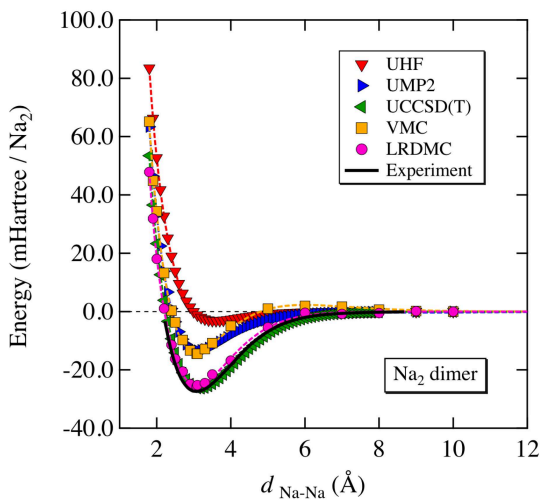}
 \caption{
PES
of the sodium dimer. The dashed lines show the obtained Murrell-Sorbie function. The error bars are
within the markers. The obtained LRDMC PES agrees with the CCSD(T) and experimental ones.
Reprinted with permission from {\citet{2019NAK}}, {\it J. Chem. Theory Comput.}, published by ACS in 2019.
 }
 \label{2019NAK_Na2}
\end{figure}

\vspace{1mm}
Table{~\ref{table-first-row-elements}} and Fig.{~\ref{fig-first-row-elements}} report the binding energies for the molecular dimers of the first-row atoms, evaluated using LRDMC with the JSD, JAGP, or JPf WF ans\"{a}tze as guiding functions. 
Some of them
are close or reach the chemical accuracy ({\it i.e.}, 1 kcal/mol or $\sim$0.04~eV). 
A systematic improvement going from the JSD $\to$ JAGP $\to$ JPf can be appreciated.
It follows from the employment of a more flexible parametrization of the WF, which improves both the variational energy and the quality of the nodes.
However, the improvement given by the JPf ansatz over JAGP and JSD is even more fundamental than what Table{~\ref{table-first-row-elements}} and Fig.{~\ref{fig-first-row-elements}} indicate, where the reported binding energies are evaluated as the difference between the energy of the molecule and twice  the energy of the atom. 
This evaluation does not highlight that JPf is a size-consistent ansatz, while both the JSD and the JAGP have a size-consistency issue with most of the molecules considered.\footnote{
A theory is size-consistent if the energy $E_{\text{A}+\text{B}}$ of the supersystem made of two fragments $\text{A}$ and $\text{B}$  at a sufficiently large distance is equal to the sum $E_\text{A} + E_\text{B}$ of the energy of the fragments.
When a theory is size consistent it is equivalent to take as reference energy for a interaction energy evaluation the sum of the fragments or the supersystem with the fragments far apart, whereas if the theory is not there is an arbitrary shift as large as $E_\text{A} + E_\text{B} - E_{\text{A}+\text{B}}$ in the interaction energy.
}
Let us consider the oxygen molecule as an example: 
the O$_2$ ground state is a 
spin triplet ($^3\Sigma_g^-$)
which, in both JSD and JAGP, can be described using two unpaired
electrons with the same spin. However, as we take the oxygen atoms far
apart (say, the O-O system), they should both be in the ground state
for the oxygen atom, which is also 
a spin triplet ($^3P$).
Neither JSD nor JAGP
have the flexibility to describe the O-O system 
by
the same WF parametrization used for the O$_2$,\footnote{JSD and JAGP could describe the O-O system if the number of unpaired is changed, for instance if four unpaired are used. However, this is not the right parametrization of the O$_2$ molecule.} thus $E_\text{O-O} > 2 E_\text{O}$.
In contrast, JPf has the flexibility to describe the dissociation limit correctly, so $E_\text{O-O}^\text{JPf} = 2 E_\text{O}^\text{JPf}$. 
Fig.~\ref{fig:dissociationO2} shows that JPf is indeed able to provide a reliable dissociation curve of the oxygen molecule, whereas JSD and JAGP are reliable only in proximity of the minimum.
Notice that the FN projection scheme alleviates in part the
limitations of the underlying ansatz, as both JSD and JAGP have a
smaller size-inconsistency at the LRDMC 
than at the VMC level,
but the 
correction
is incomplete and the inconsistency is sizable ($>1$
eV) for all but the JPf ansatz. 
Therefore, in general, JPf will yield a better description of the
overall potential energy surface (PES), 
including 
binding energies, vibrational frequencies, zero-point motion, transition states, and so on.


\begin{center}
\begin{table*}[htbp]
\caption{\label{table-first-row-elements} Binding energy of the
  first-row dimers calculated by \tvb\ with LRDMC and different types
  of ans\"{a}tze as a guiding function (GF), at the experimental bond
  length (see. Ref.{~\onlinecite{2008TOU}}). Js denotes the symmetric
  ({\it i.e.}, spin-independent) Jastrow factor. 
LRDMC calculations are performed with several lattice spaces $a$ and extrapolated for $a \to 0$. 
The spin-orbit coupling and the zero point energies~\cite{2009MAR} are not included in these evaluations.
}
\begin{tabular}{c|cccccccc}
\Hline
 Binding energy (eV) & Li$_2$ & Be$_2$ & B$_2$ & C$_2$ & N$_2$ & O$_2$ & F$_2$ \\
\Hline
 LRDMC (GF=JSD) & 0.976(6) & 0.144(7) & 2.83(1) & 5.74(2) & 9.67(1) & 4.94(3) & 1.27(1) \\
 LRDMC (GF=JsAGPs) & 0.9812(12) & -0.0270(12) & 2.66(1) & 6.01(1){\footnotemark[1]} & 9.91(1){\footnotemark[1]} & 5.06(2){\footnotemark[1]} & 1.56(1)  \\
 LRDMC (GF=JPf) & 1.0580(12) & 0.0304(22) & 2.75(1) & 6.31(1){\footnotemark[1]} & 9.97(1){\footnotemark[1]} & 5.127(6){\footnotemark[1]} & 1.64(1) \\
 Est. exact & 1.06(4){\footnotemark[2]} & 0.1153(3){\footnotemark[3]} & 2.91(6){\footnotemark[4]} & 6.43(2){\footnotemark[5]} & 9.902(3){\footnotemark[5]} & 5.233(3){\footnotemark[5]} & 1.693(5){\footnotemark[5]}  \\
\Hline
\end{tabular}
\footnotetext[1]{See Ref.~{\onlinecite{2019GEN3}}.}
\footnotetext[2]{See Ref.~{\onlinecite{1989AZI}}.}
\footnotetext[3]{See Ref.~{\onlinecite{2009MER}}.} 
\footnotetext[4]{See Ref.~{\onlinecite{1991STE}}.} 
\footnotetext[5]{See Ref.~{\onlinecite{2005LAI}}.} 
\end{table*}
\end{center}

\begin{center}
\begin{figure}[htbp]
 \includegraphics[width=3.4in]{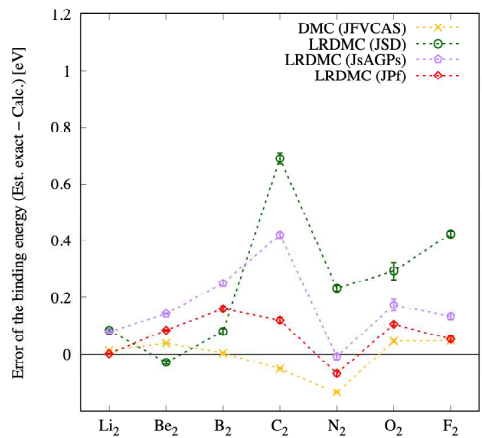}
 \caption{
Binding energies of the first-row dimers at their experimental bond length (see. Table.{~\ref{table-first-row-elements}).}
The vertical axis shows the difference between the LRDMC and the
estimated exact binding energies. The results obtained from the
DMC using a Jastrow correlated full-valence-CAS as trial WF (DMC(JFVCAS)) are taken from Ref.{~\onlinecite{2008TOU}}.
}
 \label{fig-first-row-elements}
\end{figure}
\end{center}

\begin{figure}[htbp]
 \includegraphics[width=3.4in]{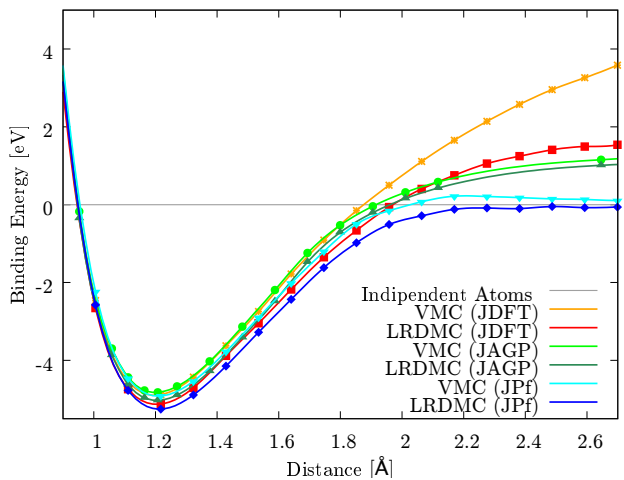}
 \caption{\label{fig:dissociationO2}
O$_2$ dissociation curve as obtained with \tvb~ at the variational (VMC) and fixed-node projected (LRDMC) levels of theory and using JSD with DFT orbitals (JDFT), JAGP and JPf WF ans\"atze. 
}
\end{figure}

\vspace{1mm}
The applications of \tvb\ 
are not limited
to single point evaluations.
As discussed in Sec.{~\ref{derivatives}}, 
a distintive feature of \tvb\, among most QMC codes is the efficent
evaluation of atomic forces. 
This 
capability 
has been extensively exploited to investigate several properties of the PES, such as vibrational properties, and to perform geometry optimizations and molecular dynamics simulations at finite temperature. 
%
It is important to remember that QMC atomic forces are, as the QMC energy, affected by a stochastic error.
In a first attempt\cite{2012ZEN} to evaluate vibrational frequencies of a triatomic molecule (namely, water), a grid of points close to the guessed minimum geometry was used to evaluate energy and forces.
The results were used to fit the PES around the minimum with a truncated Taylor expansion, yielding evaluations of the structural minimum and of both the harmonic and anharmonic vibrational frequencies (using the second-order perturbation theory).
It was observed that the 
inclusion of the estimates of forces
reduces the stochastic error on the vibrational frequencies 
by around an order of magnitude with respect to the fit using only the energy evaluations (on the same grid and  using the same sampling size on each single point evaluation).
This corresponds to about a 25 times larger speedup.\footnote{Indeed, in order to improve the precision by an order of magnitude the sampling (so, the computational cost) has to be increased by two orders of magnitude. Considering that the evaluation of the energy and all the force components in \tvb\ costs only 4 times the evaluation of only the energy with the same sampling (see Sec.~\ref{derivatives}), the speedup is 100/4=25.}  
Moreover, it was observed that the optimal grid is a tradeoff between maximizing the precision on the vibrational properties and minimizing the errors on the fit. 
A systematic study of the quality of the variational WF ans\"atze in relation to the vibrational (and other) properties of the water molecule is reported in \citet{2013ZEN}.
Afterward, Y. Luo {\it et al.}\cite{2014LUO, 2015LUO} automated the
process of finding the optimal grid by performing 
fits on samples
generated by a finite temperature molecular dynamics, and applied the approach to H$_2$O, H$_2$S, SO$_2$, NH$_3$, and PH$_3$.
The obtained vibrational frequencies are in good agreement with the experimental values. 

\vspace{1mm}
\citet{2015ZEN} applied the second-order LD to study the structural properties of liquid water. They obtained radial distribution functions in good agreement with recent neutron scattering and X-ray experiments, especially for the position of the oxygen-oxygen peak, which is very sensitive to the quality of the water description and, in particular to the inclusion of dispersive vdW forces. This is a stringent test for {\it ab initio} approaches as the shape of the peak can be directly compared with neutron scattering data.

\vspace{1mm}
F. Mouhat {\it et al.} developed a path integral LD formalism
particularly suited for QMC and applied it to the Zundel ion
(H$_5$O$_2^+$), with the aim of studying nuclear quantum
effects{\cite{2017MOU}}. The outcome of their study shows
how essential is the inclusion of nuclear quantum effects to 
properly deal with proton transfer in water and aqueous systems, even
at room temperature.

\subsection{Weakly bonded systems and non-covalent interactions}
\label{vdW}
An important application for QMC is the description of 
systems interacting through weak van der Waals (vdW) dispersive interactions, 
which are fundamental in supramolecular chemistry, layered and porous
materials, molecular crystals, adsorption of molecules on surfaces,
and so on. 
vdW interactions play a major role also in non-covalent
intra-molecular bonds, which result from the interplay of forces of
different nature, including
electrostatic contributions.
An accurate
evaluation of the
overall strength of these non-covalent bonds
requires each component of the interaction to be modeled correctly. 
This is a big challenge for all {\it ab initio} approaches, and only a few methods prove reliable.
QMC and CCSD(T) are considered to yield very accurate 
estimates, and there is a general agreement among the predictions from
these two theories. Thus, they are typically used to produce benchmark values.
Traditionally,
mean-field 
schemes
have 
difficulties in describing dispersive interactions, 
because they 
arise from
non-local electron correlations, 
completely missed by local density functionals.
However, in the
context of DFT, there has been an intensive effort in developing 
new exchange correlation functionals{~\cite{2010GRI}} to include 
vdW interactions. This is usually done in a semi-empirical way,
often building on the
benchmark provided by CCSD(T) or QMC.

\vspace{1mm}
The non-covalent interactions are typically a tiny fraction of the
total energy of 
a compound.
Similarly to the cases discussed in Sec.~\ref{sec:molecules}, also in
the evaluation of weak interactions it is important to use a size-consistent approach.
However, very often, the molecules form a closed shell, and most of the times the intermolecular interactions are among fragments having zero spin.
In this case, a single Slater determinant is a size-consistent ansatz, and so is the JSD.
The AGP 
(without Jastrow)
is not size-consistent, due to the presence of unphysical charge
fluctuations~\cite{2007SOR, 2017BRE}, but a very flexible ({\it i.e.},
large) Jastrow factor allows 
damping
the charge fluctuations and
recovering 
the size-consistency in the JAGP.
This is of course true also in a subsequent FN-DMC projection,
which is indeed equivalent to 
applying 
an infinitely
flexible Jastrow factor to the trial WF.
At the variational level, 
the JSD ansatz occasionally leads 
to better results than the JAGP.
Even though JAGP leads to a
lower total energy, it is indeed difficult to use a Jastrow factor large enough
to damp any charge fluctuation. So, there is a worse error cancellation
when energy differences are considered in JAGP. 
This was observed, for instance, in the water dimer,\cite{2015ZEN}
although at the FN level JSD and JAGP lead to the same 
binding
energy.
It should also be considered that in the widely used DMC
scheme\cite{1993UMR2} a size-consistency issue was present in any
simulation with finite time step $\tau$, until the modification of the
``standard'' algorithm has been introduced by \citet{2016ZEN2}, whose
solution removes almost completely the size-consistency error. At
variance with DMC, the LRDMC algorithm does not suffer from this bias,
and it is always size consistent. 

\vspace{1mm}
The DMC scheme is commonly used to project a single Slater determinant
WF where only the Jastrow factor has
been optimized, while the antisymmetric part is filled with DFT orbitals{\cite{2010NEM}} (JDFT).
The resulting FN energy 
is usually biased by the nodal
surface, kept frozen from previous DFT or lower level mean-field
calculations. 
Optimizing the Jastrow factor and determinant together, which is
routinely done in \tvb, improves significantly the total
energy. However, in the evaluation of non-covalent interactions, there is
not yet a clearcut indication of the advantage of the full
optimization over JDFT. This is due to the fact that the error
cancellation plays a major role. 
Indeed, it is expected that, for non-covalent interactions, JDFT has
almost the same bias in the interacting and non-interacting systems.
When pseudopotentials are used, DLA\cite{2019ZEN} can be used to
remove any bias given by the Jastrow optimization.

\vspace{1mm}
\tvb\ with JDFT, JSD, and JAGP guiding functions 
has
been applied to several interesting systems, such as molecular
hydrogen adsorbed on benzene{~\cite{2008BEA}}, 
benzene dimers{~\cite{2007SOR}},
graphite layers{~\cite{2009SPA}}, 
water or methanol molecules adsorbed on a clay surface (see Fig.~\ref{fig:kaolinite})~\cite{2016ZEN},
metalic clusters{~\cite{2009NIS}}, and 
to evaluate the 
cohesive energies of B$_2$O$_3$ polymorphs{~\cite{2019FER}}.

\begin{figure}[htbp]
\centering
\includegraphics[width=3.4in]{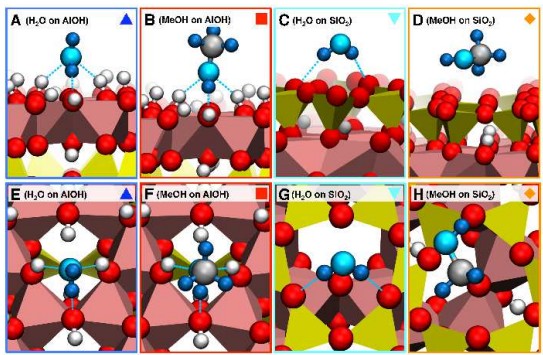}
\caption{
Adsorption of water and methanol on the hydroxyl-terminated and the silicate-terminated faces of kaolinite (side view in first row, top view in second row). The adsorbed molecule on kaolinite is depicted in cyan and gray and the H-bonds are represented by the blue dashed lines. 
Reprinted with permission from {\citet{2016ZEN}}, {\it J. Phys. Chem. C}, published by ACS in 2016.
}
\label{fig:kaolinite}
\end{figure}

\subsection{Strongly correlated and superconducting materials}
Another important QMC application 
is the study of 
strongly correlated materials,  
that represents a challenge for any theoretical method, because in several 
cases a fully consistent description remains, so far, elusive.
In this context, QMC methods have been widely used for solving,  
as accurately as possible, strongly correlated lattice Hamiltonians, like, for instance, the
Hubbard model. Real-space \emph{ab initio}
QMC methods could fill the gap between realistic Hamiltonians and lattice
models, by providing an accuracy similar to the one reached in lattice
Hamiltonians, while keeping the complexity of real materials.
\tvb\, has been originally conceived in this perspective, with the idea
that one of the most flexible WF ans\"atze tested on
correlated lattice models\cite{1998CAL2,1999LUC,2000BEC,2001LUC,2002SOR}, namely the
generalized version of the resonating valence bond (RVB)
WF, could be
translated into a successful variational WF for \emph{ab
  initio} systems. One of the most appealing features of the RVB WF is
that it  allows several broken-symmetry phases, such as
antiferromagnetic insulators and superconductors, which are present
- and sometimes coexisting - in
the extremely rich phase diagrams of strongly correlated materials.
As introduced in Sec.{~\ref{wavefunction}}, the 
RVB WF employed in \tvb\, 
takes the static and dynamic correlation effects into
consideration beyond the commonly used 
Jastrow-Slater determinant (JSD) WF.
Therefore, the code is expected to describe strongly correlated
systems
more accurately. 
In particular, among the various families of correlated WFs
which go beyond the simplest JSD form, the JAGP/JPfaffian families are
very suitable to describe \emph{extended} systems, as they keep a
compact and still strongly correlated form even in the thermodynamic
limit. Indeed, as detailed
in Sec.{~\ref{wavefunction}}, the AGP part is the particle number conserving version
of the BCS WF. In the same way, the JAGP variational form
can be seen as an efficient  formulation of a  RVB WF. 
Thus, superconducting materials including
cuprates{~\cite{1986BED}} and 
iron pnictides and selenides{~\cite{2008KAM}} are
prominent applications of the JAGP/RVB WFs. 
By running extensive VMC calculations for CaCuO$_2$, a parent compound of cuprate 
high-temperature superconductors, M. Marchi {\it et 
  al.} validated the JAGP description by correctly finding the 
expected $d$-wave symmetry of the pairing function{~\cite{2011MAR}}.
M. Casula and
S. Sorella applied the RVB WF to FeSe, one of the iron-based
high-temperature superconductors{~\cite{2013CAS}}, whose electronic
structure and pairing mechanism have been unclear and intensively debated. 
They determined the symmetry of the
superconducting order parameter and the size of its gap entirely from
first principles, by analyzing the AGP pairing function
(Fig.{~{\ref{2013CAS_FeSe}}}).
B. Busemeyer {\it et al.}
applied DMC and LRDMC  to study the structural and magnetic properties of the
normal state of FeSe
under pressure, and reported that collinear spin
configurations are energetically more favorable than other spin
patterns, such as ferromagnetic,
checkerboard, and staggered dimer, over a large range of pressures{~\cite{2016BUS}}. 
  

\begin{figure}[htbp]
 \centering
 \includegraphics[width=7.6cm]{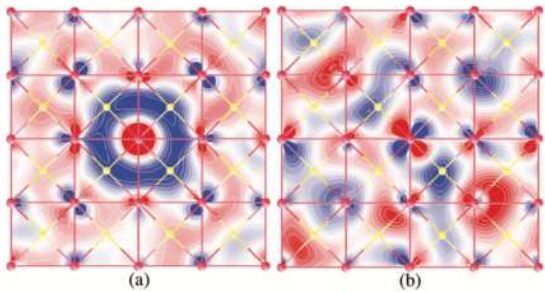}
 \caption{(a) Even-even and (b) odd-even components of the AGP pairing
   function of FeSe at 0 GPa obtained by VMC calculations, where even
   and odd refer to the parities of the orbitals to the reflection
   through the Fe plane. The contour plots show $\Phi \left(
     {{{\mathbf{R}}_{{\text{center}}}},{\mathbf{r}}} \right)$ with
   ${{\mathbf{R}}_{{\text{center}}}}$ set to be the iron lattice site
   at the center of the supercell, while ${\mathbf{r}}$ spans the
   plane defined by the 4 $\times$ 4 lattice. Red (yellow) balls are
   iron (selenium) sites. Arbitrary units of blue (red) intensity
   indicate negative (positive) regions with corresponding
   magnitude. Reprinted with permission from {\citet{2013CAS}}
   {\it Phys. Rev. B}, published by APS in 2013.
 } 
 \label{2013CAS_FeSe}
\end{figure}


\subsection{Aromatic and honeycomb lattice compounds}\label{sec:aromatic}

Aromatic compounds are also successful examples of the RVB theory,
because  they are characterized by  resonating C-C bonds that
can be efficiently represented by 
antisymmetrized singlet pairing functions 
({\it i.e.}, antisymmetrized product of geminals, or AGP).
M. Casula {\it et al.} applied the AGP
WF combined with a Jastrow factor to the simplest aromatic
compound, that is, the benzene molecule (C$_6$H$_6$) for the first
time{~\cite{2004CAS, 2005CAS2}}. 
They reported that the inclusion of the resonance between the two
possible Kekul{\'e} states significantly lowers the VMC energy
compared with the one obtained by a single Kekul{\'e} WF. 
They also showed
the importance of adding a three-body Jastrow factor to improve the
AGP description, just in the  spirit of the RVB framework.
By selectively switching on the intersite resonances, they proved that 
the Dewar contributions ({\it i.e.}, including the third
nearest neighbor carbons) improves the description of the resonating
valence bond. 
%
C. Genovese {\it et al.} have applied a more general AGP WF,
{\it i.e.}, the Pfaffian, to the benzene
molecule{~\cite{2019GEN3}}. They reported that the obtained
atomization energies are reasonably consistent with the experimental
value at the LRDMC level. On the other hand, when starting from JDFT,
{\it i.e.}, from frozen DFT nodes,
the
obtained atomization energy is severely underestimated in most  cases, 
implying that
the optimization of the nodal surface at the VMC level is essential to
obtain a correct value for the atomization energy.  
%
As mentioned in Sec.~\ref{vdW},
 S. Sorella {\it et al.} applied
the RVB WF to calculate the binding energy of the
face-to-face and displaced parallel benzene dimers{~\cite{2007SOR}},
and obtained the binding energy very close to the experimental value
in the displaced parallel case.
%
N. Dupuy and M. Casula applied the 
same RVB 
WF to study the ground-state properties of the oligoacene
series, from anthracene up to the nonacene.{~\cite{2018DUP}} They
found that the ground state obtained by the RVB WF has a
weak diradical or polyradical instabilities until the nonacene,
which is in contrast with the results previously found by lower-level
theories, such as Hartree-Fock, 
hybrid functionals, and CASSCF in a
restricted active space. 
\tvb\ has been applied not only to simple aromatic molecules but also
to honeycomb lattice compounds, in order to study the role of RVB
correlations, in other words, to investigate 
whether the resonance energy is sizable also in extended systems. 
M. Marchi {\it et al.} studied the nature of chemical bonds in
graphene using the RVB WF{~\cite{2011MAR}}, and found that
the RVB energy gain becomes extremely small in the thermodynamic limit
{\it i.e.}, in the infinite lattice. 
Conversely, S. Sorella {\it et al.} applied the RVB WF to
 isotropically strained graphene{~\cite{2018SOR}}
(Fig.{~\ref{2018SOR1_graphene}}) and found that the RVB effect is
crucial in the Kekul{\'e}-like dimerized (DIM) phase that has been proposed to 
become stable under the isotropic strain (Fig.{~\ref{2018SOR2_graphene}}). They also
reported that the JAGP  energy gain with  respect to JSD  is negligible in the
perfect honeycomb structure - as previously found by  M. Marchi {\it
  et al.} - but is extremely important in the DIM phase.  

\begin{figure}[htbp]
 \centering
 \includegraphics[width=7.6cm]{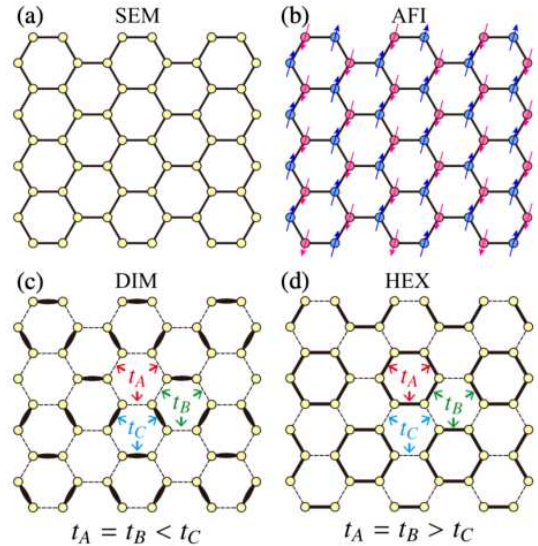}
 \caption{The graphene structures studied in 
Ref.~{\citenum{2018SOR}}. (a) SEM honeycomb, semimetallic; (b) AFI
honeycomb antiferromagnetic insulator; (c) DIM dimerized
Kekul{\'e}-like insulator; (d) HEX distorted hexagonal
insulator. There are two carbons per unit cell in (a) and (b), six in
(c) and (d). $t_A$, $t_B$, and $t_C$ schematically denote different
hopping integrals magnitudes. Reprinted with permission from
{\citet{2018SOR}}, {\it Phys. Rev. Lett.}, published by APS in 2018. 
 }
 \label{2018SOR1_graphene}
\end{figure}

\begin{figure}[htbp]
 \centering
 \includegraphics[width=3.4in]{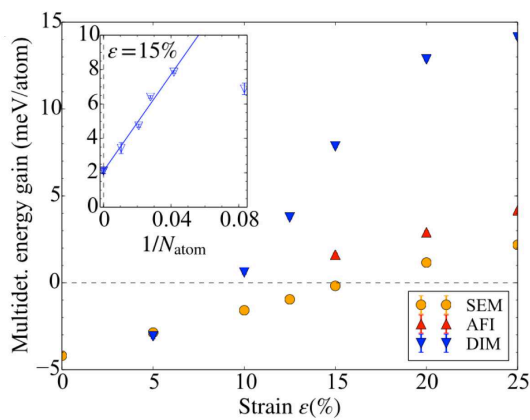}
 \caption{Energy gains by the resonance effect, measured by the energy
   per atom difference between the single determinant ansatz
   (Jastrow-Slater determinant WF) and the corresponding
   multideterminant JAGP WF. The largest energy gain occurs
   in the DIM state, underlining its resonating valence-bond nature,
   actually increasing for large strain $\varepsilon$. Small negative
   values at small strain are finite-size effects. Inset: finite-size
   scaling of this correlation energy gain in the DIM state at
   $\varepsilon$ = 15\%. Reprinted with permission from {\citet{2018SOR}}, 
   {\it Phys. Rev. Lett.}, published by APS in 2018.
 }
 \label{2018SOR2_graphene}
\end{figure}


\subsection{Organic compounds: diradicals and conjugated systems}
Diradical and conjugated organic compounds are other interesting
applications of the RVB WF, because both static and 
dynamic
correlation effects play important roles in these systems.
A. Zen {\it et al.} found that the RVB WF gives a correct torsion barrier of the ethylene (C$_2$H$_4$) and the triple-singlet gap in the methylene (CH$_2$){~\cite{2014ZEN}}. 
They also investigated the charge-transfer and diradical 
nature
of the electronic states of the retinal minimal model
(penta-2,4-dieniminium cation) at the VMC and LRDMC 
levels
{~\cite{2015ZEN2}}. They 
proved that
the dynamical electronic correlation is 
key
to get a reliable ground state energy surface in 
proximity of the
conical intersection between the electronic ground- and 
first
excited-state of the molecule. Their obtained energy landscapes
significantly different  from  the CASSCF ones, inverting the relative
stability of several torsional paths, and similar to what was obtained by
other correlated approaches. 
M. Barborini {\it et al.} applied VMC and LRDMC to calculate the
torsion barriers of 1,3-butadiene and the ring-opening barrier of
cyclobutene{~\cite{2012BAR2}}, which are in good agreement with the
experimental results.  
M. Barborini and E. Coccia investigated the potential energy curves of
the two spin states of tetramethyleneethane (TME) molecule and the two
anionic states of TME as a function of the torsion of the central
dihedral angle{~\cite{2015BAR}}. Through {\it ab initio} geometrical
optimizations at the VMC level, they proposed possible structural
interconversions between the states, which are in agreement with the
ion photoelectron spectroscopy experiments. 

\vspace{1mm}
The bond length alternation (BLA), namely, the 
impact on the geometry of the
difference between the single
and double carbon bonds, is one of the key structural descriptors in conjugated organic compounds.
M. Barborini {\it et al.} investigated the effects of the static and
dynamical electronic correlations on the BLA of the
1,3-butadiene{~\cite{2015BAR2}} using a VMC-JAGP ansatz and other quantum chemistry calculations. They found inconsistency between BLAs obtained by CCSD(T)-CBS (the most accurate) and VMC-JAGP calculations, but the reason for the discrepancy is still unclear.
They also investigated the structural properties of polyacetylene chains H-(C$_2$H$_2$)$_{n}$-H up to $N$ = 12 acetylene units{~\cite{2015BAR3}}, in which they revealed that the BLA obtained by the extrapolation to $n$ $\to$ $\infty$ is 0.0910(7) \AA, which is compatible with the experimental data.
E. Coccia {\it et al.} applied VMC to study geometries of the retinal
model, 
the chromophore of the rhodopsin involved in the mechanism of vision,
and discussed the effects of the  electronic correlation on the
BLA{~\cite{2012COC2, 2012COC3}} by comparing with DFT and quantum
chemistry methods. Wave-function and geometry optimizations of the
retinal 
model
were also performed in the presence of the protein field, classically described via electrostatic and mechanical interactions: the complex environment dramatically affects the BLA of the chromophore, which consistently increases with respect to the gas-phase case.
They also performed VMC geometry optimization of the peridinin chromophore in order to verify the interplay between the BLA and the optical properties of a large conjugated moiety, involved in the light-harvesting step of photosynthesis in the peridinin-chlorophyll-protein complex{~\cite{2014COC}}. Comparison with DFT and wave-function approaches shows that the combined application of RVB-based VMC and Bethe-Salpeter methods for geometry and excitations, respectively, gives an accurate estimation of the electronic transition energies of peridinin.
They recently applied VMC optimization to obtain a ground state
structure of keto-1, enol-1, and enol-2 forms of
oxyluciferin{~\cite{2017COC}}, which is used to calculate the $S^1$
$\leftarrow$ $S^0$ absorption energy 
based on the Bethe-Salpeter formalism.
E. Coccia {\it et al.} carried out a systematic basis-set analysis on
polarizability, quadrupole moment and electronic density of C$_2$H$_2$
molecule, with the aim to test VMC and LRDMC calculations in the
presence of an external electric field{~\cite{2012COC}}. In
particular, a relatively small hybrid (Gaussian + Slater functions)
basis set was seen 
to be able to properly reproduce reference values.


\subsection{Metal-organic complexes}
Recently, \tvb\, has also been used to study metal-organic complexes. 
S. Chu {\it et al.} studied free-energy reaction barriers of water
splitting reactions, using a simplified computational model based on
the cobalt ion{~\cite{2016CHU}}
(Fig.{~\ref{2016CHU_watersp_cycle}}). They found that the total
free-energy differences of the water oxidation,
computed
at the QMC level, fairly agree with the experimental reference values, 
in a surprisingly better way 
than  the  corresponding CCSD(T) ones. 
M. Barborini and L. Guidoni applied VMC to tackle the geometry
optimization of a Fe$_2$S$_2$(SH)$_2$$^-$ model complex (High-Spin and
Broken Symmetry states){~\cite{2016BAR2}} based on the extended
broken-symmetry (EBS) approach. 
The number of applications is still limited in this category, but the
recent development of sophisticated transition-metal pseudopotentials
will make it more feasible in the near future. 
\begin{figure}[htbp]
 \centering
 \includegraphics[width=7.6cm]{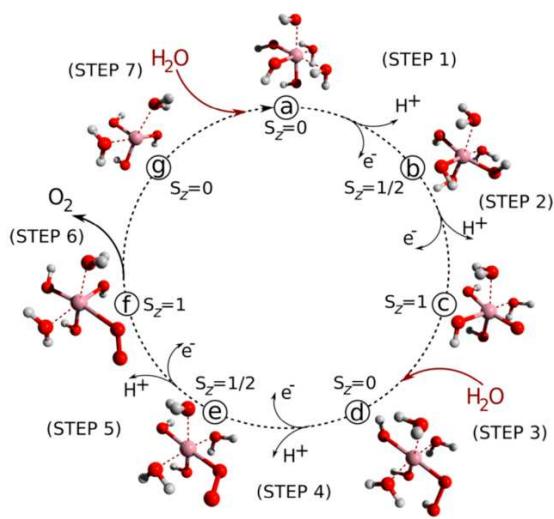}
 \caption{Reaction cycle involving the cobalt ion model, studied in Ref.~{\citenum{2016CHU}} by VMC and LRDMC. The pink, red, and white balls represent the cobalt, oxygen, and hydrogen atoms, respectively. Reprinted with permission from {\citet{2016CHU}}, {\it J. Chem. Theory Comput.}, published by ACS in 2016.
 }
 \label{2016CHU_watersp_cycle}
\end{figure}


\subsection{Crystals polymorphism}
{\it Ab initio} electronic calculations 
allow one to study the
phase stability between crystals polymorphs, by comparing the
enthalpies or free energies, and 
by computing
their equation of states (EOSs).
As mentioned in the introduction, DFT's predictive power strongly
depends on the choice of the exchange and correlation functionals,
sometimes causing 
qualitatively different 
phase
stability predictions.
On the other hand, VMC and LRDMC calculations do not suffer from this drawback in principle.
S. Sorella {\it et al.} studied the pressure-induced metal-insulator
transition in bulk silicon through 
EOS calculations 
{~\cite{2011SOR}}. They obtained a reasonable transition pressure at
room temperature by VMC, 
however
still a little far from the experimental value.
N. Devaux {\it et al.} studied 
the $\alpha$-$\gamma$ phase transition in 
elemental 
cerium at the VMC and LRDMC levels{~\cite{2015DEV}}. They 
revealed 
that the volume-collapse transition could be understood as a
conventional first-order transition of electronic origin, induced by
the $p$-$f$ hybridization, 
and by the strong local repulsion affecting the $f$ states. 
H. Hay {\it et al.} investigated the effect of 
long-range van der Waals forces 
in 
the
SiO$_2$ polymorphs using DFT and QMC{~\cite{2015HAY}}. Their LRDMC
calculations showed that the obtained energy differences in
quartz-cristobalite and quartz-stishovite agree well with the
experimental values within sub-chemical accuracy, implying that QMC is
a promising method in describing both conventional and high-pressure
SiO$_2$ polymorphs. 
Notice that, at present, only the zero-temperature energy (without the
entropic term) is available at the VMC and LRDMC levels. Therefore,
the entropic effects are estimated by DFT-PBE
calculations{~\cite{2011SOR}}. 
In the near future, it will be possible to estimate the entropic
corrections directly within the VMC framework, via phonon calculations
under the quasi-harmonic approximation.  
%
As described in Sec.~{\ref{periodic_system}}, considering 
twist
boundary conditions
 is essential to correct the one-body finite-size errors for
periodic systems, especially in a metallic case. Although the twisted
average approach is commonly used to correct the error, one can also
use the exact special twist (EST) method{\cite{2016DAG}} developed by M. Dagrada
{\it et al.} to save some computational cost. 
EST works even in realistic systems, namely, the
bcc-hydrogen, the bcc-lithium, and the silicon in the high-pressure
$\beta$-tin phase{\cite{2016DAG}}. 


\subsection{One-dimensional chains}
One-dimensional chains of atoms  have been intensively studied based 
on both
effective and {\it ab initio} Hamiltonians, because they embody many
important 
open
questions
in  modern condensed matter physics, despite their simplicity.
The finite or infinite hydrogen chains are  ideal systems for this
purpose, because it is not necessary to use a pseudopotential or to
consider the relativistic effects.
They are therefore benchmark systems.
L. Stella {\it et al.} investigated the metal-insulator transition in
the hydrogen chain by explicitly calculating the complex
polarization{~\cite{2011STE}}, as shown in
Sec.{~\ref{postprocessing}}. They 
found
that the model 
has not a metal-insulator transition and always behaves as an insulating 
1D Hubbard model at half-filling, at least for a large enough interatomic distance ($R>1$).
M. Motta {\it et al.} (Simons Collaboration on the Many-Electron Problem){~\cite{2017MOT}} undertook a comprehensive benchmark study in
the finite and infinite linear hydrogen chains using 
state-of-the-art many-body methods, including VMC and LRDMC with
extrapolations to the thermodynamic and the complete-basis-set
limits. They provide accurate potential energies curves online, which
will be very useful for benchmarking other methods. They have recently
investigated the ground states of the hydrogen-chain more in details,
namely, the insulator-to-metal transition, dimerization, and magnetic
phases{~\cite{2019MOT}}. They revealed a fascinating phase diagram,
with several emergent quantum phases,  depending on the interproton
distance. 


\subsection{Liquid and solid hydrogen/helium}

Hydrogen behavior at high pressures is still not fully understood and subject of intense research. In 1935, Wigner and Huntington predicted its metallization upon compression{~\cite{1935WIG}}, while Ashcroft also proposed its high-$T_{\text{c}}$ superconductivity{~\cite{1968ASH}}. 
Besides its fundamental interest as the simplest realistic condensed matter system in Nature, calculating its equilibrium properties remains of fundamental importance for planetary science applications{~\cite{2005GUI}}.
This system is particularly challenging for DFT-based simulations,  due to the presence of strong correlation effects and a subtle interplay between structural and electronic phase transitions.
A straightforward way to study a phase diagram under pressure and at finite temperature  in QMC is to apply the Langevin molecular dynamics described in Sec.{~\ref{molecular_dynamics}}. First-order phase transitions can be identified from a discontinuous behavior in the calculated equation of state, and of pair correlation functions.
C. Attaccalite and S. Sorella have formulated a second-order LD
suitable for QMC simulation for the first time, as described in
Sec.{~\ref{molecular_dynamics}}.
They
applied it to assess the relative stability of solid and liquid hydrogen at high pressure and room temperature{~\cite{2008ATT}}. 
%
Years later, in a series of works, 
G. Mazzola {\it et al.} applied this technique to study the proposed
liquid-liquid transition (LLT) from a molecular insulating 
to
an atomic conducting 
phase of hydrogen{~\cite{2014MAZ,2015MAZ}}.
G. Mazzola and S. Sorella have also formulated the accelerated
first-order LD{~\cite{2017MAZ}}, as described in
Sec.{~\ref{molecular_dynamics}}, to study such 
an
elusive phase transition with properly equilibrated simulations{~\cite{2017MAZ}}.
Finally, under the same framework the first QMC molecular dynamics simulations
 of an hydrogen-helium mixture has been performed at Jupiter's interior conditions{~\cite{2018MAZ}}. The study revealed that mixing He has a significant influence on the H metallization pressure as shown in Fig.{~{\ref{2018MAZ_H_He}}}, and provided  useful benchmark for the equation of states currently adopted by the planetary science community.
M. Dagrada {\it et al.} revealed that their developed EST technique
also works well in the LD simulation for 
liquid hydrogen, but a perfect agreement  with the twisted-average
technique is achieved only when relatively large supercells are
used{~\cite{2016DAG}}. Therefore, a careful 
finite size scaling analysis
is needed when the EST technique is employed, though
the thermodynamic limit extrapolation is much smoother at the special
twist, saving some computational cost. 
Moreover, EST allows for the
determinant/AGP optimization.

\begin{figure}[htbp]
 \centering
 \includegraphics[width=7.6cm]{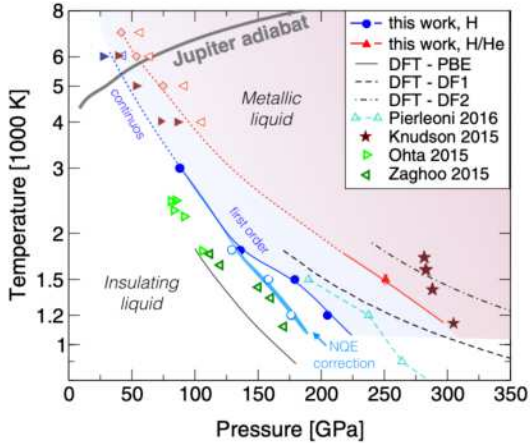}
 \caption{Phase Diagram of dense hydrogen and a hydrogen-helium mixture. ``This work" refers to the results obtained by VMC-Langevin dynamics simulations. The shaded areas represent the liquid-liquid transitions (LLT) between the insulating-molecular and the metallic-atomic fluids. The solid symbols refer to their QMC LLTs for the pure hydrogen (blue circles) and for the hydrogen-helium mixture (red triangle), and the solid blue and red lines indicate first order LLTs. The empty (solid) left (right) triangles correspond to simulations displaying a clear atomic (molecular) behavior, on the other hand, the red diamonds represent an intermediate behavior. A calculated Jupiter's adiabat calculated is shown as a gray line. Pure hydrogen first-order LLT predicted by QMC (cyan) and DFT (solid and dashed lines) are also shown. The light green and dark green triangles and the brown stars represent experimental results. Reprinted with permission from {\citet{2018MAZ}}, {\it Phys. Rev. Lett.}, published by APS in 2018.
 }
 \label{2018MAZ_H_He}
\end{figure}


\subsection{Excited states}
\tvb\, 
allows studying
not only the ground-state electronic 
properties 
but also excited states.
It is trivial to obtain  excitation properties 
when they are characterized by  different particles or 
different
spin projection,  because these quantities are clearly conserved.
For example, M. Barborini {\it et al.} calculated the $^{3}B_{1u}$ $\leftarrow$ $^{1}A_{g}$ vertical triplet excitation of the ethylene molecule and its adiabatic excitation $^{3}A_{1}$ $\leftarrow$ $^{1}A_{g}${\cite{2012BAR}}.
One can also deal with nontrivial cases based on a flexible symmetry-adapted RVB WF within its MO representation, which was proposed by N. Dupuy {\it et al.} in 2015{~\cite{2015DUP}}. Instead of dealing with a linear combination of symmetry-adapted configuration state functions, they construct a symmetry-adapted RVB WF that accurately describes the targeted symmetry.  
They reported that their developed constrained (fixed-rank) minimization of the geminal expansion is stable without symmetry contamination of the starting excited state symmetry; as a result, the vertical ionization energies and the electron affinities of anthracene obtained by VMC and LRDMC calculations agree with the experimental values.

 
\subsection{Connection to experimental spectroscopies}
M. Barborini {\it et al.} proposed a new {\it ab initio} approach to
calculate quasiparticle WFs (QPWF) of isolated systems and 
resolved
in momentum space, based on the correlated sampling technique. They applied the new approach to simulate Scanning Tunneling Microscopy (STM) images{~\cite{2016BAR}} and Angle-Resolved Photoemission Spectroscopy (ARPES){~\cite{2018BAR}}. 
The STM image of [CuCl$_4$]$^{2-}$ obtained by this approach is shown in Fig.~{\ref{2016BAR_STM}}.
It reveals that the electronic correlation has  a significant effect on the
QPWF 
in this system.
Indeed, the resulting
STM image clearly differs from
that obtained by uncorrelated methods ({\it i.e.}, the Hartree-Fock
counterpart).
In general, the electronic calculation should be useful to interpret
the experimental results theoretically.
We believe therefore that
\tvb\ can become useful also for state-of-the-art research in this
field. 

\begin{figure}[htbp]
 \centering
 \includegraphics[width=7.6cm]{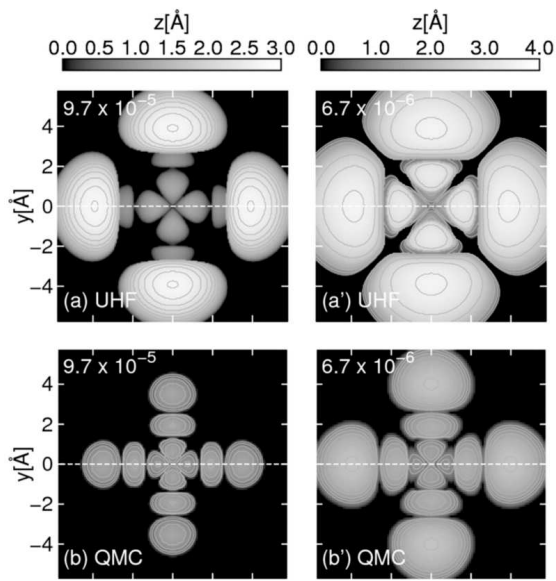}
 \caption{The STM images of in-plane [CuCl$_4$]$^{2-}$ simulated using UHF and QMC calculations. The left column shows the isosurfaces of the square modulus of the eQPWF for a fixed value of probability density of 9.7 $\times$ 10$^{-5}$ \AA$^{-3}$, while in the right column the isosurfaces take the value of 6.7 $\times$ 10$^{-6}$ \AA$^{-3}$. The top panels ($a$ and $a'$) display the LUMO+1 UHF molecular orbitals, while the middle panels ($b$ and $b'$) display the QMC images. Reprinted with permission from {\citet{2016BAR}}, {\it J. Chem. Theory Comput.}, published by ACS in 2016.}
 \label{2016BAR_STM}
\end{figure}


%
%

\section{Python module:\tvbg}
\label{gturbo}
In general, a computational study 
involves
many complicated
operations, such as preparing input files, searching for optimal
parameters, performing benchmark studies, transferring 
information 
coming from previous calculations 
to a subsequent simulation's input file, and analyzing
output files. 
Automatizing
these operations can significantly
save researcher's burden, avoid trivial human errors, and is
beneficial for accelerating the distribution 
of a computational package
to  a much
wider  community. We are now developing a python-based workflow system
suitable for \tvb\, named \tvbg, like Nexus{\cite{2016KRO2}} and
QMC-SW{\cite{2019KON}}.  
The workflow system 
has been 
implemented 
in
Python~3 in an object-oriented fashion, so the modules are highly
extensible. Notice that, since \tvb\, is an all-in-one package, any
other commercial code is not necessary to run the code. There are two
main modules, turbo-genius-serial.py and turbo-genius-sequential.py,
which manage other modules/classes. The former 
allows one
to
perform a single job specified by an option ({\it e.g.}, {\it
  turbo-genius-serial.py -j makefort10}), and the latter 
does a
sequential job ({\it e.g.}, {\it makefort10.x $\to$ convertfort10mol.x
  $\to$ prep.x $\to$ turborvb.x}). 
By using \tvbg,
one can also 
analyze
a simulation output.
For example, 
one can plot
an optimization
history, 
average the
variational parameters after optimization, and
search
for an optimal number of equilibration steps and a
reblocking length.\cite{1989FLY}  
\tvbg\,
is
still at an early stage 
of development.
However, its main features are already there.
In the near future,
the \tvbg\ modules will be capable of
sophisticated \tvb\, jobs automatizations,
as well as setting optimal
parameters depending on 
target systems.

\section{Conclusions}

In this paper, we have reviewed  \tvb, an {\it ab initio} quantum
Monte Carlo package featuring two well established QMC algorithms for
electronic structure: variational Monte Carlo, and diffusion Monte
Carlo. 
The beginning of the \tvb\ development dates back to 2003,  with the
Ph.D thesis project of  M. Casula supervised by S. Sorella.
Since then,
\tvb\ has rapidly 
grown up,
and it has been applied to the study of several molecular and condensed matter systems, from benzene to high-temperature superconducting 
materials.
%
Central to this package are variational parameterizations of correlated electronic WFs in the  product form 
$\Psi  =  \Phi _\text{AS} * \exp J$, 
made of
an antisymmetric part ($\Phi _\text{AS}$), and a Jastrow factor ($J$).

%
%
%
\vspace{1mm}
Concerning the 
antisymmetric part,
\tvb\ implements five 
ans\"atze (in order of decreasing
variational flexibility):
%
$\rm(\hspace{.18em}i\hspace{.18em})$
the Pfaffian (Pf),
$\rm(\hspace{.08em}ii\hspace{.08em})$ 
the Pfaffian with constrained number of molecular orbitals (Pfn),
$\rm(i\hspace{-.08em}i\hspace{-.08em}i)$
the Antisymmetrized Geminal Power (AGP),
$\rm(i\hspace{-.08em}v\hspace{-.06em})$
the Antisymmetrized Geminal Power 
with constrained number of molecular orbitals (AGPn),
and 
$\rm(\hspace{.06em}v\hspace{.06em})$
the single Slater determinant (SD).
%
%
Notably, the user can freely navigate between these ans\"atze.
Indeed,
the
package 
includes
conversion modules, 
which allow the user to choose the most suitable WF form for the
target system.

%
\vspace{1mm}
All the above ans\"atze  can be optimized by  state-of-the-art algorithms, namely, the stochastic reconfiguration and the linear method,
thus achieving highly accurate variational and fixed-node energies,
with a computational cost remaining at the single-determinant level,
thanks to efficient algorithmic developments.

\vspace{1mm}
The 
stochastic 
optimization of  many  variational parameters 
(so far up to the order of $10^5$)
is feasible in \tvb\ thanks to 
an efficient evaluation of energy derivatives using the 
AAD.
This 
algorithmic scheme,
which 
gives the code an original architecture,
drastically
decreases 
the computational cost 
of
ionic forces calculations,
%
keeping them of order $N^3$.
Thus, it 
paves
the way for efficient structural optimizations and molecular dynamics simulations at the VMC level of theory.
Indeed,  by means of the  \tvb\  package, it is at present  possible to
perform QMC molecular dynamics simulations of dense hydrogen (with up
to 256 ions) and liquid water (with up to 64 water molecules), with a
remarkably large  number of atoms, not far from the  DFT state of the
art. 
Thanks to \tvb, nuclear quantum effects are also accessible in a
correlated electronic environment provided by the VMC WF.

\vspace{1mm}
Large scale calculations are possible with \tvb\ not  only because of
cutting-edge algorithmic developments but also 
by virtue of 
an efficient
parallelization 
based on
a hybrid MPI and OpenMP
protocol,  that is 
ideal for employing GPU accelerators 
as well, 
recently available in the most 
advanced  HPC infrastructures.

%
%
\vspace{1mm}
A python wrapper, named \tvbg, has recently been developed to make
the code user-friendly, and to allow both beginners and professional
users to handle it more efficiently.
We will continue developing \tvb\ and \tvbg\ to implement new QMC
algorithms and to extend 
their
range of
applications 
for the current developers and expected new upcoming contributors
and users.
\section{Data availability}
The data that supports the findings of this study are available within this article and the cited references.
\tvb\, is available from S.S.'s web site [https://people.sissa.it/$\sim$sorella/] upon request.

\section{Acknowledgements}
The authors acknowledge all contributors for the development of the
code \tvb\ over the last twenty years. We would especially like to
acknowledge 
very fruitful collaborations with
F. Becca,
G. Carleo,
C. Cavazzoni,
N. Devaux,
N. Dupuy,
L. Guidoni,
H. Hay, 
R. Hlubina,
Y. Iqbal,
R. Maezono,
F. Mouhat,
A. Parola,
K. Seki,
T. Shirakawa,
L. Spanu,
L. Stella,
L.F. Tocchio,
and S. Yunoki.

%
%
\vspace{1mm}
K.N. is grateful for computational resources from PRACE project No.~2019204934 and those from the facilities of Research Center for Advanced Computing Infrastructure at Japan Advanced Institute of Science and Technology (JAIST).
K.N. also acknowledges a financial support from the Simons Foundation and that from Grant-in-Aid for Scientific Research on Innovative Areas (No.~16H06439).
C.A. acknowledges funding from the European Union Seventh Framework Program under grant agreement No.~785219 Graphene Core~2 and COST Action TUMIEE CA17126.
M.C. is grateful to the French grand \'equipement national de calcul
intensif  (GENCI) for the computational time provided through these
years under the project number 0906493, for the funded PRACE projects Nos.
2012061116, 2015133179, 2016163936, and for the access to the Hokusai and
K-computer granted by the Institute of Physical and Chemical Research
(RIKEN).
Y.L. was supported by the Argonne Leadership Computing Facility, which is a U.S. Department of Energy Office of Science User Facility operated under contract DE-AC02-06CH11357.
A.Z.'s work is sponsored by the Air Force Office of Scientific
Research, Air Force Material Command, U.S. Air Force, under Grant
No.~FA9550-19-1-7007.
S.S. acknowledges a financial support from  
PRIN 2017BZPKSZ and computational resources from PRACE project No.~2019204934.
S.S is also grateful for his wife L. Urgias for bearing the too many weekends spent for the development of \tvb.

\vspace{1mm}
We dedicate this paper to the memory of P.W. Anderson, remembering him as one of the most influential contributors to physics and chemistry of the past century, and in particular for deeply inspiring this work with his RVB theory of matter.

\section{References}

\bibliographystyle{apsrev4-1}
\bibliography{./references.bib}

%
%

\end{document}